\documentclass[preprint,preprintnumbers,aps,superscriptaddress,nofootinbib,tightenlines,floatfix]{revtex4}

\usepackage{amsmath,amssymb}
\usepackage{graphicx}
\usepackage{bm}
\usepackage{comment}
\usepackage{color}

\newcommand{\beq}{\begin{equation}}
\newcommand{\eeq}{\end{equation}}
\newcommand{\bea}{\begin{eqnarray}}
\newcommand{\eea}{\end{eqnarray}}

\newcommand{\gsim}{\raisebox{-0.7ex}{$\stackrel{\textstyle >}{\sim}$ }}

\def\OMIT#1{{}}

\def\qslash{q\hskip-0.5em /}

\newcommand{\lsim}{\raisebox{-0.7ex}{$\stackrel{\textstyle <}{\sim}$ }}

\newcount\hour \newcount\hourminute \newcount\minute 
\hour=\time \divide \hour by 60
\hourminute=\hour \multiply \hourminute by 60
\minute=\time \advance \minute by -\hourminute
\newcommand{\mydate}{\ \today \ - \number\hour :\number\minute}

\begin{document}

\begin{figure}[!t]

  \vskip -1.5cm
  \leftline{\includegraphics[width=0.25\textwidth]{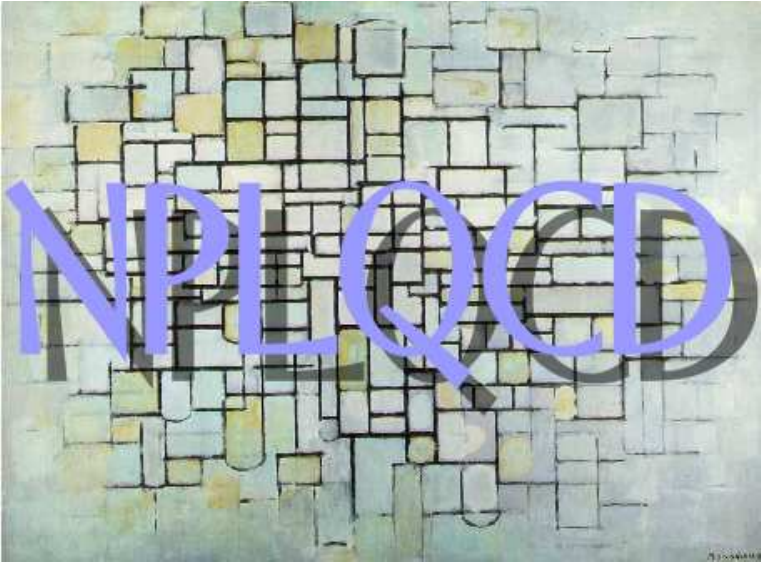}}
\end{figure}

\preprint{\vbox{ 
\hbox{UNH-11-2} 
\hbox{JLAB-THY-11-1364}
\hbox{NT@UW-11-04}
\hbox{IUHET-558}
\hbox{NT-LBNL-11-009}
\hbox{UCB-NPAT-11-006}}}

\title{
High Statistics Analysis using Anisotropic Clover Lattices: (IV)
Volume Dependence of Light Hadron Masses}

\author{S.R.~Beane} 

\affiliation{Albert Einstein Zentrum f\"ur Fundamentale Physik,
Institut f\"ur Theoretische Physik,
Sidlerstrasse 5,
CH-3012 Bern, Switzerland}
\affiliation{Department of Physics, University
  of New Hampshire, Durham, NH 03824-3568, USA}

\author{E.~Chang}
\affiliation{Dept. d'Estructura i Constituents de la Mat\`eria. 
Institut de Ci\`encies del Cosmos (ICC),
Universitat de Barcelona, Mart\'{\i} Franqu\`es 1, E08028-Spain}

\author{W.~Detmold} 
\affiliation{Department of Physics, College of William and Mary, Williamsburg,
  VA 23187-8795, USA}
\affiliation{Jefferson Laboratory, 12000 Jefferson Avenue, 
Newport News, VA 23606, USA}

\author{H.W.~Lin}
\affiliation{Department of Physics,
  University of Washington, Box 351560, Seattle, WA 98195, USA}

\author{T.C.~Luu}
\affiliation{N Division, Lawrence Livermore National Laboratory, Livermore, CA
  94551, USA}

\author{K.~Orginos}
\affiliation{Department of Physics, College of William and Mary, Williamsburg,
  VA 23187-8795, USA}
\affiliation{Jefferson Laboratory, 12000 Jefferson Avenue, 
Newport News, VA 23606, USA}

\author{A.~Parre\~no}
\affiliation{Dept. d'Estructura i Constituents de la Mat\`eria. 
Institut de Ci\`encies del Cosmos (ICC),
Universitat de Barcelona, Mart\'{\i} Franqu\`es 1, E08028-Spain}

\author{M.J.~Savage} \affiliation{Department of Physics,
  University of Washington, Box 351560, Seattle, WA 98195, USA}

\author{A.~Torok} \affiliation{Department of Physics, Indiana University,
  Bloomington, IN 47405, USA}
\author{A.~Walker-Loud}
\affiliation{Lawrence Berkeley National Laboratory, Berkeley, CA 94720, USA}

\collaboration{ NPLQCD Collaboration }

\date{\mydate}

\begin{abstract}
  \noindent The volume dependence of the octet baryon masses and
  relations among them are explored with Lattice QCD.  Calculations
  are performed with $n_f=2+1$ clover fermion discretization in four
  lattice volumes, with spatial extent $L\sim 2.0, 2.5, 3.0$ and
  $3.9~{\rm fm}$, with an anisotropic lattice spacing of $b_s\sim
  0.123~{\rm fm}$ in the spatial direction, and $b_t = b_s/3.5$ in the
  time direction, and at a pion mass of $m_\pi\sim 390~{\rm MeV}$.
  The typical precision of the ground-state baryon mass determination
  is $\lsim 0.2\%$, enabling a precise exploration of the volume
  dependence of the masses, the Gell-Mann--Okubo mass relation, and of
  other mass combinations.  A comparison of the volume dependence with
  the predictions of heavy baryon chiral perturbation theory is
  performed in both the ${\rm SU(2)}_L\otimes {\rm SU(2)}_R$ and ${\rm
    SU(3)}_L\otimes {\rm SU(3)}_R$ expansions.  Predictions of the
  three-flavor expansion for the hadron masses are found to describe
  the observed volume dependences reasonably well.  Further, the
  $\Delta N\pi$ axial coupling constant is extracted from the volume
  dependence of the nucleon mass in the two-flavor expansion, with
  only small modifications in the three-flavor expansion from the
  inclusion of kaons and $\eta$'s.  At a given value of $m_\pi L$, the
  finite-volume contributions to the nucleon mass are predicted to be
  significantly smaller at $m_\pi\sim 140~{\rm MeV}$ than at
  $m_\pi\sim 390~{\rm MeV}$ due to a coefficient that scales as $\sim
  m_\pi^3$.  This is relevant for the design of future ensembles of
  lattice gauge-field configurations.  Finally, the volume dependence
  of the pion and kaon masses are analyzed with two-flavor and
  three-flavor chiral perturbation theory.

\end{abstract}
\maketitle
\tableofcontents
\vfill\eject
%

\section{Introduction}
\label{sec:intro}
\noindent
The calculation of the properties and interactions of light nuclei is
a major goal of Lattice QCD.  While Lattice QCD calculations at the
physical light-quark masses, including strong isospin breaking and
electroweak interactions, are a number of years in the future,
precision calculations of hadron masses are being performed today in
the isospin limit and without electroweak interactions over a range of
light-quark masses.  The masses of the baryons and nuclei are in the
GeV energy regime, but the typical excitation energies and binding
energies found in light nuclei are in the MeV energy regime.  This
hierarchy presents a significant challenge for Lattice QCD
calculations as correlation functions must be determined with
exceptionally high precision in order to obtain statistically
significant energy differences that yield nuclear excitation and
binding energies.

A source of systematic uncertainty in the extraction of scattering
parameters and nuclear binding energies is the volume dependence of
the hadron masses themselves.  Given that the deuteron binding energy
is $B_D\sim 2.2~{\rm MeV}$, an accurate Lattice QCD calculation of
this energy will require that the nucleon mass be known to a precision
of $\Delta M_N\ll 1~{\rm MeV}$.  This includes the contribution from
the finite lattice volume.  Further, the exponential volume
corrections to L\"uscher's eigenvalue
relation~\cite{Hamber:1983vu,Luscher:1986pf,Luscher:1990ux} are also
required to be small~\cite{Bedaque:2006yi,Sato:2007ms}.  In our recent
calculation of the H-dibaryon binding energy~\cite{Beane:2010hg}, the
volume dependence of the $\Lambda$-baryon mass was presented, and it
was concluded that the standard rule-of-thumb, $m_\pi L \gsim 2\pi$,
is in fact necessary at a pion mass of $m_\pi\sim 390~{\rm MeV}$ in
order for the $\Lambda$ finite-volume mass shift to be much smaller
than the observed binding energy.

In this work, which is a continuation of our high statistics Lattice
QCD explorations~\cite{Beane:2009kya,Beane:2009gs,Beane:2009py}, we
present results for the volume dependence of the masses of the baryons
in the lowest-lying SU(3)-flavor octet, and of relations among them,
calculated with four ensembles of $n_f=2+1$ anisotropic clover
gauge-field configurations at a pion mass of $m_\pi\sim 390~{\rm MeV}$
with a spatial lattice spacing of $b_s\sim 0.123~{\rm fm}$, an
anisotropy of $\xi=3.5$ and with cubic volumes of spatial extent
$L\sim 2.0, 2.5, 3.0$ and $3.9~{\rm fm}$.  The volume dependence of
the pion and kaon masses are also determined.  Having the multiple
lattice volumes with all of the other parameters fixed is critical to
fully understanding the volume dependence of the hadron masses and
other quantities.  In particular, lattice-spacing artifacts, which
enter at ${\cal O}(b_s^2)$, are the same in all four ensembles to very
high precision.  The results of the Lattice QCD calculations are
compared with the expectations from next-to-leading-order (NLO) chiral
perturbation theory ($\chi$PT) and heavy-baryon chiral perturbation
theory (HB$\chi$PT) with two and three flavors of active
quarks~\footnote{The leading volume dependence of the hadron masses
  arises at NLO in the chiral expansion.  The leading order (LO)
  contributions to hadron masses from local operators make vanishing
  contributions to the volume dependence.}.  While it is interesting
to compare the calculated volume dependences with the corresponding
expectations from low-energy effective field theories (EFT), perhaps
the most important reason for such a study is in order to plan for the
future production of ensembles of lattice gauge-field configurations.
An interesting result of this study is that while the octet baryons
experience significant and quantifiable finite-volume corrections, the
pion and the kaon, whose masses are determined at the $0.1\%$ level,
show effectively no finite-volume effects. That is, the pions and
kaons are in the infinite-volume regime for the pion mass and the
range of volumes that we explore.  These results are consistent with
the expectations derived from $\chi$PT and HB$\chi$PT.

The first realistic attempt to determine the coefficients of
counterterms in the chiral Lagrangian from the volume dependence of
the nucleon mass was performed in Ref.~\cite{Ali Khan:2003cu}.  The
coefficients in the ${\rm SU(2)}_L\otimes {\rm SU(2)}_R$ chiral
Lagrangian without dynamical $\Delta$'s were constrained by the
results of $n_f=2$ Lattice QCD calculations using the clover
discretization with $m_\pi\gsim 550~{\rm MeV}$ and with lattices of
spatial extent $L\lsim 2.2~{\rm fm}$.  The precision of these
calculations was substantially lower than in the present work,
nonetheless, nontrivial constraints were found on the values of
coefficients in the chiral Lagrangian at NNLO in the expansion by
using phenomenologically determined values for the NLO constants.
These constraints should be viewed only as a demonstration of the
method, given the large pion masses.

This paper is organized as follows. In section~\ref{sec:FVchipt}, we
formulate finite-volume correction formulas for octet-baryon masses to
NLO in ${\rm SU(2)}_L\otimes {\rm SU(2)}_R$ and ${\rm SU(3)}_L\otimes
{\rm SU(3)}_R$ HB$\chi$PT.  Section~\ref{sec:details} gives a concise
description of the specific Lattice QCD calculations that are used in
the present finite-volume study. In section~\ref{sec:Bvol}, we analyze
the octet-baryon finite-volume effects, first (in
subsection~\ref{subsec:Simple}) using a simple, intuitive description,
and then (in subsection~\ref{subsec:TreeLevelChiPT}) using
HB$\chi$PT. In subsection~\ref{subsec:Relations}, various combinations
of baryon masses are likewise analyzed. In section~\ref{sec:Mvol}, we
consider the finite-volume dependence of the pion and kaon masses and
in section~\ref{sec:Conclusions} we conclude.

\section{Finite-Volume Chiral Perturbation Theory }
\label{sec:FVchipt}

\subsection{The Nucleon in  ${\rm SU(2)}_L\otimes {\rm SU(2)}_R$ HB$\chi$PT}

\noindent 
As the Goldstone bosons are the lightest hadrons, $\chi$PT is the
appropriate tool to develop systematic expansions to describe
finite-volume
effects~\cite{Gasser:1986vb,Gasser:1987ah,Gasser:1987zq}.  The crucial
observation is that if the hadronic system is in a sufficiently large
volume, then the infinite-volume chiral Lagrangian can be used to
calculate finite-volume corrections, with no further operators
required~\footnote{ The finite-volume corrections are related to
  forward scattering
  amplitudes~\cite{Luscher:1985dn,Colangelo:2005cg,Colangelo:2010ba}.  For
  instance~\cite{Colangelo:2005cg},
\begin{eqnarray}
M_N(m_\pi, L) - M_N(m_\pi, \infty)& = &
M_N\ 
{3\varepsilon_\pi^2\over 4\pi^2}
\ \sum_{{\bf n}\ne {\bf 0}}\ 
{1  \over  |{\bf n}|   m_\pi  L}\ 
\left[\ 
2\pi\varepsilon_\pi g_{\pi N}^2 
 e^{-|{\bf n}|  m_\pi  L\ \sqrt{1-\varepsilon_\pi^2} }
\right.
\nonumber\\
&& \left.\ 
 -\ 
\int_{-\infty}^\infty\ dy\ \tilde D^+(y)\  e^{-|{\bf n}|  m_\pi  L\
  \sqrt{1+y^2}}
\ \right]
\ +\ {\mathcal O}(e^{-\overline{M} L})
\ \ \ ,
\label{eq:formalvol}
\end{eqnarray}
where $\varepsilon_\pi = m_\pi/(2 M_N)$,
$\overline{M}\geq\sqrt{3/2}m_\pi$ and 
$ \tilde D^+(y) = M_N\ D^+(im_\pi y,0)$,
which is related to 
forward $\pi N$ scattering via
\begin{eqnarray}
T(\pi^a(q)+N(p)\rightarrow \pi^{a^\prime}(q^\prime)+N(p^\prime)) & = & 
\delta_{aa^\prime} T^+\ +\ {1\over 2}\left[\tau_{a^\prime},\tau_a\right]\ T^-
\nonumber\\
T^{\pm} & = & \overline{u^\prime}
\left[\ 
D^{\pm}(\nu,t)\ -\ {1\over 4M_N} \left[\qslash , \qslash\right]\ B^{\pm}(\nu,t)
\ \right] u
\ \ \ .
\label{eq:Tmat}
\end{eqnarray}
The strong coupling between the nucleon and pion is $g_{\pi N}$, which
is related to the axial coupling constant via the chiral expansion
$g_{\pi N} = g_A M_N\sqrt{2}/f_\pi + ...$ where $f_\pi\sim 132~{\rm
  MeV}$.  Evaluating eq.~(\ref{eq:formalvol}) at NLO in HB$\chi$PT
recovers the perturbative result of eq.~(\ref{eq:FVexpMN}).}.  At NLO
in HB$\chi$PT, the finite-volume corrections to the nucleon mass arise
from one-loop self-energy diagrams with nucleon and $\Delta$
intermediate states and, in the limit of exact isospin symmetry, are given by~\cite{Beane:2004tw,Ali
  Khan:2003cu}~\footnote{ The substitutions $w\rightarrow
  \lambda/m_\pi$, $z\rightarrow \Delta/m_\pi$,
  $\overline{\beta}(w,{y\over x})\rightarrow \beta_\Delta/m_\pi$,
  $y\rightarrow \Delta L$, $x\rightarrow m_\pi L$, and
  $F_\Delta^{(FV)}(x,y) \rightarrow -{\cal K}(\Delta)/( 3\pi m_\pi^3)$
  recover the expressions given in Ref.~\cite{Beane:2004tw}.  }
\begin{eqnarray}
\delta M_N
& = & 
M_N(m_\pi, L) - M_N(m_\pi, \infty) \ ;
\nonumber\\
& = & 
{9 m_\pi^3 g_A^2  \over 8\pi f_\pi ^2}\  
F_N^{(FV)}(m_\pi L)
\ +\ 
{m_\pi^3 g_{\Delta N\pi}^2  \over \pi f_\pi ^2}\  
F_\Delta^{(FV)}(m_\pi L, \Delta_{\Delta N} L) \ ,
\label{eq:FVexpMN}
\end{eqnarray}
where the loop functions are defined to be
\begin{eqnarray}
F_N^{(FV)}(x) & = &   \frac{1}{6}\sum_{{\bf n}\ne {\bf 0}}\ { e^{-|{\bf n}|  x } \over
  |{\bf n}| x}
\ =\ 
{e^{-x}\over x}\ +\ 2 \ {e^{-\sqrt{2}x}\over \sqrt{2} x}\ +\
{4\over 3} \ {e^{-\sqrt{3}x}\over \sqrt{3} x}\ +\ ... \  ;
\label{eq:FVexpAFdef1}\\
F_\Delta^{(FV)}(x, y) & = &    
{1\over {3\pi}}\sum_{{\bf n}\ne {\bf 0}}\ 
\int_0^\infty dw \ 
\overline\beta(w,{y\over x})\ 
\left[ 
 \overline\beta(w,{y\over x})\ K_0\left( |{\bf n}| x  \overline\beta(w,{y\over x}) \right)
 - 
{ K_1\left( |{\bf n}|  x \overline\beta(w,{y\over x}) \right)\over  |{\bf n}| x}
\right] ;
\nonumber\\
\label{eq:FVexpAFdef2}
\end{eqnarray}
where $\overline\beta(w,z)\ = \ \sqrt{ w^2 + 2 z w + 1}$,
$\Delta_{\Delta N} = M_\Delta - M_N$, and the $K_n(z)$ are modified
Bessel functions.  In the limit $\Delta\rightarrow 0$,
$F_\Delta^{(FV)}(m_\pi L, \Delta L)\rightarrow F_N^{(FV)}(m_\pi L)$.
For asymptotically large lattice volumes, it is expected that only the
leading contributions in the sums in eq.~(\ref{eq:FVexpAFdef1}) and
eq.~(\ref{eq:FVexpAFdef2}) will be relevant.

\subsection{The Hyperons in  ${\rm SU(2)}_L\otimes {\rm SU(2)}_R$ HB$\chi$PT}

\noindent It is straightforward to compute the finite-volume
corrections to the hyperon masses at NLO in  ${\rm SU(2)}_L\otimes {\rm SU(2)}_R$
HB$\chi$PT. One finds, in a generalization of eq.~(\ref{eq:FVexpMN}),
that
\begin{eqnarray}
\delta M_\Lambda
& = & 
{3 \ g_{\Lambda\Sigma}^2\  m_\pi^3\over 8\pi f_\pi^2} 
F_\Delta^{(FV)}( m_\pi L, \Delta_{\Sigma\Lambda} L)
\ +\ 
{3  g_{\Sigma^*\Lambda\pi}^2 \ m_\pi^3\over 2\pi f_\pi^2} 
F_\Delta^{(FV)}( m_\pi L, \Delta_{\Sigma^*\Lambda} L)\ ;
\label{eq:FVLam}
\\
\delta M_\Sigma
& = & 
{g_{\Lambda\Sigma}^2\  m_\pi^3\over 8\pi f_\pi^2} 
F_\Delta^{(FV)}( m_\pi L, \Delta_{\Lambda\Sigma} L)
\  + \
{3  g_{\Sigma\Sigma}^2\  m_\pi^3\over 4\pi f_\pi^2} 
F_N^{(FV)}( m_\pi L) \nonumber\\
&&\qquad\qquad\qquad\qquad\qquad\qquad
\  + \
{g_{\Sigma^*\Sigma\pi}^2 \ m_\pi^3\over 2\pi f_\pi^2} 
F_\Delta^{(FV)}( m_\pi L,\Delta_{\Sigma^*\Sigma} L) \ ;
\label{eq:FVSig}
\\
\delta M_\Xi
& = & 
{9  g_{\Xi\Xi}^2\  m_\pi^3\over 8\pi f_\pi^2} F_N^{(FV)}( m_\pi L)
\ +\ 
{3  g_{\Xi^*\Xi\pi}^2 \ m_\pi^3\over 4\pi f_\pi^2} F_\Delta^{(FV)}( m_\pi L,
\Delta_{\Xi^*\Xi} L) \ ,
\label{eq:FVXi}
\end{eqnarray}
where $\Delta_{AB}\equiv M_B - M_A$.  For definitions of the various
axial couplings in terms of chiral Lagrangian operators, see
Ref.~\cite{Tiburzi:2008bk}.  In this formulation of hyperon
finite-volume effects, the contributions from kaon and $\eta$ loops
are in the coefficients of local operators, and therefore do not
contribute until higher orders in the chiral expansion (as the
finite-volume effects arise from pion loops). This implies that for
quark masses such that $m_\pi< m_K$ (but not $m_\pi\ll m_K$) there is
a region where the two-flavor chiral expansion of infinite-volume
quantities will converge but finite-volume effects will not be
accounted for systematically in the two-flavor expansion.

\subsection{The Baryon Octet in  ${\rm SU(3)}_L\otimes {\rm SU(3)}_R$ HB$\chi$PT}

\noindent 
In addition to the contributions to the volume dependence from higher
orders in the ${\rm SU(2)}_L\otimes {\rm SU(2)}_R$ chiral expansion,
there are contributions from quantum fluctuations of the nucleon into
strange hadrons, and of the hyperons into other members of the baryon
octet.  For instance, in addition to the $\pi N$ and $\pi\Delta$
intermediate states that give finite volume contributions to the
nucleon mass, intermediate states such as $\Lambda K$ or $N\eta$ also
contribute.  Such fluctuations give rise to a volume dependence that
scales as $\sim m_K^2 \ e^{-m_K L}/L$ or $\sim m_\eta^2 \ e^{-m_\eta
  L}/L$.  In the Lattice calculations presented in this paper,
$m_K/m_\pi\sim 1.4$, and as a result, such contributions are naively
expected to be of the same magnitude as the $\sim m_\pi^2 \
e^{-\sqrt{2} m_\pi L}/L$ contributions, which are suppressed compared
with the leading volume dependence.  As the axial-couplings between
the nucleons and pions are of order one, and the couplings to strange
intermediate states are generically small, such strange contributions
are expected to be small. The explicit contributions to the
octet-baryon mass shifts, written in terms of the SU(3)-symmetric
axial couplings $D$, $F$ and ${\cal C}$~\cite{Jenkins:1991ts} are for
the nucleon,
\begin{eqnarray}
\delta M_N^{(K,\eta)} 
& = & 
(D-F)^2
\  {9 m_K^3\over 8\pi f_K^2}\ 
F^{(FV)}_\Delta (m_K L, \Delta_{\Sigma N} L)
\ +\ 
(D+3F)^2
\  {m_K^3\over 8\pi f_K^2}\ 
F^{(FV)}_\Delta (m_K L, \Delta_{\Lambda N} L)
\nonumber\\ 
& + & 
(D-3F)^2
\  {m_\eta^3\over 8\pi f_\eta^2}\ 
F^{(FV)}_\Delta (m_\eta L, 0)
\ +\ 
{\cal C}^2
\  {m_K^3\over 4\pi f_K^2}\ 
F^{(FV)}_\Delta (m_K L, \Delta_{\Sigma^* N} L)
\ \ \ , 
\label{eq:FVNsu3}
\end{eqnarray}
and for the hyperons,
\begin{eqnarray}
\delta M_\Lambda^{(K,\eta)} 
& = & 
(D+3F)^2 {m_K^3\over 4\pi f_K^2}  \ 
F_\Delta^{(FV)}( m_K L, \Delta_{N\Lambda} L)
\ +\ 
(D-3F)^2 { m_K^3\over 4\pi f_K^2}  \ 
F_\Delta^{(FV)}( m_K L, \Delta_{\Xi\Lambda} L)
\nonumber\\
& + & 
D^2 {m_\eta^3\over 2\pi f_\eta^2}  \ 
F_\Delta^{(FV)}( m_\eta L, 0)
\ + \ 
{\cal C}^2 {m_K^3\over 2\pi f_K^2}  \ 
F_\Delta^{(FV)}( m_K L, \Delta_{\Xi^*\Lambda} L)
\ \ \ ,  \label{eq:FVLamsu3}
\end{eqnarray}
\begin{eqnarray}
\delta M_\Sigma^{(K,\eta)} 
& = & 
(D-F)^2 {3  m_K^3\over 4\pi f_K^2}  \ 
F_\Delta^{(FV)}( m_K L, \Delta_{N\Sigma} L)
\ +\ 
(D+F)^2 { 3 m_K^3\over 4\pi f_K^2}  \ 
F_\Delta^{(FV)}( m_K L, \Delta_{\Xi\Sigma} L)
\nonumber\\
& + & 
D^2 {m_\eta^3\over 2\pi f_\eta^2}  \ 
F_\Delta^{(FV)}( m_\eta L, 0)
\ +\ 
{\cal C}^2 {m_\eta^3\over 4\pi f_\eta^2}  \ 
F_\Delta^{(FV)}( m_\eta L, \Delta_{\Sigma^*\Sigma} L)
\nonumber\\
& + & 
{\cal C}^2 {2 m_K^3\over 3 \pi f_K^2}  \ 
F_\Delta^{(FV)}( m_K L, \Delta_{\Delta\Sigma} L)
\ +\ 
{\cal C}^2 {m_K^3\over 6\pi f_K^2}  \ 
F_\Delta^{(FV)}( m_K L, \Delta_{\Xi^*\Sigma} L)
\ \ \ ,  \label{eq:FVSigsu3}
\end{eqnarray}
\begin{eqnarray}
\delta M_\Xi^{(K,\eta)} 
& = & 
(D+F)^2 {9  m_K^3\over 8\pi f_K^2}  \ 
F_\Delta^{(FV)}( m_K L, \Delta_{\Sigma\Xi} L)
\ +\ 
(D-3F)^2 { m_K^3\over 8\pi f_K^2}  \ 
F_\Delta^{(FV)}( m_K L, \Delta_{\Lambda\Xi} L)
\nonumber\\
& + & 
(D+3 F)^2 {m_\eta^3\over 8\pi f_\eta^2}  \ 
F_\Delta^{(FV)}( m_\eta L, 0)
\ +\ 
{\cal C}^2 {m_\eta^3\over 4\pi f_\eta^2}  \ 
F_\Delta^{(FV)}( m_\eta L, \Delta_{\Xi^*\Xi} L)
\nonumber\\
& + & 
{\cal C}^2 {m_K^3\over 4\pi f_K^2}  \ 
F_\Delta^{(FV)}( m_K L, \Delta_{\Sigma^*\Xi} L)
\ +\ 
{\cal C}^2 {m_K^3\over 2\pi f_K^2}  \ 
F_\Delta^{(FV)}( m_K L, \Delta_{\Omega\Xi} L)
\ \ \ . \label{eq:FVXisu3}
\end{eqnarray}
In the limit of exact SU(3) symmetry, the SU(2) axial couplings introduced
above are related to the SU(3) couplings via:
\begin{eqnarray} 
&& g_A  =  D+F
\ \ ,\ \ 
g_{\Delta N \pi}={\cal C}
\ \ ,\ \ 
g_{\Lambda\Sigma} = 2 D
\ \ \ ,\ \ \
g_{\Sigma^*\Lambda\pi} = {\cal C}/\sqrt{2} \ ,
\nonumber\\
&& 
g_{\Sigma\Sigma}  =  2F
\ \ \ ,\ \ \ \ 
g_{\Sigma^*\Sigma\pi}  =  {\cal C}/\sqrt{3}
\ \ \ ,\ \ \ \ 
g_{\Xi\Xi}=D-F
\ \ \ ,\ \ \ 
g_{\Xi^*\Xi\pi}={\cal C}/\sqrt{3}
\ \ \ .
\label{eq:su3couplings}
\end{eqnarray}
Adding the finite-volume modifications 
in eqs.~(\ref{eq:FVNsu3})-(\ref{eq:FVXisu3})
to those in eq.~(\ref{eq:FVexpMN}) 
and eqs.~(\ref{eq:FVLam})-(\ref{eq:FVXi}) gives the full 
NLO  ${\rm SU(3)}_L\otimes {\rm SU(3)}_R$ HB$\chi$PT finite-volume
contributions to the baryons in the lowest-lying octet.

\section{Details of the Lattice QCD Calculation}
\label{sec:details}
\noindent Anisotropic gauge-field configurations have proven useful
for the study of hadronic
spectroscopy~\cite{Dudek:2009qf,Bulava:2009jb,Lin:2008pr,Edwards:2008ja},
and, as the calculations required for studying multi-hadron systems
rely heavily on spectroscopy, we have put considerable effort into
calculations using ensembles of gauge fields with clover-improved
Wilson fermion actions with anisotropic lattice spacing that have been
generated by the Hadron Spectrum Collaboration (HSC).  In particular,
the $n_f=2+1$ flavor anisotropic clover Wilson
action~\cite{Okamoto:2001jb,Chen:2000ej} with stout-link
smearing~\cite{Morningstar:2003gk} of the spatial gauge fields in the
fermion action with a smearing weight of $\rho=0.14$ and $n_\rho=2$
has been used.  The gauge fields entering the fermion action are not
smeared in the time direction, thus preserving the ultra-locality of
the action in the time direction.  Further, a tree-level
tadpole-improved Symanzik gauge action without a $1\times 2$ rectangle
in the time direction is used. 

The present calculations are performed on four ensembles of
gauge-field configurations with $L^3\times T$ of $16^3\times 128$,
$20^3\times 128$, $24^3\times 128$ and $32^3\times 256$ lattice sites,
with a renormalized anisotropy $\xi=b_s/b_t=3.5$ where $b_s$ and $b_t$
are the spatial and temporal lattice spacings, respectively. The
spatial lattice spacing of each ensemble is $b_s = 0.1227\pm
0.0008~{\rm fm}$~\cite{Lin:2008pr} giving spatial lattice extents of
$L\sim 2.0, 2.5, 3.0$ and $3.9~{\rm fm}$ respectively.  The same input
light-quark mass parameters, $b_t m_l = -0.0840$ and $b_t m_s =
-0.0743$, are used in the production of each ensemble, giving a pion
mass of $m_\pi\sim 390~{\rm MeV}$.  The relevant quantities to assign
to each ensemble that determine the impact of the finite lattice
volume are $m_\pi L$ and $m_\pi T$, which for the four ensembles are
$m_\pi L \sim 3.86, 4.82, 5.79$ and $7.71$ respectively, and $m_\pi T
\sim 8.82, 8.82, 8.82$ and $17.64$ respectively.

Multiple light-quark propagators were calculated on each configuration
in the four ensembles. The source locations were chosen randomly in an
effort to minimize correlations among propagators.  On the $\{
16^3\times 128$, $20^3\times 128$, $24^3\times 128$, $32^3\times
256\}$ ensembles, an average of $\{224$, $364$, $180$, $174\}$
propagators were calculated on each of $\{2001$, $1195$, $2215$,
$774\}$ gauge-field configurations, to give a total number of $\sim
\{4.5$, $4.3$, $3.9$, $1.3 \}\times 10^5$ light-quark propagators,
respectively.
\begin{figure}[!ht]
  \centering
     \includegraphics[width=0.49\textwidth]{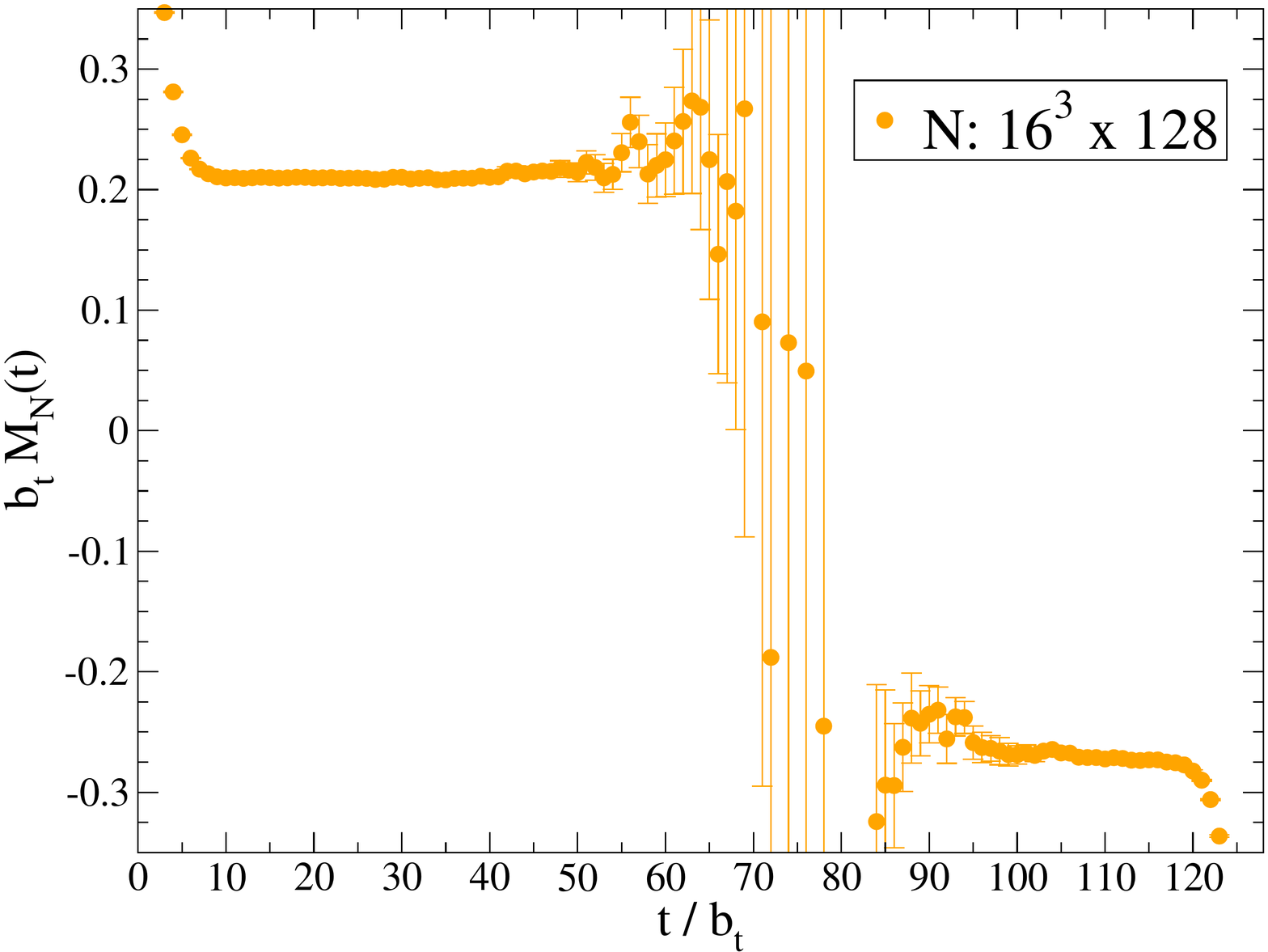}
     \includegraphics[width=0.49\textwidth]{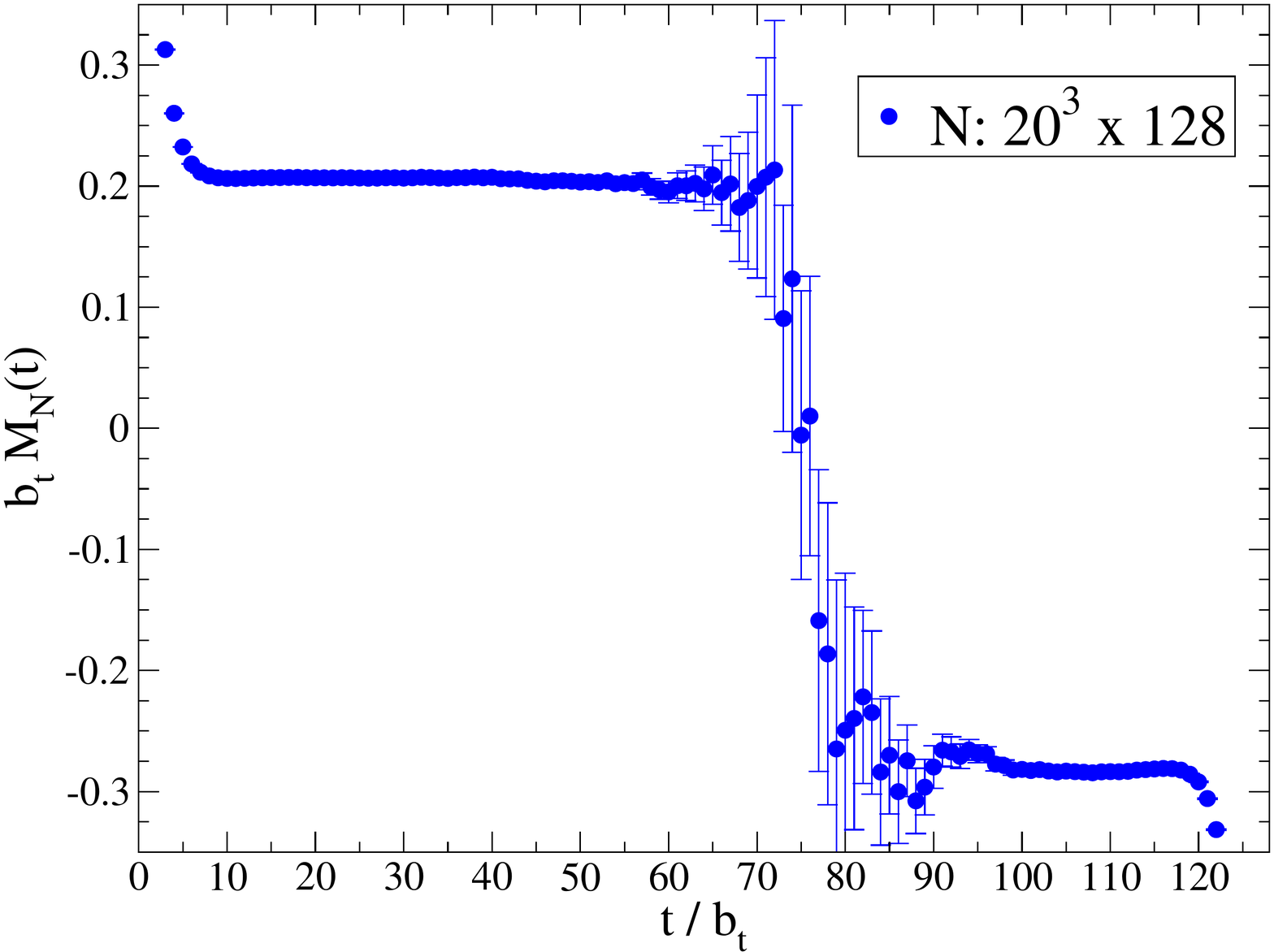}\\
     \includegraphics[width=0.49\textwidth]{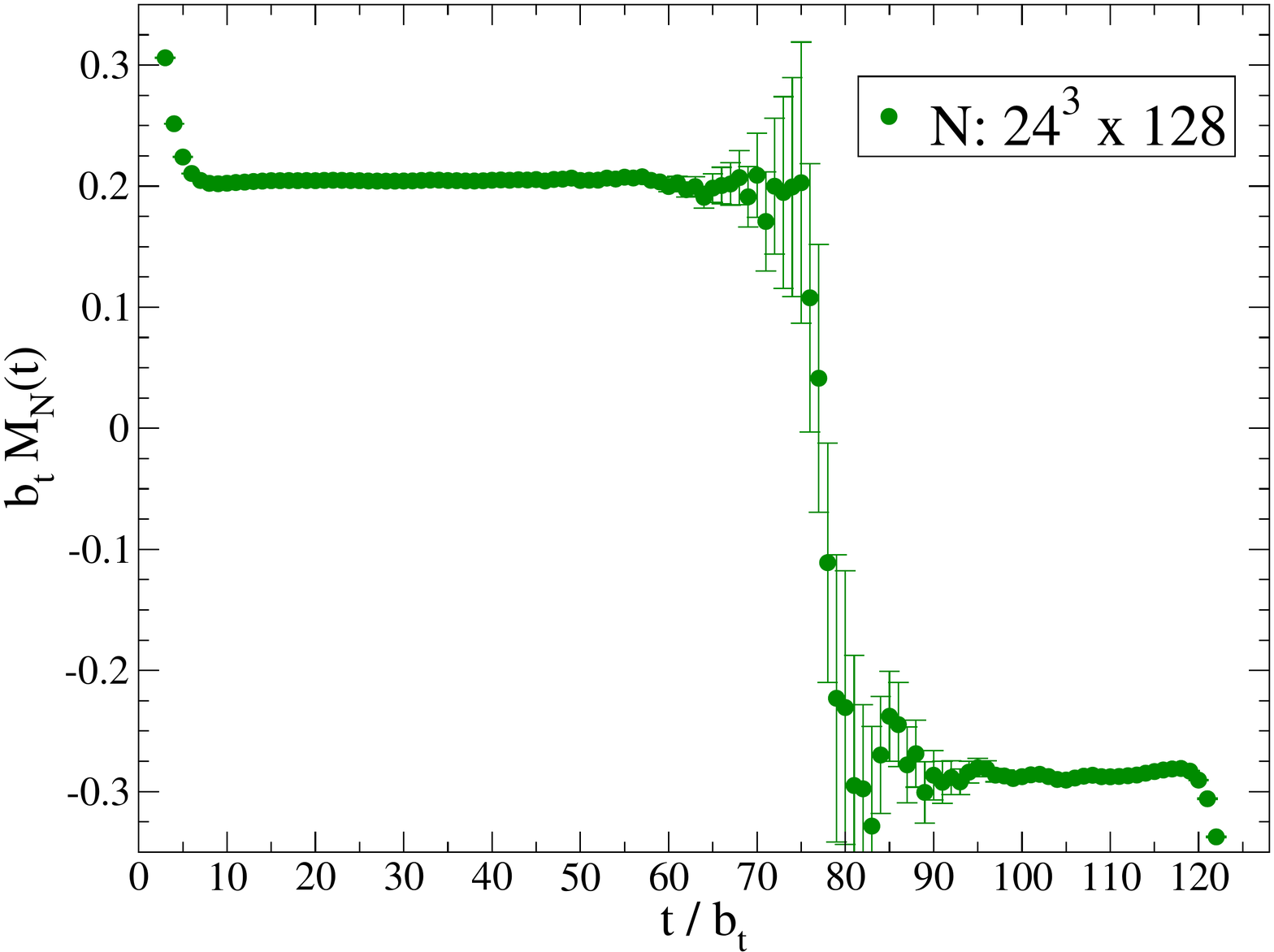}
     \includegraphics[width=0.49\textwidth]{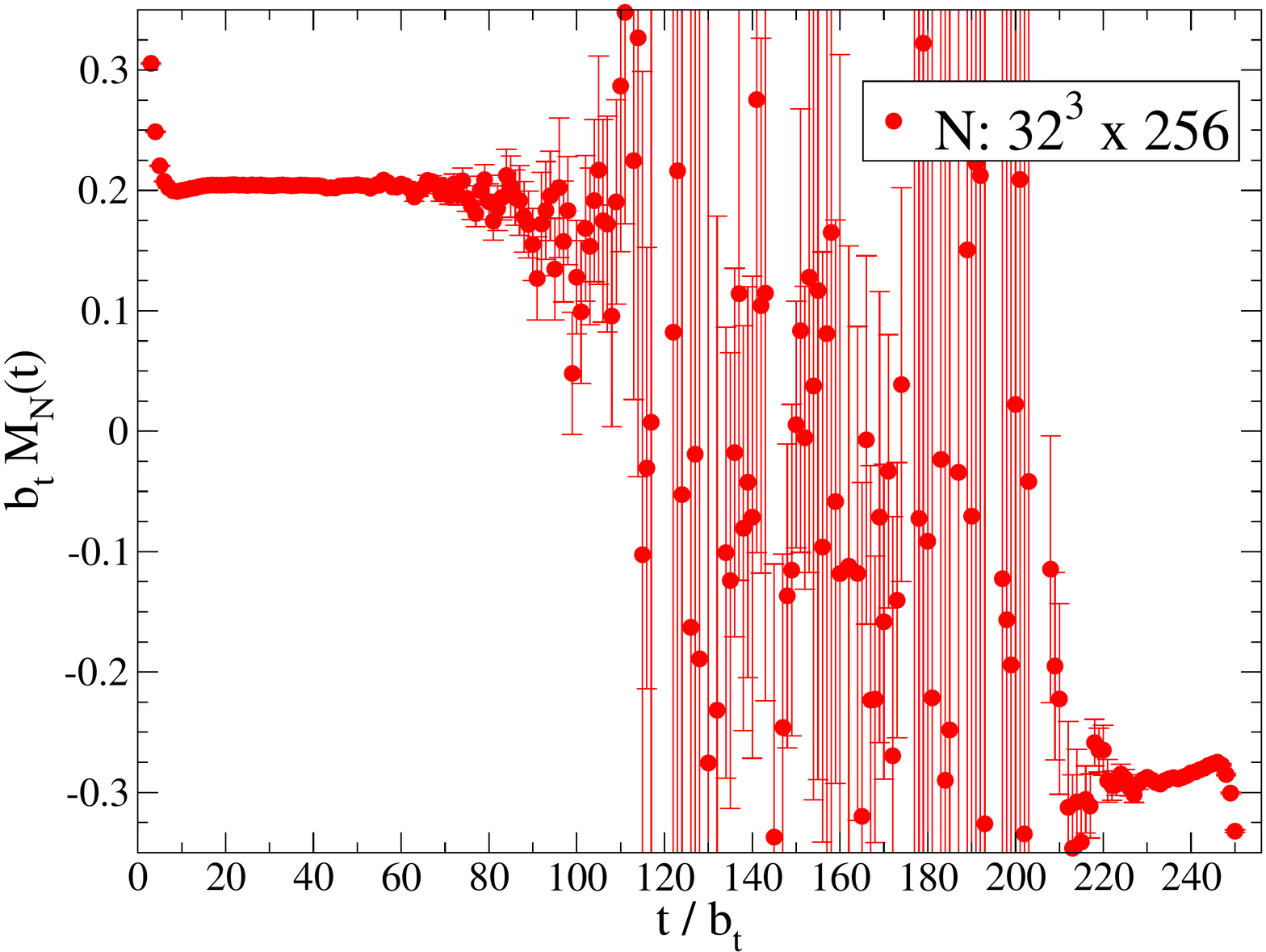}
     \caption{The nucleon EMP's obtained in the four lattice 
volumes.
Note that the temporal extent of the $32^3\times 256$ ensemble is twice that of
the other three ensembles.
}
  \label{fig:empNfull}
\end{figure}
The EMP's for the nucleon on all of the time slices of
each lattice ensemble are shown in
fig.~\ref{fig:empNfull}.
They provide an indication of the precision of the
present calculations.

\section{The Volume Dependence of the Baryon Masses}
\label{sec:Bvol}
\noindent
The baryon masses that were extracted from 
the Lattice QCD calculations in the four different lattice
volumes are given in table~\ref{tab:LQCDbaryonmasses}. A detailed discussion of
the fitting methods used 
in the analysis of the correlation functions
is given in Refs.~\cite{Beane:2009kya,Beane:2009gs,Beane:2009py,Beane:2010em}.
\begin{table}[!ht]
  \caption{Results from the Lattice QCD calculations in the 
four lattice volumes.
  }
  \label{tab:LQCDbaryonmasses}
  \begin{ruledtabular}
    \begin{tabular}{c||cccc}
      $L^3\times T$  &  $16^3\times 128$ &  $20^3\times 128$ &  $24^3\times 128$ &
      $32^3\times 256$  \\
      \hline
      $L~({\rm fm})$ & $\sim$ 2.0 &  $\sim$2.5 &  $\sim$3.0 &  $\sim$3.9 \\
      $m_\pi L$ & 3.888(20)(01) & 4.8552(84)(35) & 5.799(16)(04) & 7.7347(74)(91)\\
      $e^{-m_\pi L}$ & $\sim $0.0205 & $\sim $0.0078 & $\sim $0.0030 & $\sim $0.00044\\
      ${1\over m_\pi L}\ e^{-m_\pi L}$ & $\sim $$5.3\times 10^{-3}$ & $\sim $$1.6\times
      10^{-3}$ & $\sim $$5.2\times 10^{-4}$ & $\sim $$5.7\times 10^{-5}$\\
      ${1\over (m_\pi L)^{3/2}}\ e^{-m_\pi L}$ & $\sim $$2.7\times 10^{-3}$ & $\sim $$7.4\times
      10^{-4}$ & $\sim $$2.2\times 10^{-4}$ & $\sim $$2.1\times 10^{-5}$\\
      $m_\pi T$ & 8.89(16)(01) & 8.878(54)(22) & 8.836(85)(02) & 17.679(59)(73)\\
      $e^{-m_\pi T}$ & $\sim $$1.38\times 10^{-4}$   & $\sim $$1.39\times 10^{-4}$   & 
$\sim $$1.45\times 10^{-4}$ & $\sim $$2.10\times 10^{-8}$    \\
\hline
      $M_N$ (t.l.u.) &  0.21004(44)(85) & 0.20682(34)(45) & 0.20463(27)(36) & 0.20457(25)(38)\\
      $M_\Lambda$ (t.l.u.)  &  0.22446(45)(78) & 0.22246(27)(38) & 0.22074(20)(42) & 0.22054(23)(31)\\
      $M_\Sigma$ (t.l.u.)  & 0.22861(38)(67) & 0.22752(32)(43) & 0.22791(24)(31) & 0.22726(24)(43)  \\
      $M_\Xi$ (t.l.u.) & 0.24192(38)(63) & 0.24101(27)(38) & 0.23975(20)(32) & 0.23974(17)(31) \\
      \hline
    \end{tabular}
  \end{ruledtabular}
\end{table}

\subsection{A Simplistic Analysis}
\label{subsec:Simple}

\begin{figure}[!ht]
  \centering
     \includegraphics[width=0.49\textwidth]{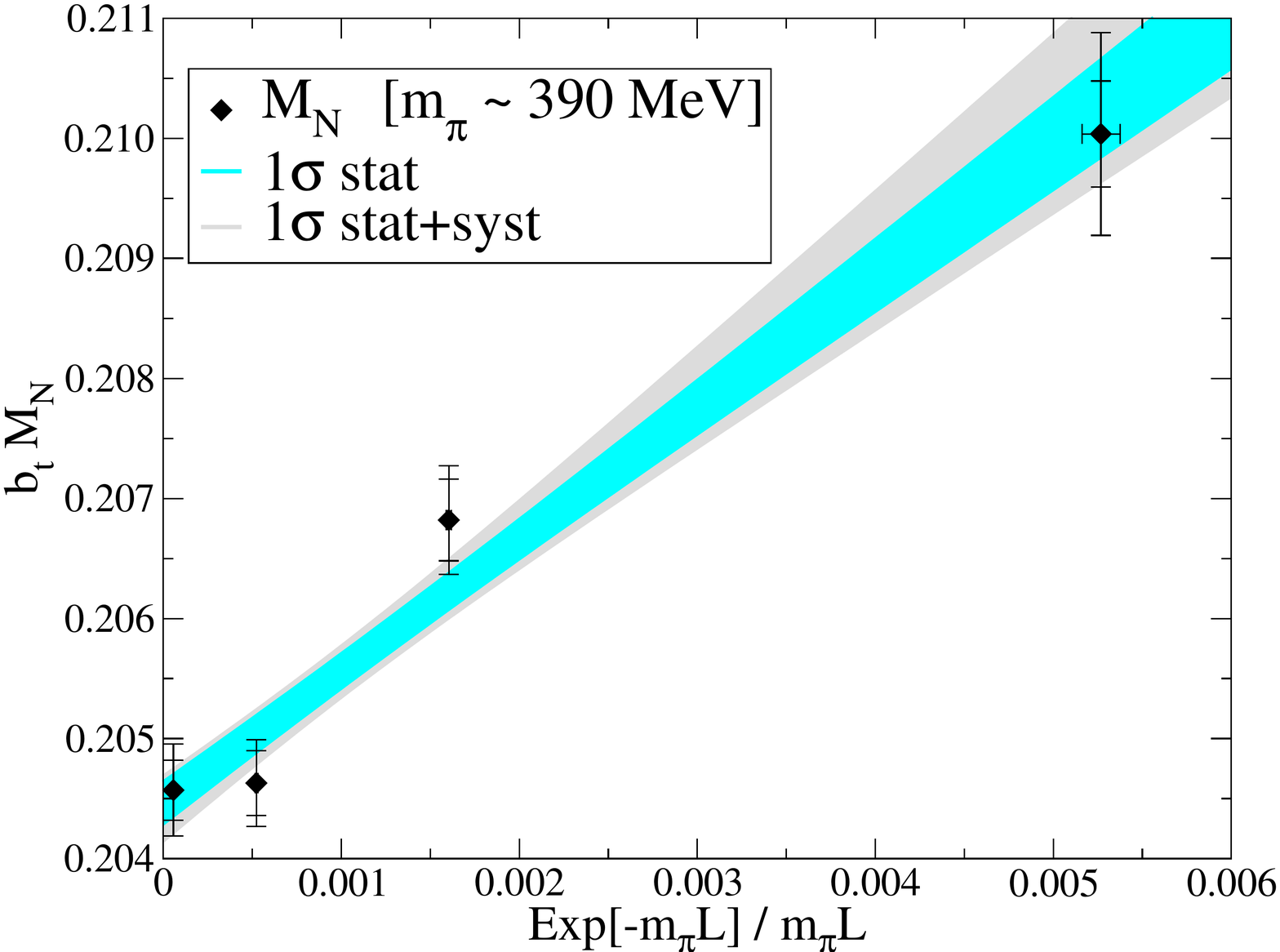}\ 
\includegraphics[width=0.49\textwidth]{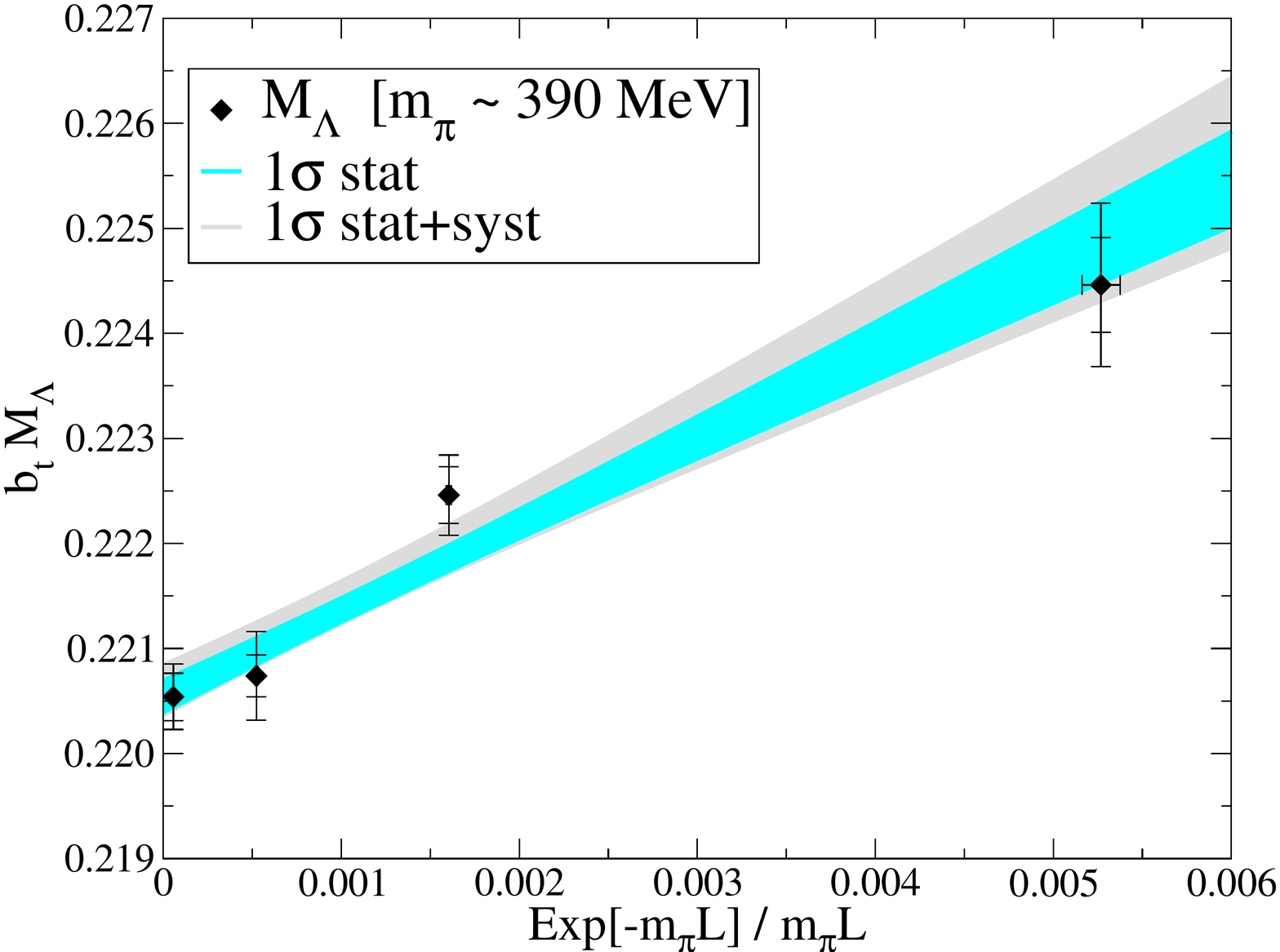} \\
     \includegraphics[width=0.49\textwidth]{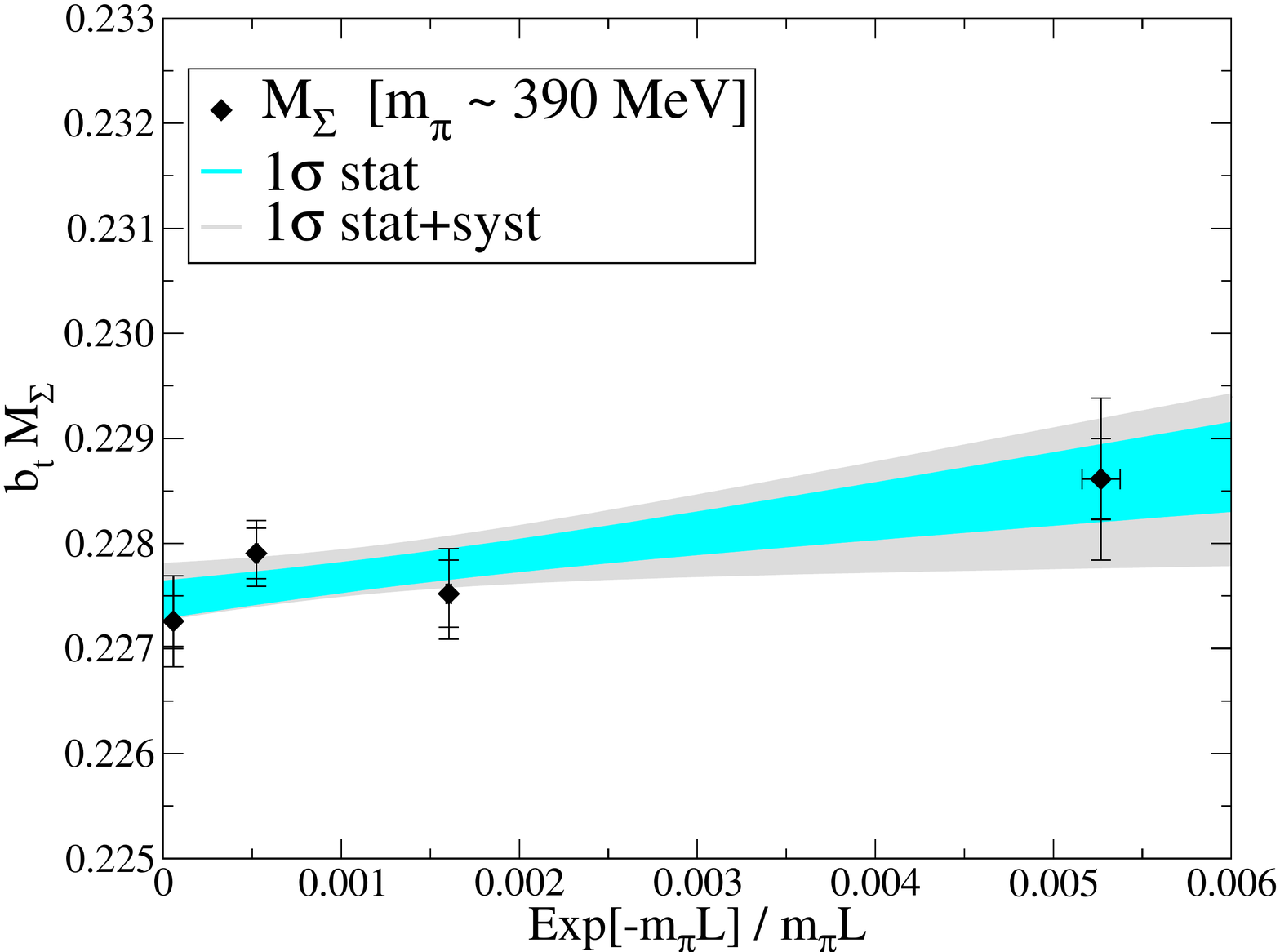}\ 
\includegraphics[width=0.49\textwidth]{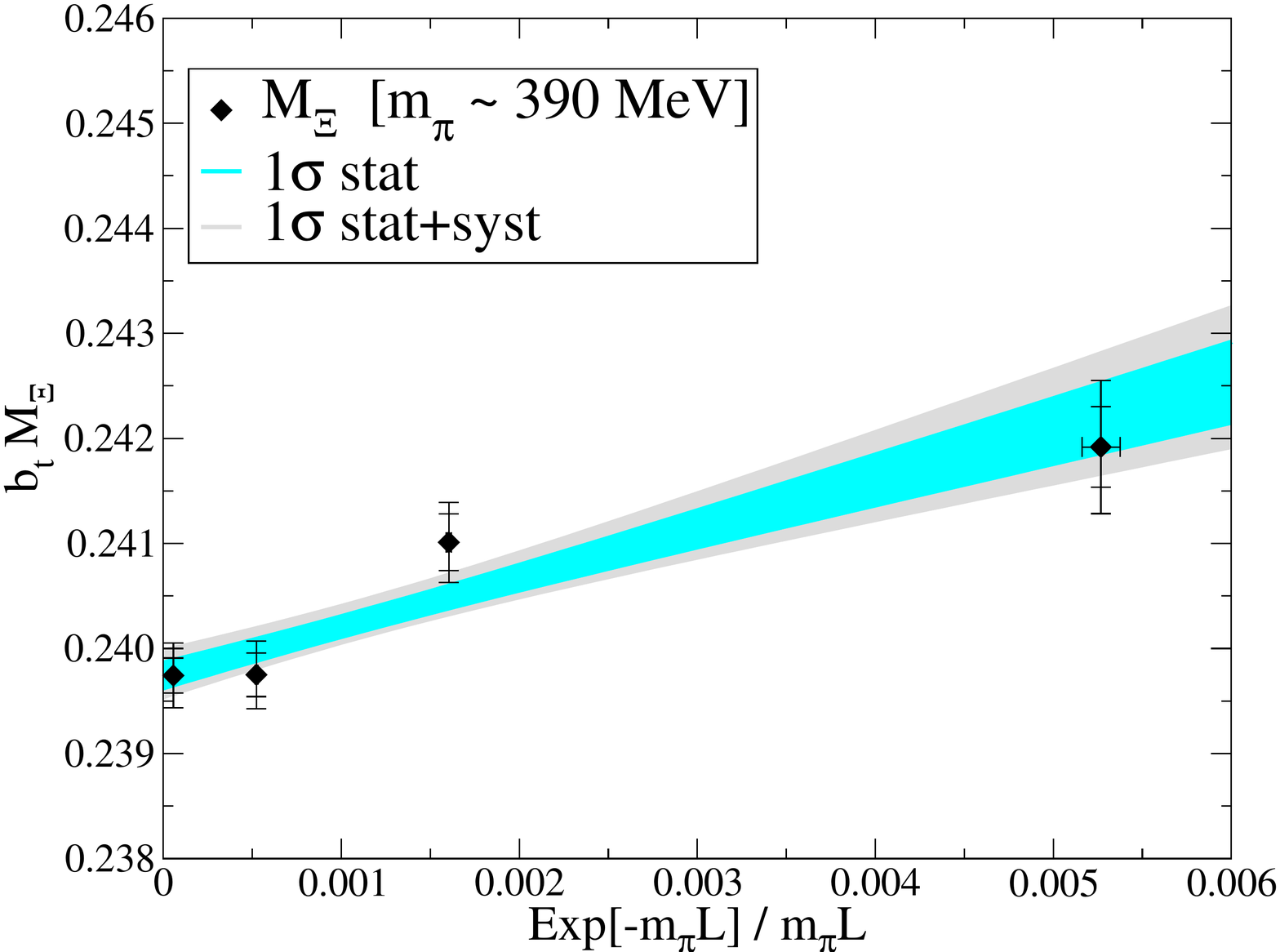}\ 
\caption{ The mass of the nucleon (upper left panel), the $\Lambda$
  (upper right panel),
the $\Sigma$ (lower left panel) and the $\Xi$ (lower right
  panel)
as a function of $e^{-m_\pi L}/(m_\pi L)$.  The points
  and associated uncertainties (blue) are the results of the Lattice
  QCD calculations.  The dark (light) shaded region corresponds to the
  $1\sigma$ statistical uncertainty (statistical and systematic
  uncertainties combined in quadrature) associated with a fit of the
  form given in eq.~(\protect\ref{eq:FVsimplefit}).  }
  \label{fig:NvolLamvol}
\end{figure}

\begin{figure}[!ht]
  \centering
     \includegraphics[width=0.65\textwidth]{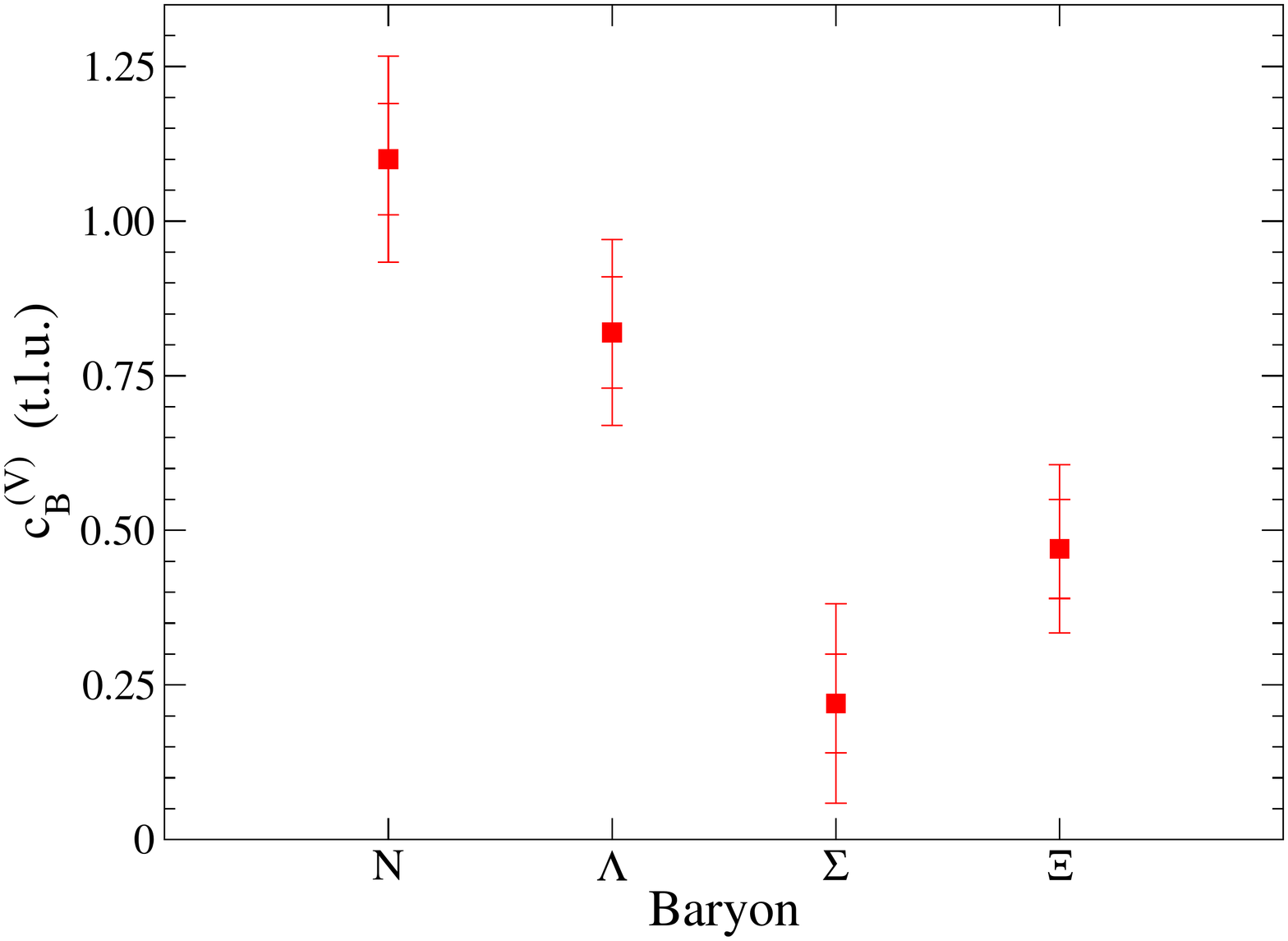}
     \caption{ The fit values of  $c_B^{(V)}$, the coefficient of 
 $e^{-m_\pi L}/(m_\pi L)$ in eq.~(\ref{eq:FVsimplefit}), given in 
table~\ref{tab:baryonmassesExtrap} for each of the octet baryons.
}
  \label{fig:cBV}
\end{figure}

\noindent It is useful to begin the analysis of the volume dependence
of the baryon masses by performing a straightforward, but only partly
motivated, fit to results of the Lattice QCD calculations given in
table~\ref{tab:LQCDbaryonmasses}. As shown previously, the volume
dependence of the mass of a given baryon can be calculated
order-by-order in HB$\chi$PT.  The formally-leading contribution to
the volume dependence of the mass of an octet baryon results from a
one-loop diagram involving a pion and an octet baryon (ignoring for
the moment the contribution from decuplet baryons).  These one-loop
contributions give rise to a volume dependence of the form given in
eq.~(\ref{eq:FVexpAFdef1}), $F_N^{(FV)}(m_\pi L)$.  In obtaining this
result, it is assumed that $m_\pi L$ is large, but significantly
smaller than $m_X L$ where $m_X$ is the mass of other mesons, such as
the kaon and the $\eta$, i.e. $m_K, m_\eta \gg m_\pi$.  In the
very-large volume limit, the finite-volume contributions are dominated
by the first term in eq.~(\ref{eq:FVexpAFdef1}).  As such, it is
useful, as a preliminary analysis, to fit a function of the form
\begin{eqnarray}
M_B^{(V)} (m_\pi L) & = & M_B^{(\infty)}\ +\ 
c_B^{(V)}\ {e^{- m_\pi  L}\over m_\pi L}
\label{eq:FVsimplefit}
\end{eqnarray}
to the results of the Lattice QCD calculations given in
table~\ref{tab:LQCDbaryonmasses}.  One should view the parameter
$c_B^{(V)}$ as providing an estimate of the strength of the axial
coupling between the pion and the baryon.  It should be stressed that
the higher-order terms, beginning with terms of order $e^{- \sqrt{2}\
  m_\pi L}/( m_\pi L)$, give a non-negligible contribution (relative
to the uncertainties) in the $16^3\times 128$ lattice volume, and a
fit using the full function in eq.~(\ref{eq:FVexpAFdef1}) leads to
slightly reduced values of $c_B^{(V)}$ compared to those determined in
the fits to eq.~(\ref{eq:FVsimplefit}).  The fits to each of the
baryon masses are shown, along with the results of the Lattice QCD
calculations, in fig.~\ref{fig:NvolLamvol}.  The same vertical scale
(but different intervals) has been used in the plots in
fig.~\ref{fig:NvolLamvol} in order for the reader to easily determine
the relative size of the volume dependence of each of the masses.
The values of the infinite-volume masses, $ M_B^{(\infty)}$, and the coefficients of the
leading volume dependences, $c_B^{(V)}$, are presented in table~\ref{tab:baryonmassesExtrap}.
\begin{table}[!ht]
  \caption{The results of linear fits, 
of the form given in eq.~(\protect\ref{eq:FVsimplefit}),
to the volume dependence of the baryon masses.
$M_B^{(\infty)}$ is the infinite-volume extrapolation of the baryon mass and 
$c_B^{(V)}$ is the coefficient of $e^{-m_\pi L}/(m_\pi L)$.  
The first uncertainty is statistical, the second is the fitting
systematic, and the third (where appropriate) is due to scale setting.
  }
  \label{tab:baryonmassesExtrap}
  \begin{ruledtabular}
    \begin{tabular}{c||cccc}
      Hadron  &  $M_B^{(\infty)}$ (t.l.u.) & $M_B^{(\infty)}$ (MeV) &  $c_B^{(V)}$
      (t.l.u.) & $c_B^{(V)}$  (MeV)   \\
      \hline
      $M_N$  &  0.20427(17)(19) & 1149.8(1.0)(1.1)(7.5)& 1.15(09)(14) &
      $6.47(51)(78)(04)\times 10^3$\\
      $M_\Lambda$ &  0.22053(15)(21) & 1241.2(0.9)(1.1)(8.1)& 0.83(09)(12) &
      $4.64(53)(69)(03)\times 10^3$\\
      $M_\Sigma$ &  0.22744(17)(22) & 1280.3(1.0)(1.1)(8.3)& 0.21(09)(13)&
      $1.19(48)(71)(08)\times 10^3 $\\
      $M_\Xi$ &  0.23972(13)(18) & 1349.4(0.8)(1.1)(8.8)& 0.47(08)(11) &
      $2.62(44)(60)(02)\times 10^3$\\
      \hline
    \end{tabular}
  \end{ruledtabular}
\end{table}
The nucleon is found to have the largest volume
dependence. 
As the nucleon is comprised of light valence quarks only, 
it is expected to couple most strongly to pions, 
which should dominate its finite-volume modifications in the
large-volume limit.  It is expected that baryons with more strange
quarks exhibit less volume sensitivity, and  
that the finite-volume mass shifts to the baryons, $\delta(FV)_B$,
should naively satisfy the hierarchy
\begin{eqnarray}
\delta(FV)_N\ >\ 
\delta(FV)_\Sigma\ ,\ 
\delta(FV)_\Lambda\ >\ 
\delta(FV)_\Xi\
\ \ \ .
\label{eq:expected}
\end{eqnarray}
The fit coefficients, $c_B^{(V)}$, given in
table~\ref{tab:baryonmassesExtrap} are shown in fig.~\ref{fig:cBV},
where the expected hierarchy is approximately observed within the
uncertainties of the calculation.  The volume dependence of the
$\Sigma$ is somewhat smaller than naive expectations would suggest.

\subsection{The Octet Baryons  with NLO HB$\chi$PT}
\label{subsec:TreeLevelChiPT}

\noindent In this section we explore both ${\rm SU(2)}_L\otimes {\rm
  SU(2)}_R$ and ${\rm SU(3)}_L\otimes {\rm SU(3)}_R$ HB$\chi$PT
predictions and fits to the results of the Lattice QCD calculations.
The analyses are performed at NLO in the chiral expansion;
unfortunately they do not provide significant constraints on the
counterterms that appear beyond NLO in HB$\chi$PT. Our strategy in
these analyses is to use the octet-octet axial couplings and the
octet-decuplet mass splittings from experimental data and Lattice QCD
results, and fit the octet-decuplet axial couplings and the baryon
masses in the infinite-volume limit to the results of the Lattice QCD
calculations, given in table~\ref{tab:LQCDbaryonmasses}, using
two-flavor HB$\chi$PT. Inserting these fit values into the full
three-flavor finite-volume corrections gives a measure of the
relevance of kaon and $\eta$ loops.  The goal is to determine the
extent to which two- and three-flavor HB$\chi$PT describe the volume
dependence of the results of the Lattice QCD calculations.  And, of
course, it is of interest to determine whether any significant
constraints can be placed on the --with few exceptions, rather poorly
known-- axial coupling constants of the baryons by studying
finite-volume effects.

\subsubsection{Parameter Set}
\label{subsec:ParSet}
\noindent 
Lattice QCD calculations show that the nucleon axial coupling,
$g_A$, is essentially independent of the light-quark 
masses~\cite{Edwards:2005ym,Yamazaki:2008py,Gockeler:2011ze,Alexandrou:2010hf}, and so in
the following we use the experimentally-determined value $g_A=1.26$ as well
as 
the central values of 
$g_{\Lambda\Sigma}$, $g_{\Sigma\Sigma}$, and $g_{\Xi\Xi}$
determined from Lattice QCD calculations~\cite{Lin:2007ap}
interpolated to the appropriate pion mass, or from the tree-level SU(3)
relations between axial couplings at the physical pion mass
(which are consistent with each other)
\begin{eqnarray}
&& 
g_{\Lambda\Sigma} = 1.58(20)
\ \ ,\ \ 
g_{\Sigma\Sigma} = 0.900(30)
\ \ ,\ \ 
g_{\Xi\Xi}= 0.262(13)
\ \ \ .
\label{eq:axialcouplings}
\end{eqnarray}
While the pion decay constant, $f_\pi$, is
experimentally determined to be $f_\pi\sim 132~{\rm MeV}$ at the physical
light-quark masses, Lattice QCD
calculations have determined how it depends upon the light-quark masses,
and at $m_\pi\sim 390~{\rm MeV}$ its value is 
$f_\pi\sim 150~{\rm  MeV}$~\cite{Bazavov:2009ir,Dimopoulos:2009zza,Allton:2008pn,Beane:2006kx}.
We take the baryon mass splittings determined from octet and decuplet correlation functions calculated on
the $32^3\times 256$ ensemble\footnote{We do not
quote uncertainties on these determinations as they do not significantly affect the NLO HB$\chi$PT fits.}:
\begin{eqnarray}
&& \Delta_{\Delta N}=298~{\rm  MeV} \ \ , \ \
\Delta_{\Sigma N}=128~{\rm MeV} \ \ , \ \
\Delta_{\Lambda N}=90~{\rm MeV} \ \ , \ \
\Delta_{\Sigma^* N}=427~{\rm MeV} \ ; \nonumber \\
&&
\Delta_{\Sigma\Lambda} = 38~{\rm MeV} \ \ , \ \
\Delta_{\Sigma^*\Lambda} = 336~{\rm MeV}   \ \ , \ \
\Delta_{\Xi\Lambda}=108~{\rm MeV}  \ \ , \ \
\Delta_{\Xi^*\Lambda}=406~{\rm MeV}  \ ; \nonumber \\
&&
\Delta_{\Xi\Sigma}=69~{\rm MeV}  \ \ , \ \
\Delta_{\Sigma^*\Sigma} = 298~{\rm MeV}\ \ , \ \
\Delta_{\Delta\Sigma} = 229~{\rm MeV}  \ \ , \ \
\Delta_{\Xi^*\Sigma}=368~{\rm MeV}\ ; \nonumber \\
&&
\Delta_{\Xi^*\Xi}=298~{\rm MeV}  \ \ , \ \
\Delta_{\Sigma^*\Xi}=228~{\rm MeV}  \ \ , \ \
\Delta_{\Omega\Xi}=367~{\rm MeV} \ .
\label{eq:ODmasssplit}
\end{eqnarray}

For the  ${\rm SU(3)}_L\otimes {\rm SU(3)}_R$ HB$\chi$PT analysis, the 
SU(3)-symmetric axial couplings are fixed to the central values of the 
best-fit experimental values: $D=0.79$, $F=0.47$ and ${\cal
  C}=1.47$~\cite{FloresMendieta:1998ii}.
Further, the decay constants and masses are set to
$f_K=f_\eta=160~{\rm MeV}$ and the lattice-determined
values $m_K=544~{\rm MeV}$, and $m_\eta=587~{\rm MeV}$ (the latter determined
from $m_\pi$ and $m_K$ via the Gell-Mann--Okubo (GMO) relation), respectively.

\subsubsection{The Nucleon Mass}
\label{subsec:NucleonMass}
\begin{figure}[!ht]
  \centering
     \includegraphics[width=0.49\textwidth]{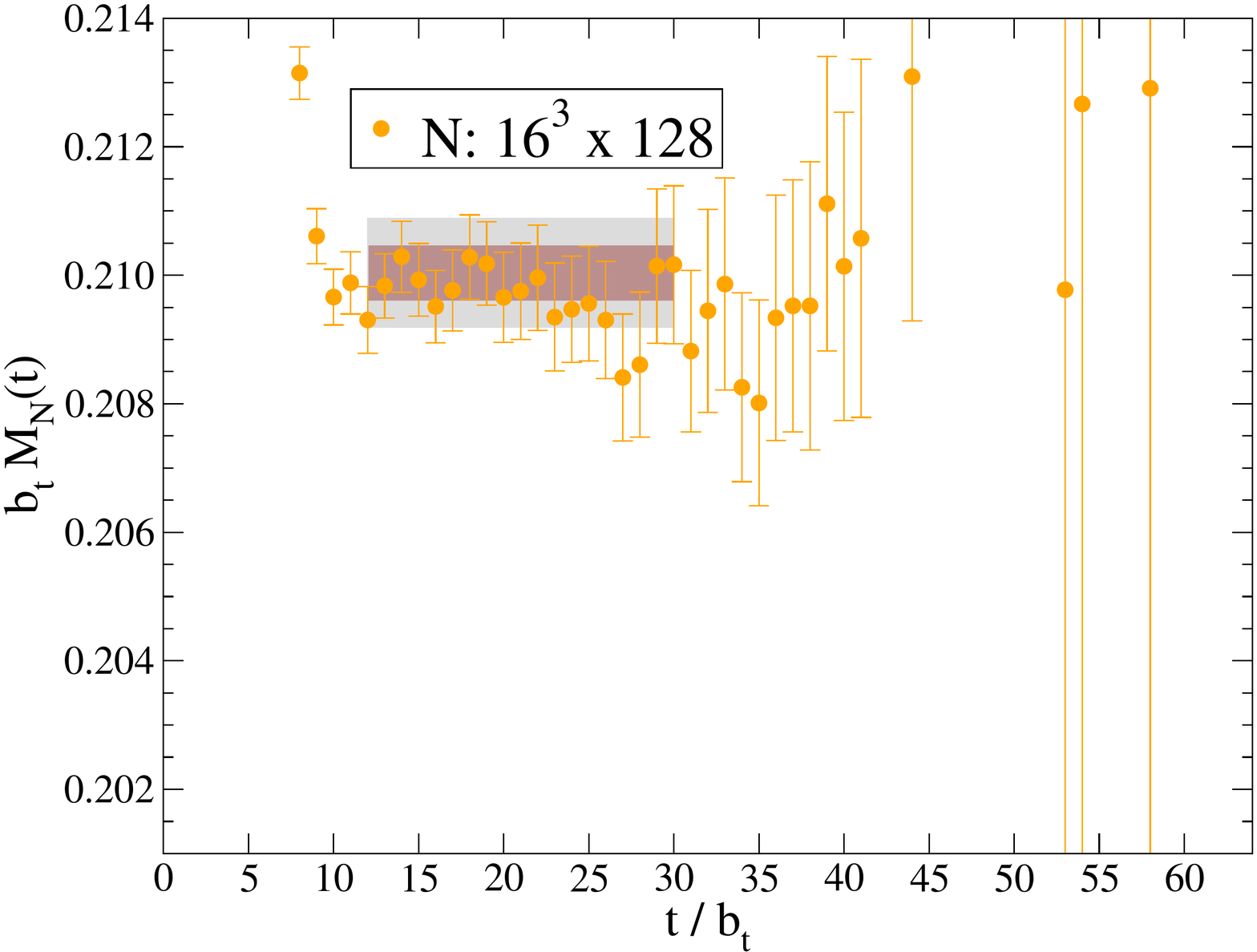} 
     \includegraphics[width=0.49\textwidth]{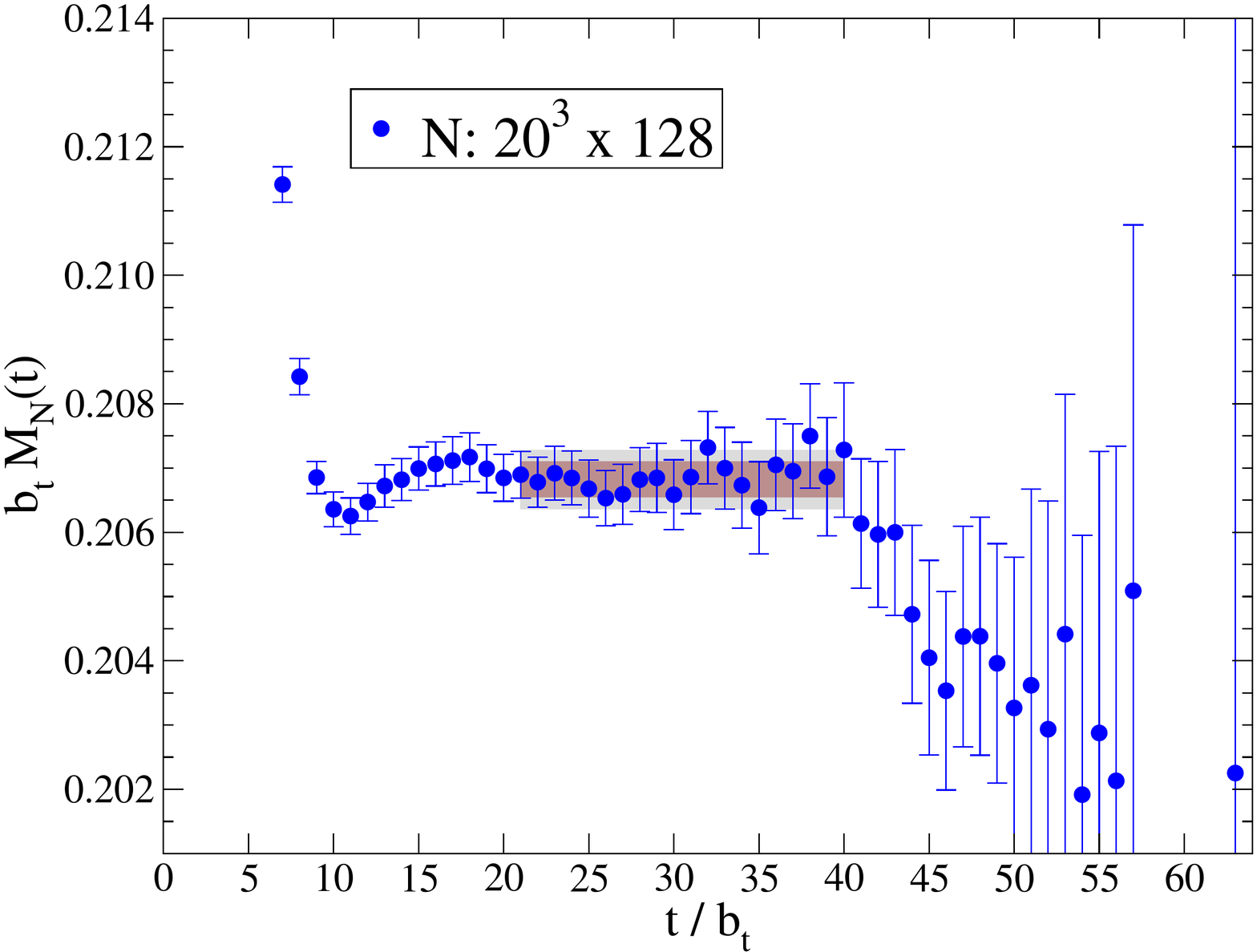} \\
     \includegraphics[width=0.49\textwidth]{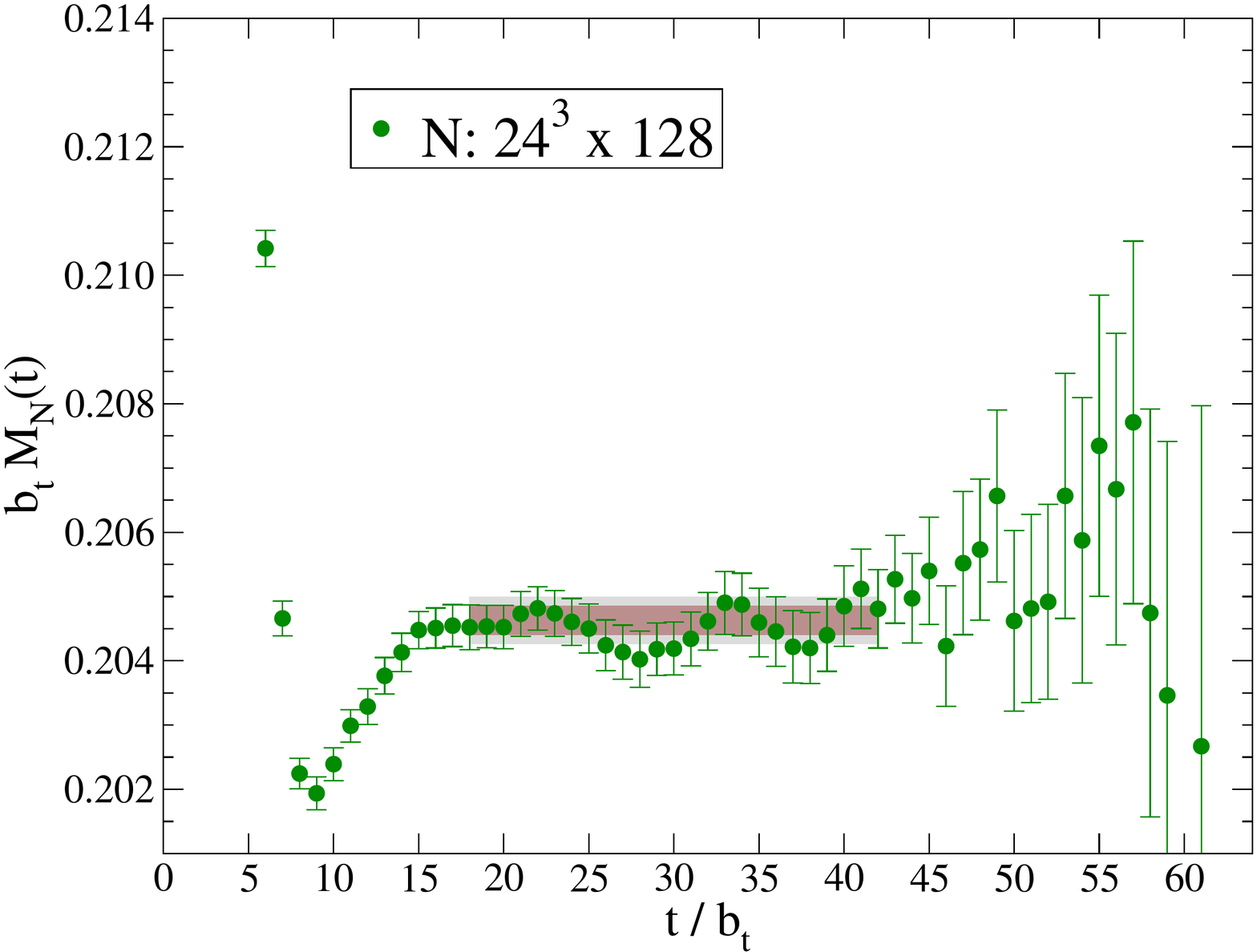} 
     \includegraphics[width=0.49\textwidth]{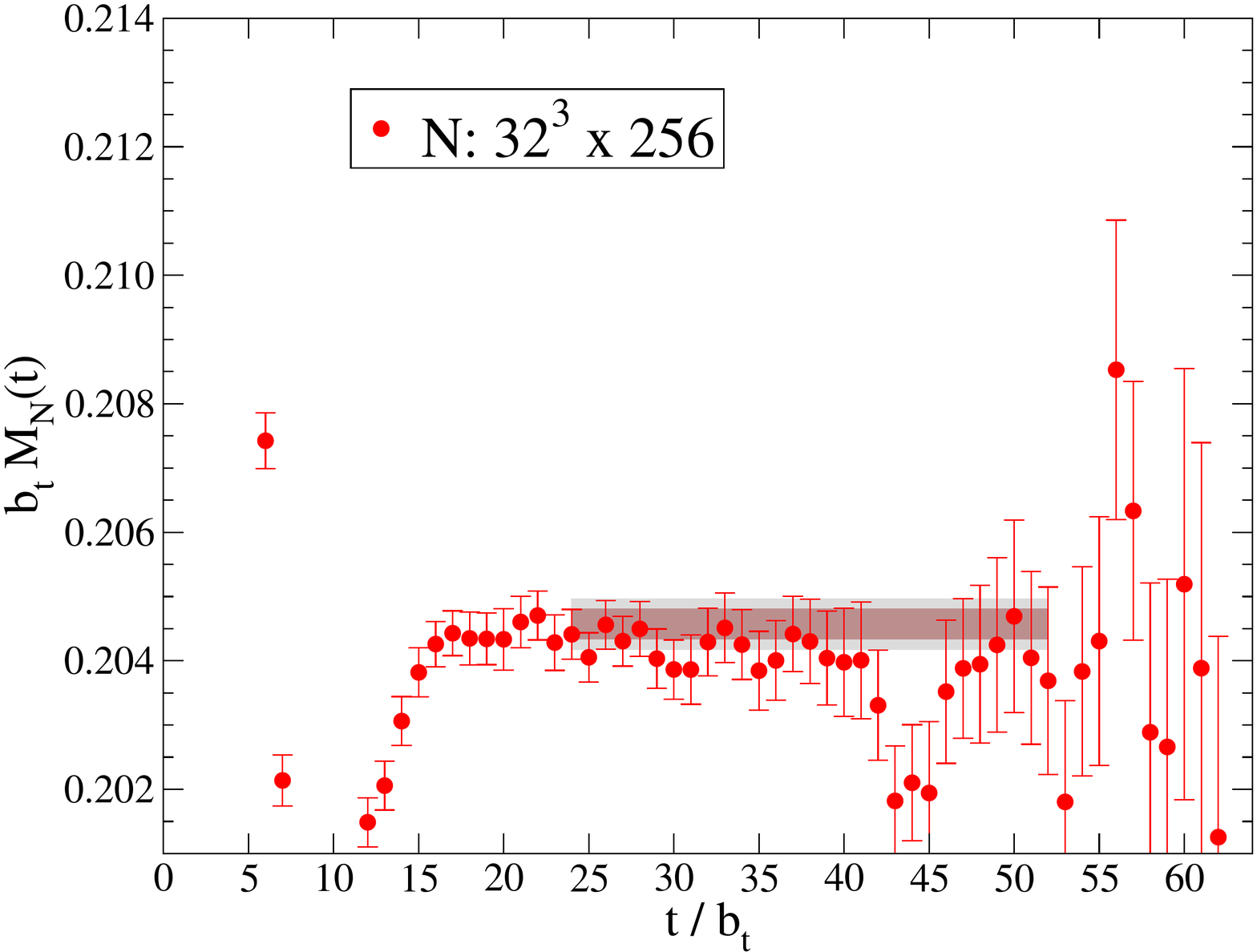}
     \caption{The Nucleon EMP's determined 
on the four lattice ensembles used in this work. 
They each result from linear combinations of different correlation functions that 
optimize the plateau of the ground state.
Note that the y-axis scale is the same in all four panels.}
  \label{fig:empN}
\end{figure}
\begin{figure}[!ht]
  \centering
     \includegraphics[width=0.9\textwidth]{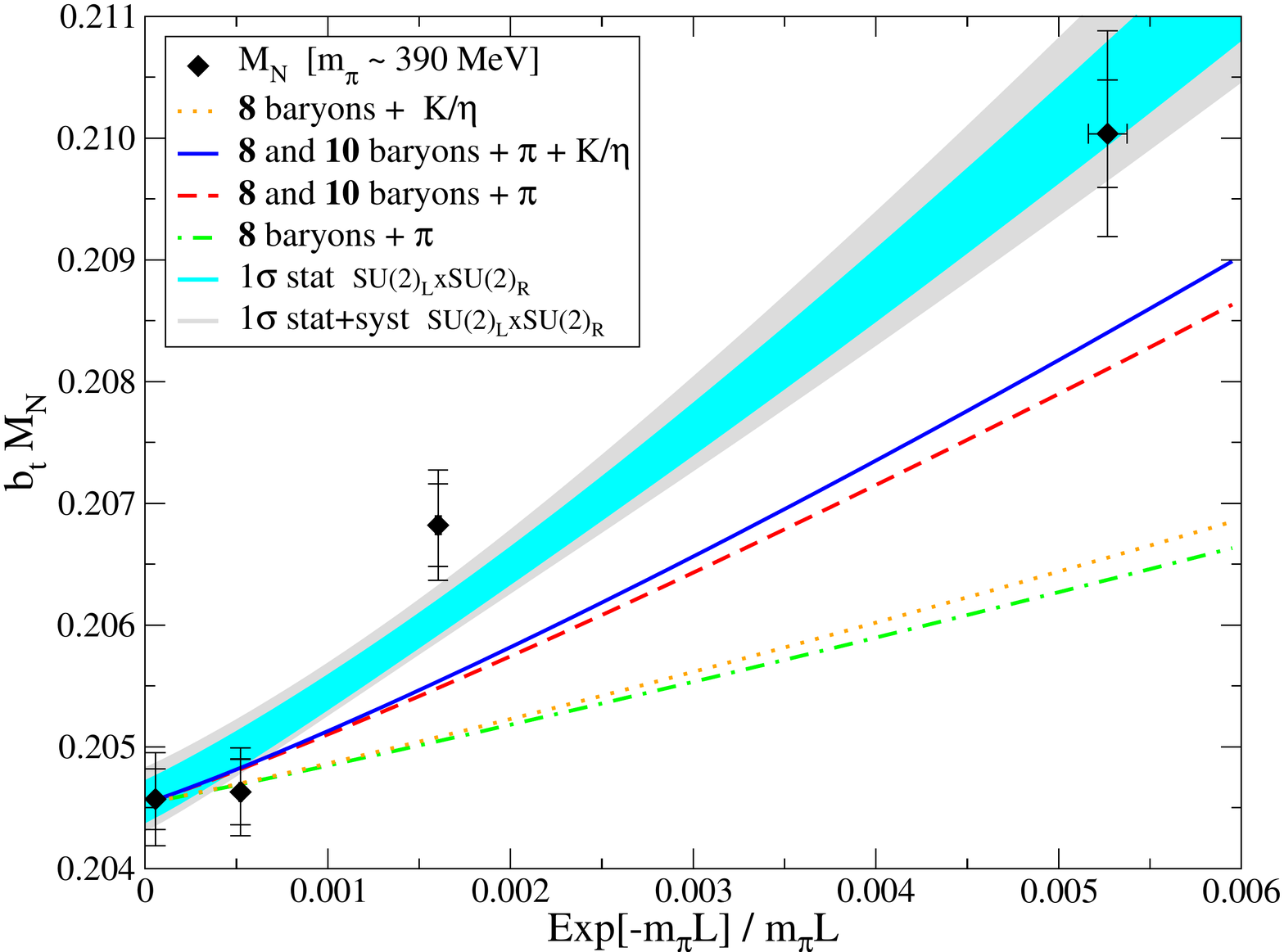}
     \caption{ The mass of the nucleon as a function of $e^{-m_\pi
         L}/( m_\pi L)$.  The dark (light) grey shaded region
       corresponds to the $1\sigma$ statistical uncertainty
       (statistical and systematic uncertainties combined in
       quadrature) resulting from fitting $M_N^{(\infty)}$ and
       $g_{\Delta N\pi}$.  Using this value of $M_N^{(\infty)}$, the
       dot-dashed curve (green) corresponds to the contribution from
       octet baryons and pions, the dotted curve (orange) corresponds
       to the contribution from octet baryons and kaons or an $\eta$,
       the dashed curve (red) corresponds to the contribution from
       octet and decuplet baryons and pions, and the solid curve
       (blue) corresponds to the contribution from octet and
       decuplet baryons and pions, kaons or an $\eta$.  }
  \label{fig:NucVolPlotsu3}
\end{figure}

\noindent 
The nucleon EMP's and fits to the mass plateaus obtained from the
results of the Lattice QCD calculations on the four lattice ensembles
are shown in fig.~\ref{fig:empN}.  With the input parameters given in
subsection~\ref{subsec:ParSet}, a two-parameter fit to the nucleon
mass data, given in table~\ref{tab:LQCDbaryonmasses}, can be performed
in two-flavor HB$\chi$PT to determine the infinite-volume value of the
nucleon mass, $M_N^{(\infty)}$, and the value of the $\Delta N\pi$
axial coupling, $g_{\Delta N\pi}$.  While there have been many
previous determinations of $g_{\Delta N\pi}$, we treat it as a fit
parameter and compare its value with the previous extractions.
Fitting the expression in eq.~(\ref{eq:FVexpMN}) to the results given
in table~\ref{tab:LQCDbaryonmasses} gives the fit-regions shown in
fig.~\ref{fig:NucVolPlotsu3}, and fit values of
\begin{eqnarray}
M_N^{(\infty)} & = & 0.20455(19)(17)~{\rm t.l.u}
\ \ ,\ \ 
|g_{\Delta N\pi}|\ =\ 2.80(18)(21)
\ \ \ .
\label{eq:Nfit}
\end{eqnarray}
The ratio of couplings
\begin{eqnarray}
{|g_{\Delta N\pi}|\over g_A} & = & 
2.22(14)(17)
\ \ \ ,
\label{eq:gg}
\end{eqnarray}
can be compared with the value of $|g_{\Delta N\pi}|/ g_A =
1.56(06)$~\cite{Hemmert:1994ky}~\footnote{The value of $|g_{\Delta
    N\pi}|/ g_A$ in Ref.~\cite{Hemmert:1994ky} has been divided by
  $\sqrt{2}$ in order to match the definition of $g_{\Delta N\pi}$
  employed in defining eq.~(\ref{eq:FVexpMN})~\cite{Beane:2004tw}.}
extracted from an analysis of experimentally-measured $\pi N$
scattering phase shifts.  The difference between these two values is
significant, but as the two extractions have been performed at two
different unphysical pion masses, little can be concluded.  On the
other hand, direct Lattice QCD calculations of $|g_{\Delta N\pi}|/
g_A$ have been performed~\cite{Alexandrou:2007xj,Alexandrou:2010uk}
over a range of pion masses~\footnote{The two methods employed in
  Ref.~\cite{Alexandrou:2010uk} suggest that there may be a relatively
  large systematic uncertainty in their value of $|g_{\Delta N\pi}|/
  g_A$ beyond that quoted.  }.  One such calculation performed with a
pion mass in the vicinity of $m_\pi\sim 390~{\rm MeV}$ gives
$|g_{\Delta N\pi}|/ g_A=1.47(19)$.  The finite-volume corrections to
the nucleon mass resulting from this value of the coupling is shown as
the dashed (red) curve in fig.~\ref{fig:NucVolPlotsu3}, and clearly
the contribution from the $N\pi$ and the $\Delta\pi$ intermediate
states constitute a large fraction of the finite-volume shift of the
nucleon mass.  Given the size of the pion mass in the present
calculations, $m_\pi/M_N\sim 0.35$, we anticipate that higher orders
in HB$\chi$PT will change the finite-volume corrections at the $\sim
30\%$ level, consistent with the difference between the results of the
lattice QCD calculations and NLO in HB$\chi$PT~\footnote{If instead of
  using $f_\pi\sim 150~{\rm MeV}$ to evaluate the NLO HB$\chi$PT
  result, the value at the physical pion mass, $f_\pi\sim 132~{\rm
    MeV}$ is used, then a value of $|g_{\Delta N\pi}|/ g_A = 1.76(18)$
  is obtained, consistent within uncertainties with the extraction
  from the matrix element of the axial
  current~\cite{Alexandrou:2007xj}.  The ambiguity in the value of the
  decay constant that is used in the NLO contribution will be
  parametrically reduced by a NNLO calculation.  }.  A
next-to-next-to-leading order (NNLO) calculation of the finite-volume
contributions to the nucleon mass in HB$\chi$PT, accompanied by more
precise Lattice QCD calculations over a range of lattice volumes and
quark masses, is required in order to improve upon this determination
of $g_{\Delta N\pi}$~\footnote{An alternative, efficient way of
  capturing the bulk of the volume dependence is to insert the forward
  $\pi$N scattering amplitude, calculated in HB$\chi$PT with explicit
  $\Delta$ degrees of freedom into
  eq.~(\ref{eq:formalvol})~\cite{Fuhrer:2004zz}. This procedure has
  been shown to work well for the $\pi$ mass volume
  dependence~\cite{Colangelo:2006mp}.}.

With these fit parameters and the parameter set previously defined in
subsection~\ref{subsec:ParSet}, the effect of kaon and $\eta$ loops
can be estimated by including the finite-volume corrections given in
eq.~(\ref{eq:FVNsu3}), the results of which are shown in
fig.~\ref{fig:NucVolPlotsu3}.  The contributions from the
strange-baryon and strange-meson intermediate states are estimated to
be small, and somewhat improve the agreement between theory and the
Lattice QCD calculation.  Including them in the fit of $g_{\Delta
  N\pi}$ to the results of the Lattice QCD calculation gives
$|g_{\Delta N\pi}|/g_A = 2.10(15)(20)$, which is to be compared with
the result in eq.~(\ref{eq:gg}).
\begin{figure}[!ht]
  \centering
     \includegraphics[width=0.85\textwidth]{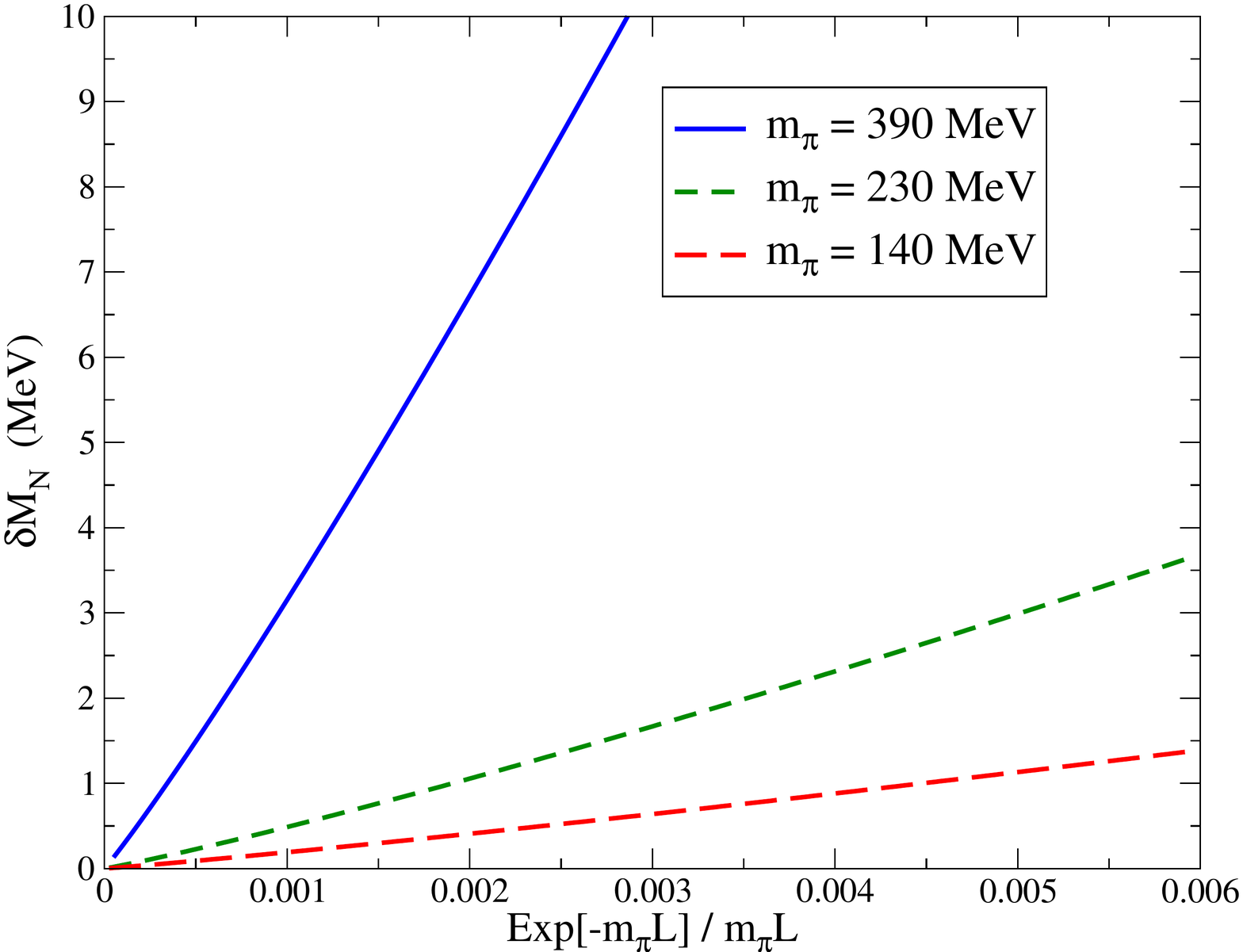}
     \caption{Estimates of the finite-volume contributions to the
       nucleon mass at NLO in HB$\chi$PT for $m_\pi= 390~{\rm
         MeV}$ (upper, blue, solid, which corresponds to the red dashed curve
in fig.~\protect\ref{fig:NucVolPlotsu3}),
       $m_\pi= 230~{\rm MeV}$ (middle, green, dotted), and
       $m_\pi= 140~{\rm MeV}$ (lower, red, dashed).  }
  \label{fig:NucDN230140}
\end{figure}

As NLO HB$\chi$PT reproduces most of the volume dependence of the
nucleon mass at $m_\pi\sim 390~{\rm MeV}$, and is expected to become
more accurate at lighter pion masses, it is useful to use the NLO
expression to estimate the size of the finite-volume contributions to
the nucleon mass at the pion masses other than the current one.  In
fig.~\ref{fig:NucDN230140} we show the finite-volume contributions to
the nucleon mass that are predicted at NLO in HB$\chi$PT for
$m_\pi\sim 390, 230$ and $140~{\rm MeV}$.  This makes clear that
finite-volume effects are expected to be significantly smaller at the
lighter pion masses for fixed $m_\pi L$.  The values of $m_\pi L$,
estimated at NLO in HB$\chi$PT, for which the finite-volume
contributions to the nucleon mass are $\delta M_N^{(FV)} = 1~{\rm
  MeV}$ at $m_\pi = 390, 230$ and $140~{\rm MeV}$ are $m_\pi L \sim
6.2, 4.7$ and $3.9$, respectively.  For $\delta M_N^{(FV)} = 100~{\rm
  keV}$, the corresponding values are $m_\pi L \sim 8.0$, $6.4$, and
$5.8$, respectively.  Given that the deuteron binding energy and
nuclear excitation energies are in the MeV regime, the estimates
indicate that $m_\pi L\gsim 2\pi$ is required at the physical pion
mass in order to eliminate contamination from this class of
exponentially-suppressed finite-volume effects that contaminate the
extraction of phase shifts and binding energies from Lattice QCD
calculations.

\subsubsection{The $\Lambda$  Mass}
\label{subsec:LMass}

\begin{figure}[!ht]
  \centering
     \includegraphics[width=0.49\textwidth]{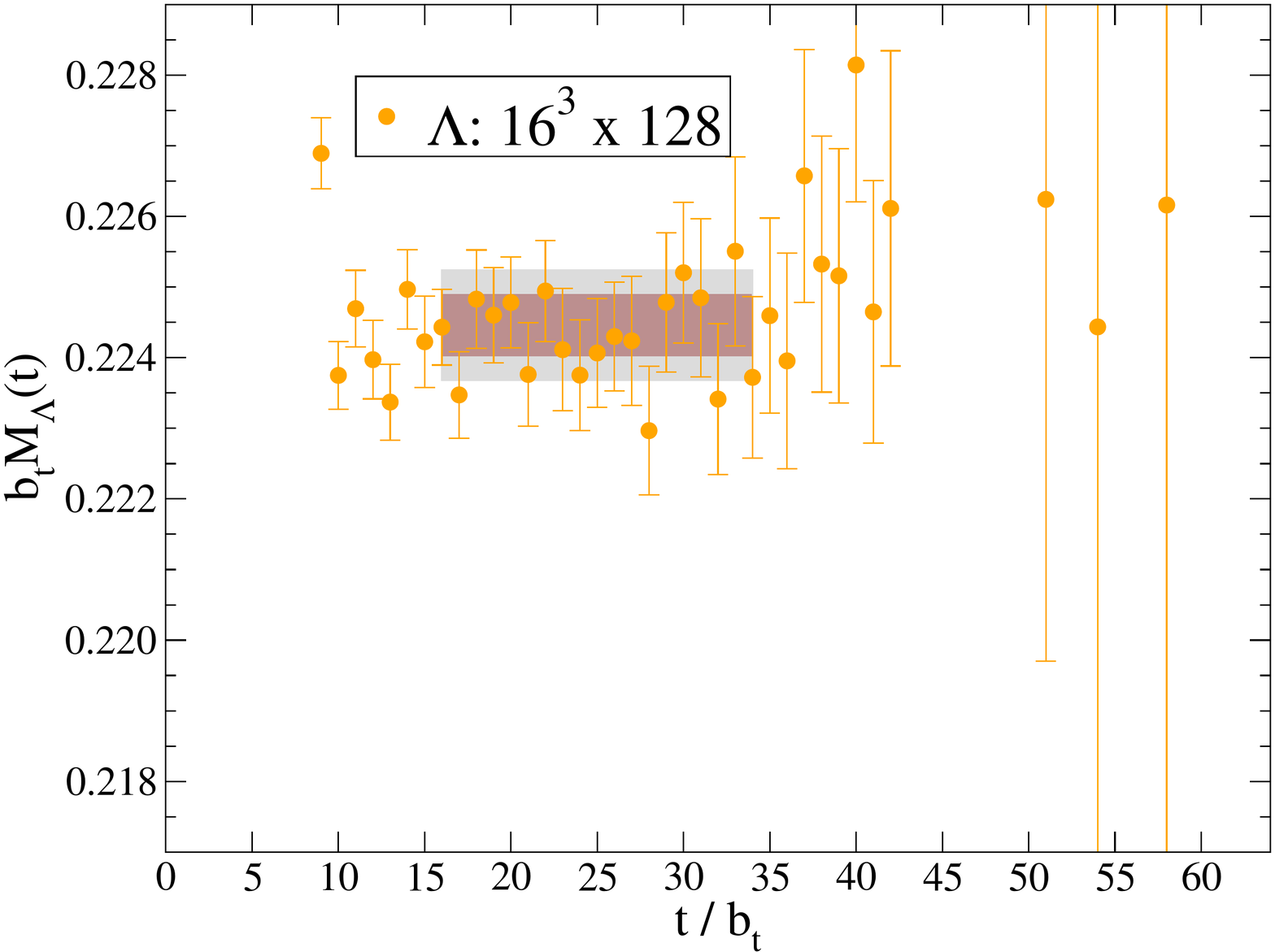} 
     \includegraphics[width=0.49\textwidth]{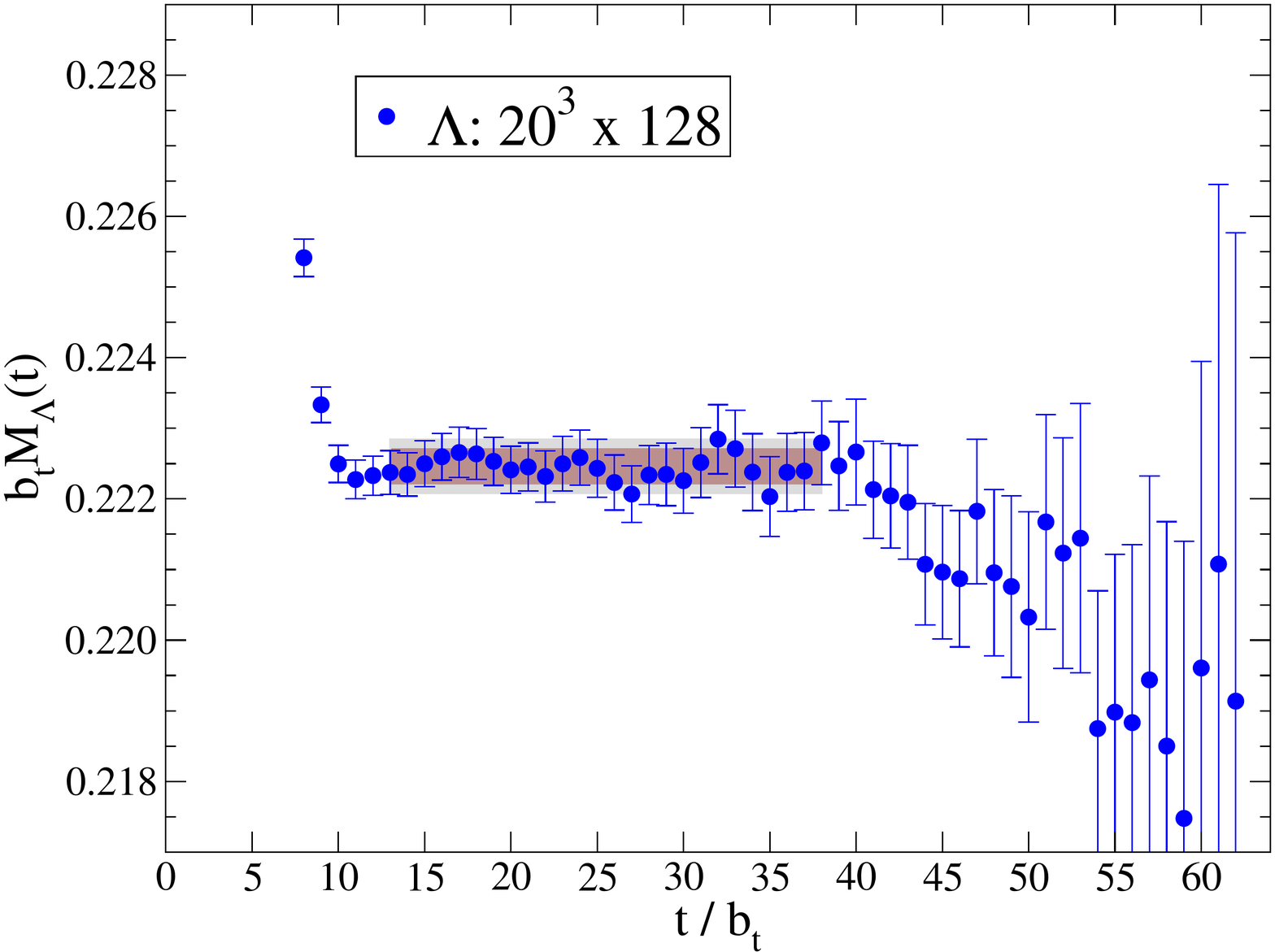} \\
     \includegraphics[width=0.49\textwidth]{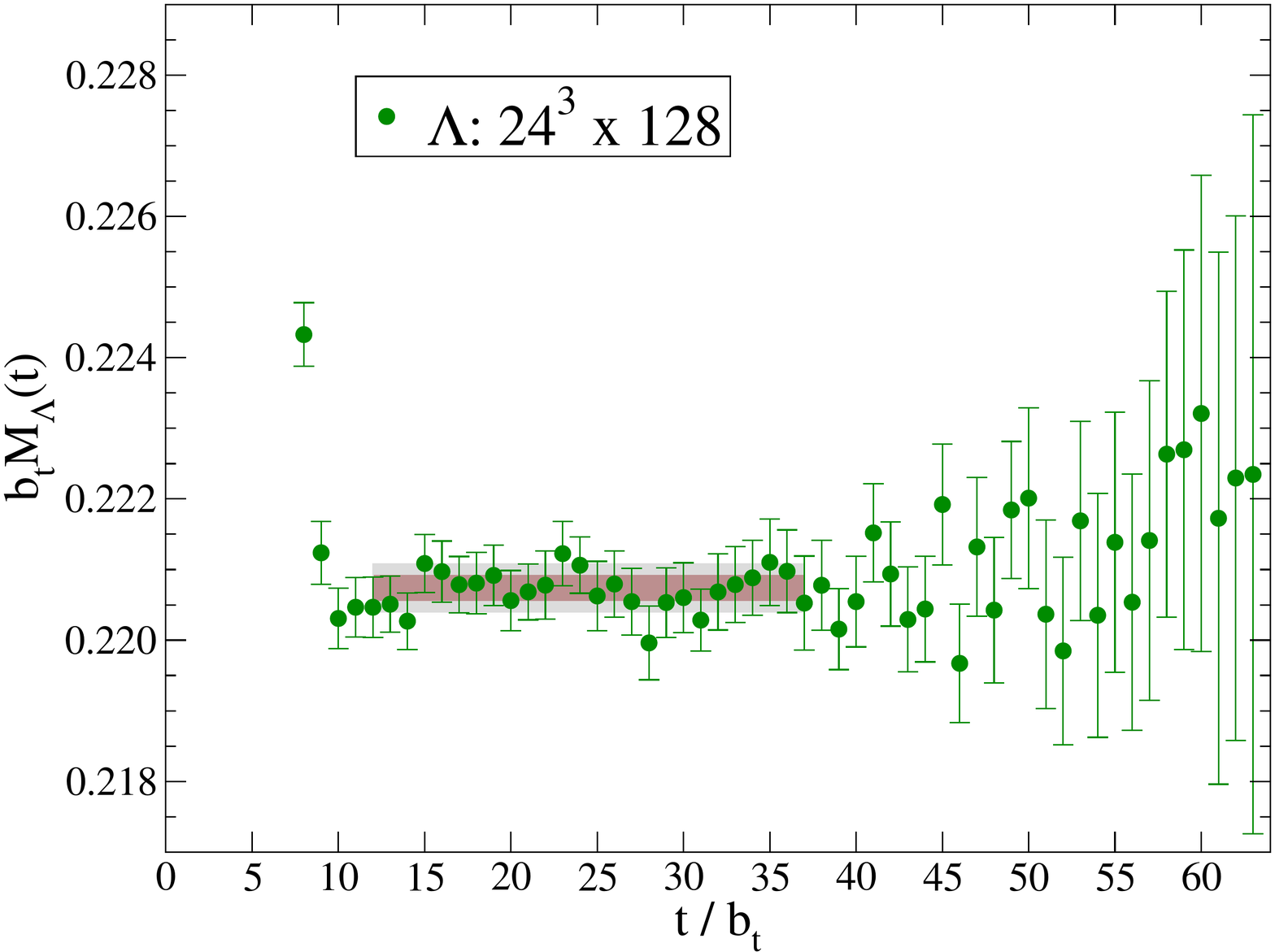} 
     \includegraphics[width=0.49\textwidth]{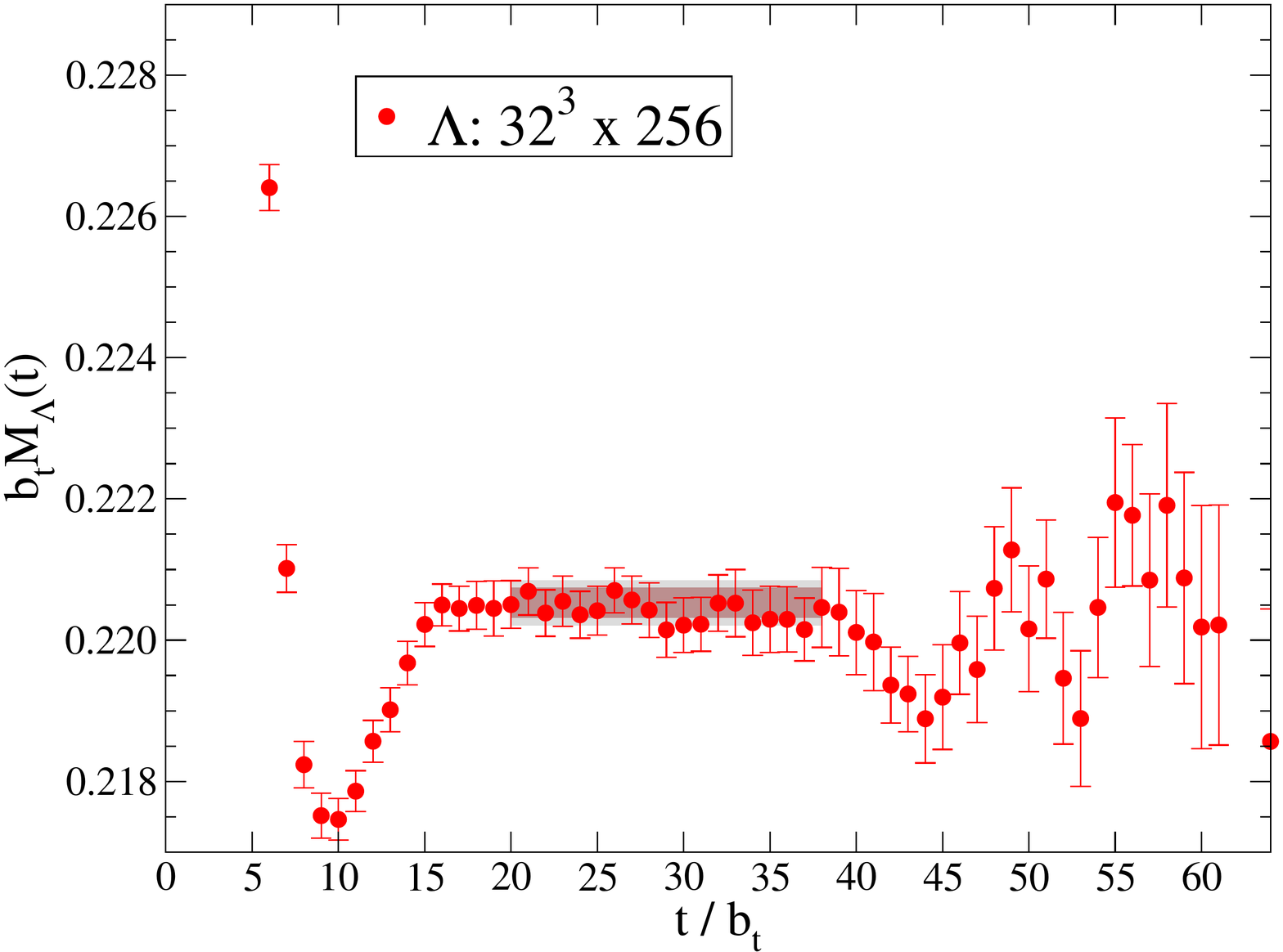}
     \caption{The $\Lambda$ EMP's determined 
on the four lattice ensembles used in this work. 
They each result from linear combinations of different correlation functions that 
optimize the plateau of the ground state.
Note that the y-axis scale is the same in all four panels.
}
  \label{fig:empLAM}
\end{figure}
\begin{figure}[!ht]
  \centering
     \includegraphics[width=0.9\textwidth]{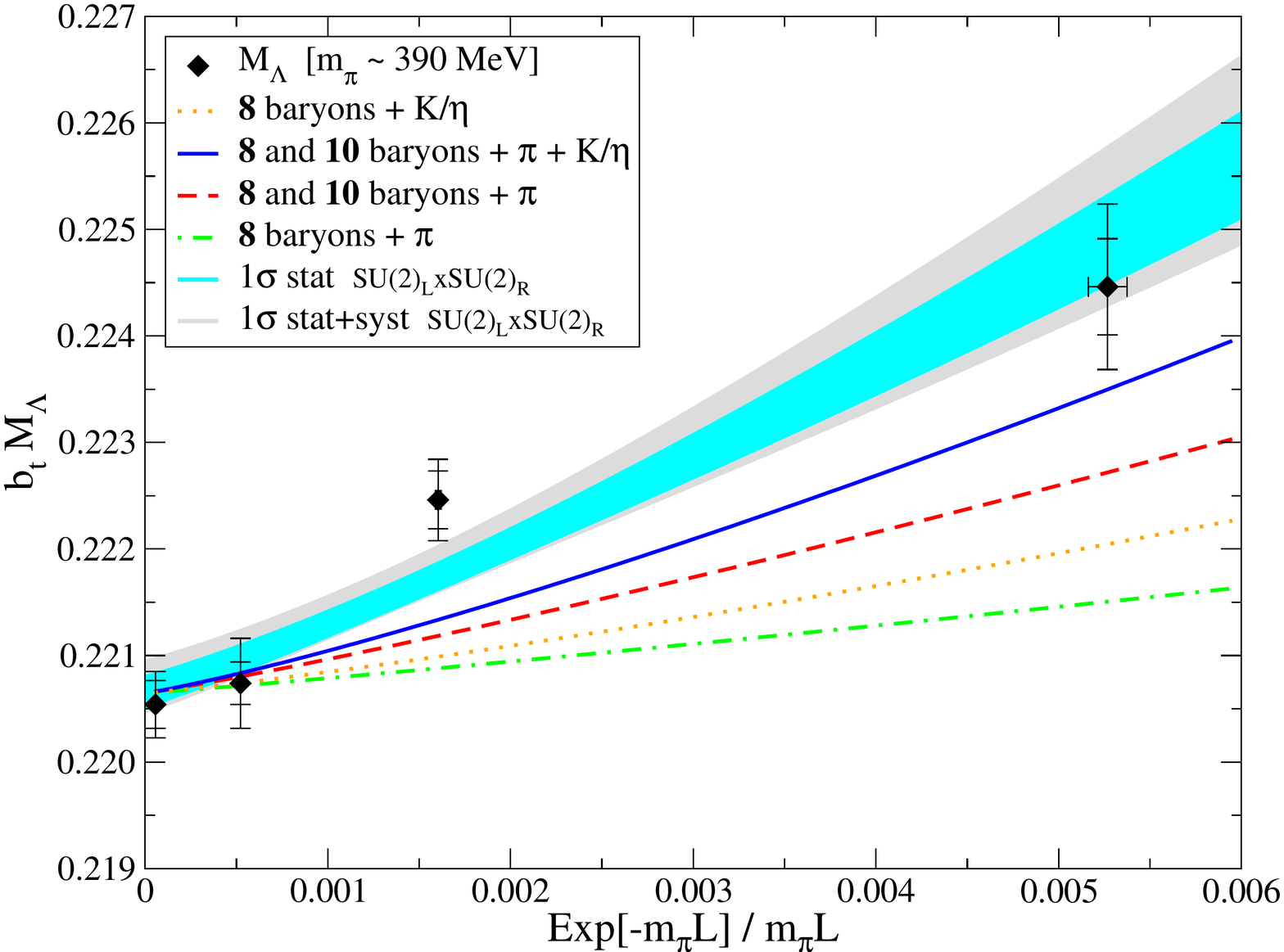}
     \caption{ The mass of the $\Lambda$ as a function of $e^{-m_\pi
         L}/( m_\pi L)$.  The dark (light) grey shaded region
       corresponds to the $1\sigma$ statistical uncertainty
       (statistical and systematic uncertainties combined in
       quadrature) resulting from fitting $M_\Lambda^{(\infty)}$ and
       $g_{\Sigma^*\Lambda\pi}$.  Using this value of
       $M_\Lambda^{(\infty)}$, the dot-dashed curve (green)
       corresponds to the contribution from octet baryons and pions,
       the dotted curve (orange) corresponds to the contribution from
       octet baryons and kaons or an $\eta$, the dashed curve (red)
       corresponds to the contribution from octet and
       decuplet baryons and pions, and the solid curve (blue)
       corresponds to the contribution from octet and
       decuplet baryons and pions, kaons or an $\eta$.  }
  \label{fig:LambdaVolPlotsu3}
\end{figure}

\noindent The $\Lambda$ EMP's and fits to the mass plateaus obtained
from the results of the Lattice QCD calculations for the four lattice
ensembles are shown in fig.~\ref{fig:empLAM}.  The fit values of the
$\Lambda$ masses in the four lattice volumes are given in
table~\ref{tab:LQCDbaryonmasses}, and are shown as the points with
uncertainties in fig.~\ref{fig:LambdaVolPlotsu3}.  The shaded regions
in fig.~\ref{fig:LambdaVolPlotsu3} show the results of the  ${\rm SU(2)}_L\otimes {\rm SU(2)}_R$
HB$\chi$PT fit to the volume dependence of the $\Lambda$ mass using
eq.~(\ref{eq:FVLam}) and the value of $g_{\Sigma\Lambda}$ given in
eq.~(\ref{eq:axialcouplings}). The fit gives
\begin{eqnarray}
M_\Lambda^{(\infty)} & = & 0.22064(15)(19)~{\rm t.l.u}
\ \ ,\ \ 
|g_{\Sigma^*\Lambda\pi}|\ =\  2.21(16)(23)
\ \ \ .
\end{eqnarray}
If instead of fitting $g_{\Sigma^*\Lambda\pi}$, flavor SU(3) symmetry
is used to relate it to $g_{\Delta N\pi}$, $g_{\Sigma^*\Lambda\pi} =
g_{\Delta N\pi}/\sqrt{2} = 1.3$, then the contribution from $\Sigma$
intermediate states and from $\Sigma$ and $\Sigma^*$ intermediate
states are shown as the dot-dashed (green) and dashed (red) curves in
fig.~\ref{fig:LambdaVolPlotsu3}, respectively.  Comparing these
expectations with the results of the Lattice QCD calculations,
manifested in the fit value of $g_{\Sigma^*\Lambda\pi}$ being $\sim
40\%$ larger than phenomenological expectations, indicates that higher
orders in two-flavor $\chi$PT are important, or that the strange quark
plays a role in the finite-volume contributions through kaons or an
$\eta$.

As with the nucleon, we can now estimate the effects of kaon and $\eta$
loops by including the finite-volume corrections given in
eq.~(\ref{eq:FVLamsu3}).  This gives rise to the curves shown in
fig.~\ref{fig:LambdaVolPlotsu3}.
(Note that the same values for the input parameters are used to
generate the curves in fig.~\ref{fig:LambdaVolPlotsu3} as are used in
the case of the nucleon).  The dot-dashed (green) and dashed (red)
curves are the contributions from pions, while the dotted (orange) and
solid (blue) curves correspond to the sum of contributions from pions,
kaons and an $\eta$.  The pions make the largest contribution to the
volume dependence of the mass of the $\Lambda$, but, unlike in the
case of the nucleon, the kaons and $\eta$ contribute significantly.
It is interesting to note that the NLO contribution in three-flavor  HB$\chi$PT
(blue curve) agrees well with the results of the Lattice QCD
calculation.  However, NLO HB$\chi$PT is expected to be modified at the
$\sim 30\%$ level by higher orders in the expansion.

\subsubsection{The $\Sigma$  Mass}
\label{subsec:SMass}

\begin{figure}[!ht]
  \centering
     \includegraphics[width=0.49\textwidth]{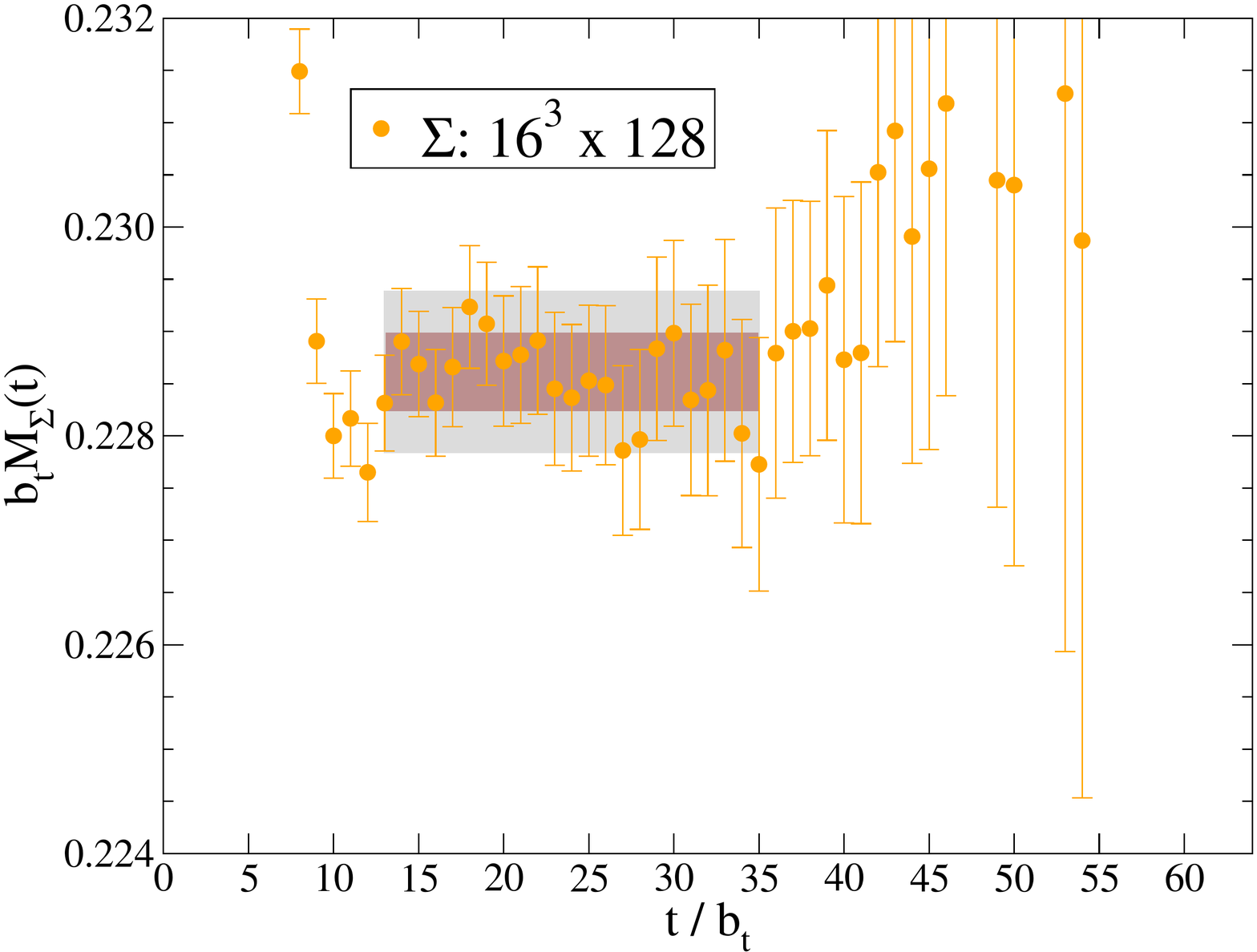} 
     \includegraphics[width=0.49\textwidth]{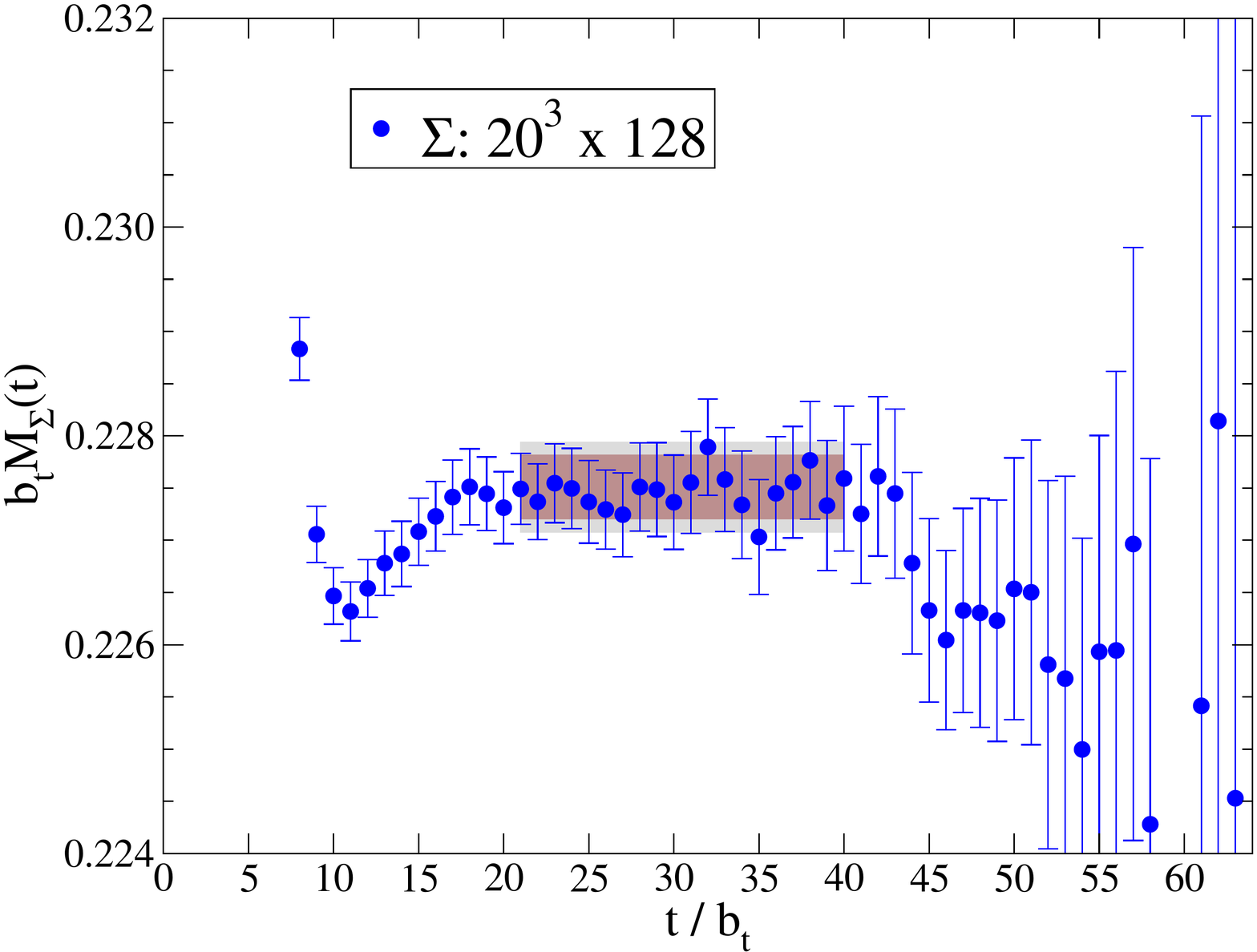} \\
     \includegraphics[width=0.49\textwidth]{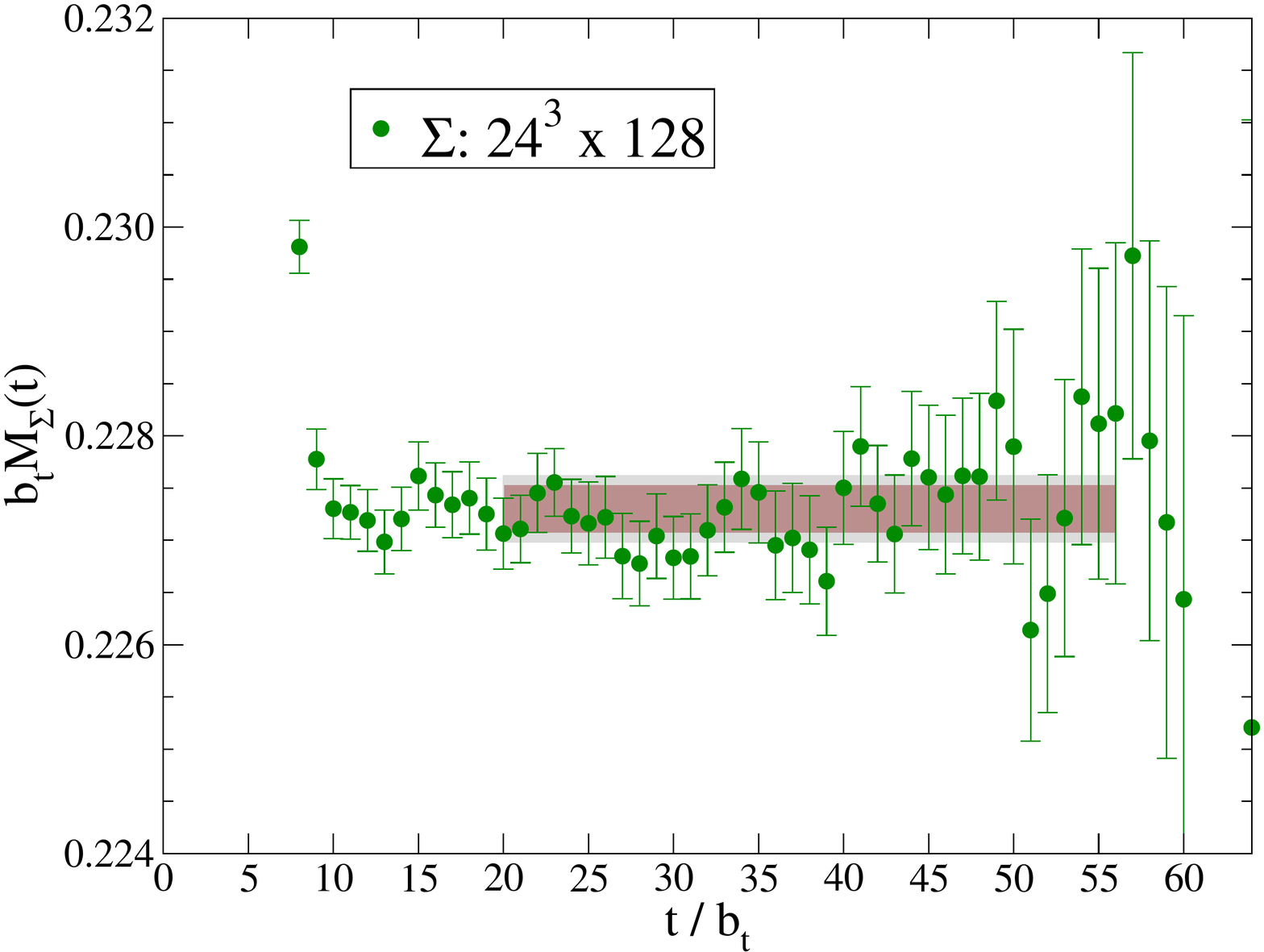} 
     \includegraphics[width=0.49\textwidth]{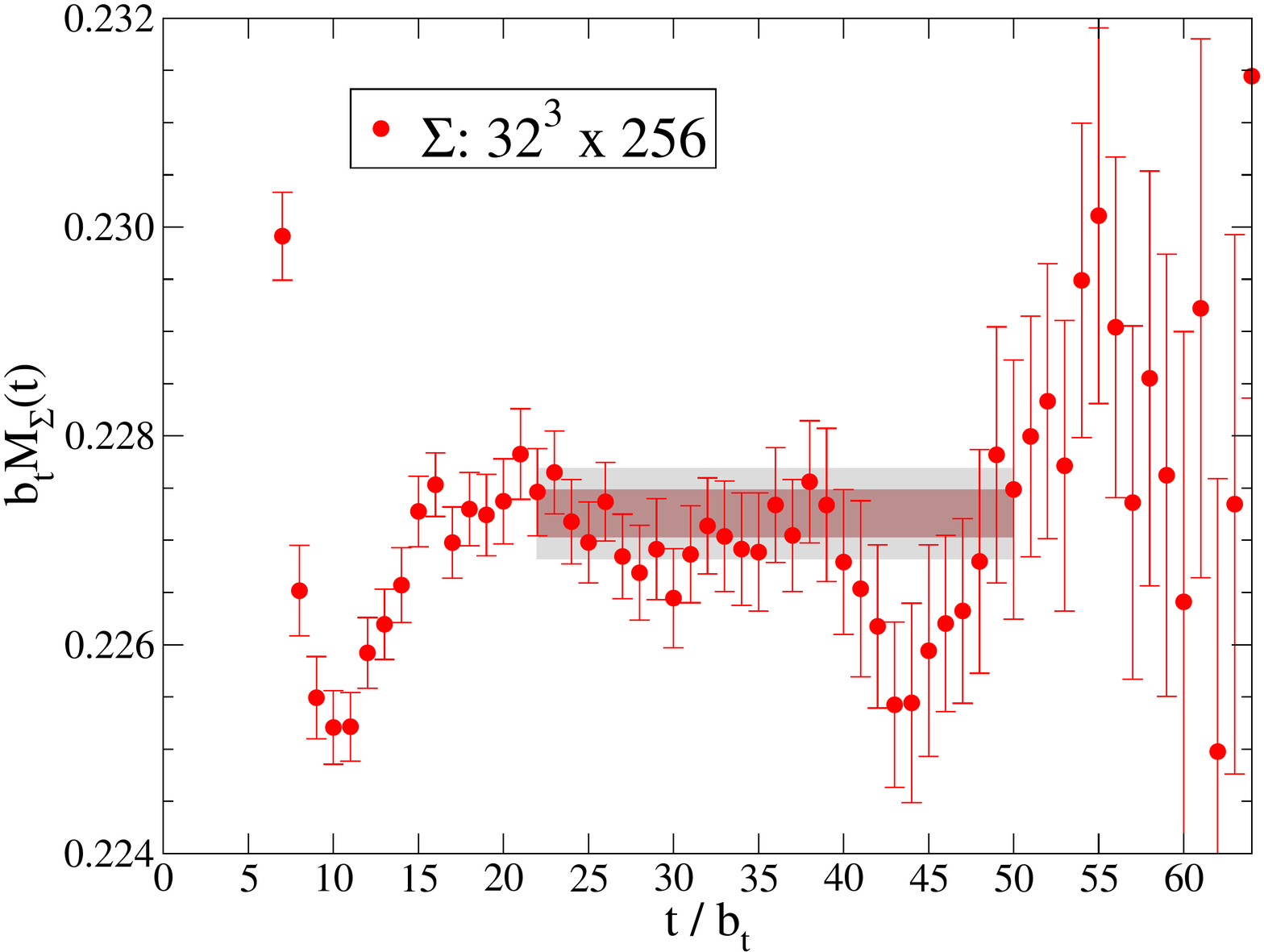}
     \caption{
The $\Sigma$ EMP's determined 
on the four lattice ensembles used in this work. 
They each result from linear combinations of different correlation functions that 
optimize the plateau of the ground state.
Note that the y-axis scale is the same in all four panels.
}
  \label{fig:empSIG}
\end{figure}
\begin{figure}[!ht]
  \centering
     \includegraphics[width=0.9\textwidth]{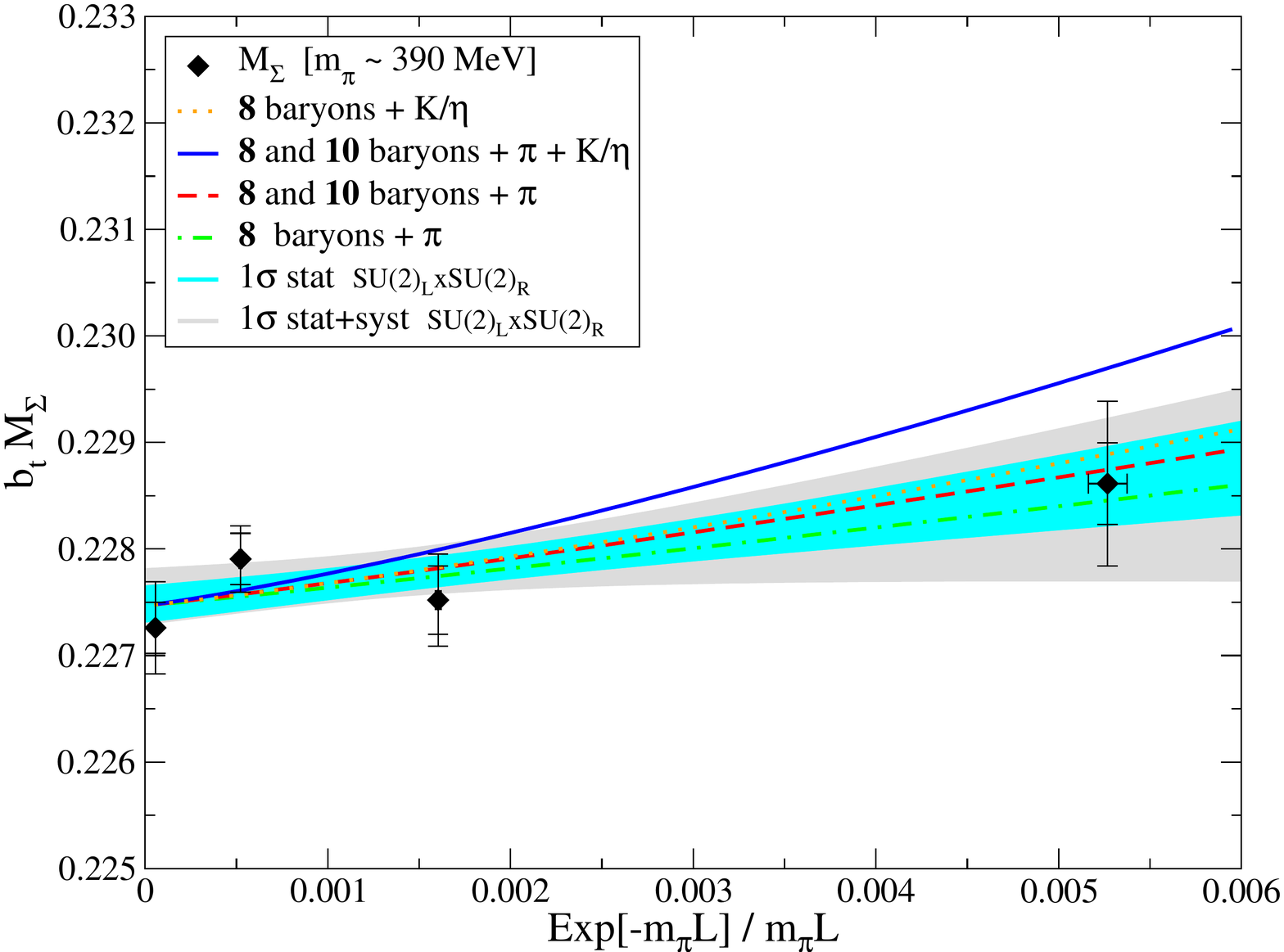}
     \caption{ The mass of the $\Sigma$ as a function of $e^{-m_\pi
         L}/( m_\pi L)$.  The dark (light) grey shaded region
       corresponds to the $1\sigma$ statistical uncertainty
       (statistical and systematic uncertainties combined in
       quadrature) resulting from fitting $M_\Sigma^{(\infty)}$ and
       $g_{\Sigma^*\Sigma\pi}$.  Using this value of
       $M_\Sigma^{(\infty)}$, the dot-dashed curve (green) corresponds
       to the contribution from octet baryons and pions, the dotted
       curve (orange) corresponds to the contribution from
       octet baryons and kaons or an $\eta$, the dashed curve (red)
       corresponds to the contribution from octet and
       decuplet baryons and pions, and the solid curve (blue)
       corresponds to the contribution from octet and
       decuplet baryons and pions, kaons or an $\eta$.  }
  \label{fig:SigmaVolPlotsu3}
\end{figure}

\noindent 
The $\Sigma$ EMP's and fits to the mass plateaus obtained from the
results of the Lattice QCD calculations on the four lattice ensembles
are shown in fig.~\ref{fig:empSIG}.  The fit values of the $\Sigma$
masses in the four lattice volumes are given in
table~\ref{tab:LQCDbaryonmasses}, and are shown as the points with
uncertainties in fig.~\ref{fig:SigmaVolPlotsu3}.  The shaded regions
in fig.~\ref{fig:SigmaVolPlotsu3} show the results of the ${\rm
  SU(2)}_L\otimes {\rm SU(2)}_R$ HB$\chi$PT fit to the volume
dependence of the $\Sigma$ mass using eq.~(\ref{eq:FVSig}) with the
axial couplings given in eq.~(\ref{eq:axialcouplings}).  The fit gives
\begin{eqnarray}
M_\Sigma^{(\infty)} & = & 0.22747(17)(19)~{\rm t.l.u}
\ \ \ \ ,\ \ \ \
|g_{\Sigma^*\Sigma\pi}|\ <\ 1.38 [1.90]
\ \ \ ,
\label{eq:sigmafit}
\end{eqnarray}
where we have quoted a $68\%$ confidence interval for $g_{\Sigma^*\Sigma\pi}$ including
statistical errors and statistical and systematic errors added in quadrature (bracketed).
If instead of fitting $g_{\Sigma^*\Sigma\pi}$, the SU(3) relation is
used, $g_{\Sigma^*\Sigma\pi} = g_{\Delta N\pi}/\sqrt{3} = 1.07$, then
the contribution from $\Sigma$, $\Lambda$ intermediate states and from
$\Sigma$, $\Lambda$ and $\Sigma^*$ intermediate states are shown as
the dot-dashed (green) and dashed (red) curves in
fig.~\ref{fig:SigmaVolPlotsu3}, respectively.  The NLO ${\rm
  SU(3)}_L\otimes {\rm SU(3)}_R$ HB$\chi$PT prediction (blue curve) is
somewhat higher than the result of the Lattice QCD calculation in the
smallest volume, but not significantly so.  It should be added that
the SU(3) value of the coupling constant, $g_{\Sigma^*\Sigma\pi} =
1.07$ is consistent with the confidence interval extracted from the
${\rm SU(2)}_L\otimes {\rm SU(2)}_R$ fit, given in
eq.~(\ref{eq:sigmafit}).

The volume dependence of the $\Sigma$ mass is found to be 
somewhat smaller than that of the $\Lambda$ mass.  
This is consistent with the prediction  of NLO  
${\rm SU(3)}_L\otimes {\rm SU(3)}_R$ HB$\chi$PT, which is
largely driven by the coupling to the decuplet intermediate states.
The difference in the couplings to the decuplet, given in 
eq.~(\ref{eq:su3couplings}), is sufficient to explain the difference 
in volume dependence.

\subsubsection{The $\Xi$ Mass}
\label{subsec:XiMass}

\begin{figure}[!ht]
  \centering
     \includegraphics[width=0.49\textwidth]{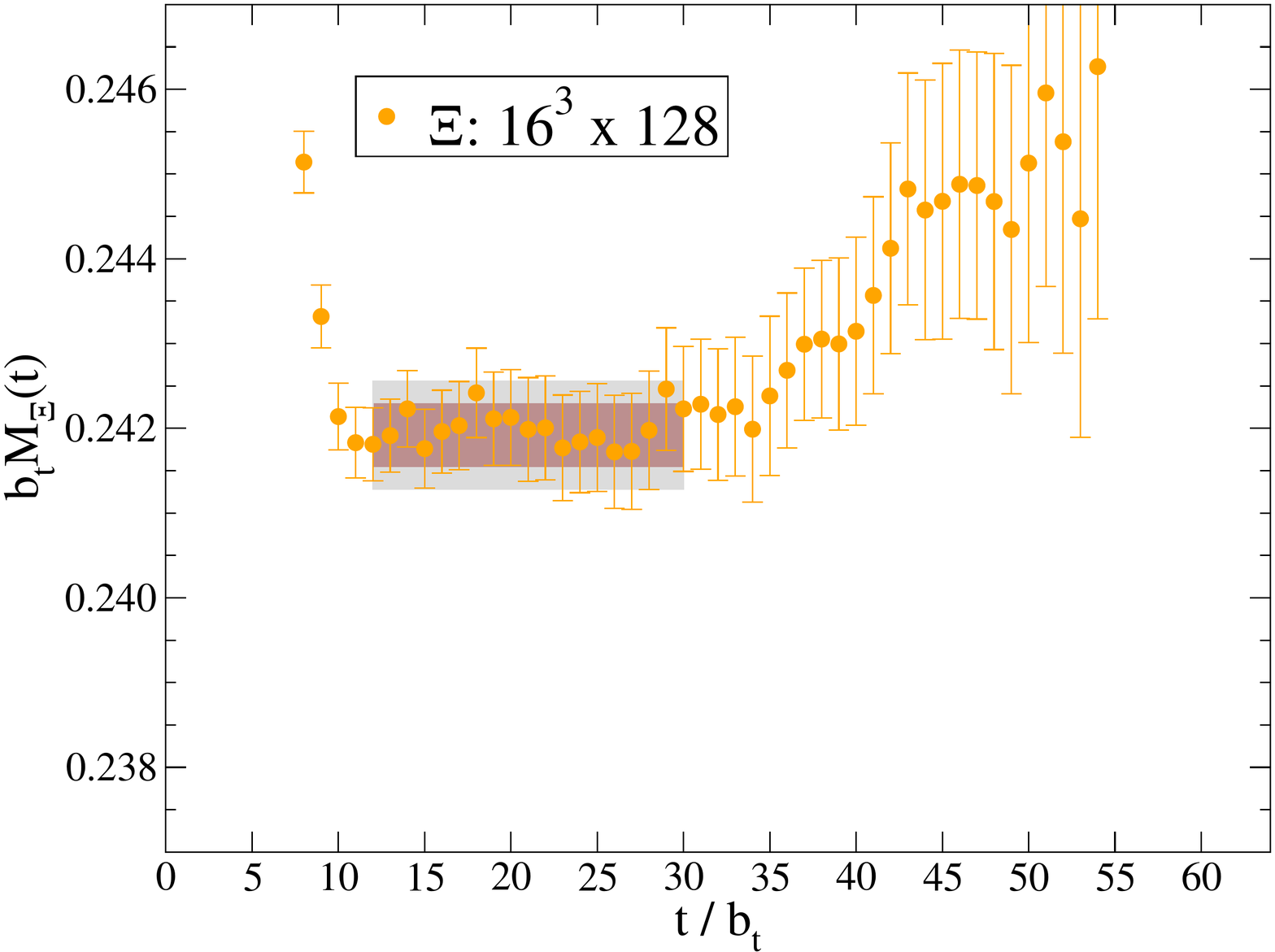} 
     \includegraphics[width=0.49\textwidth]{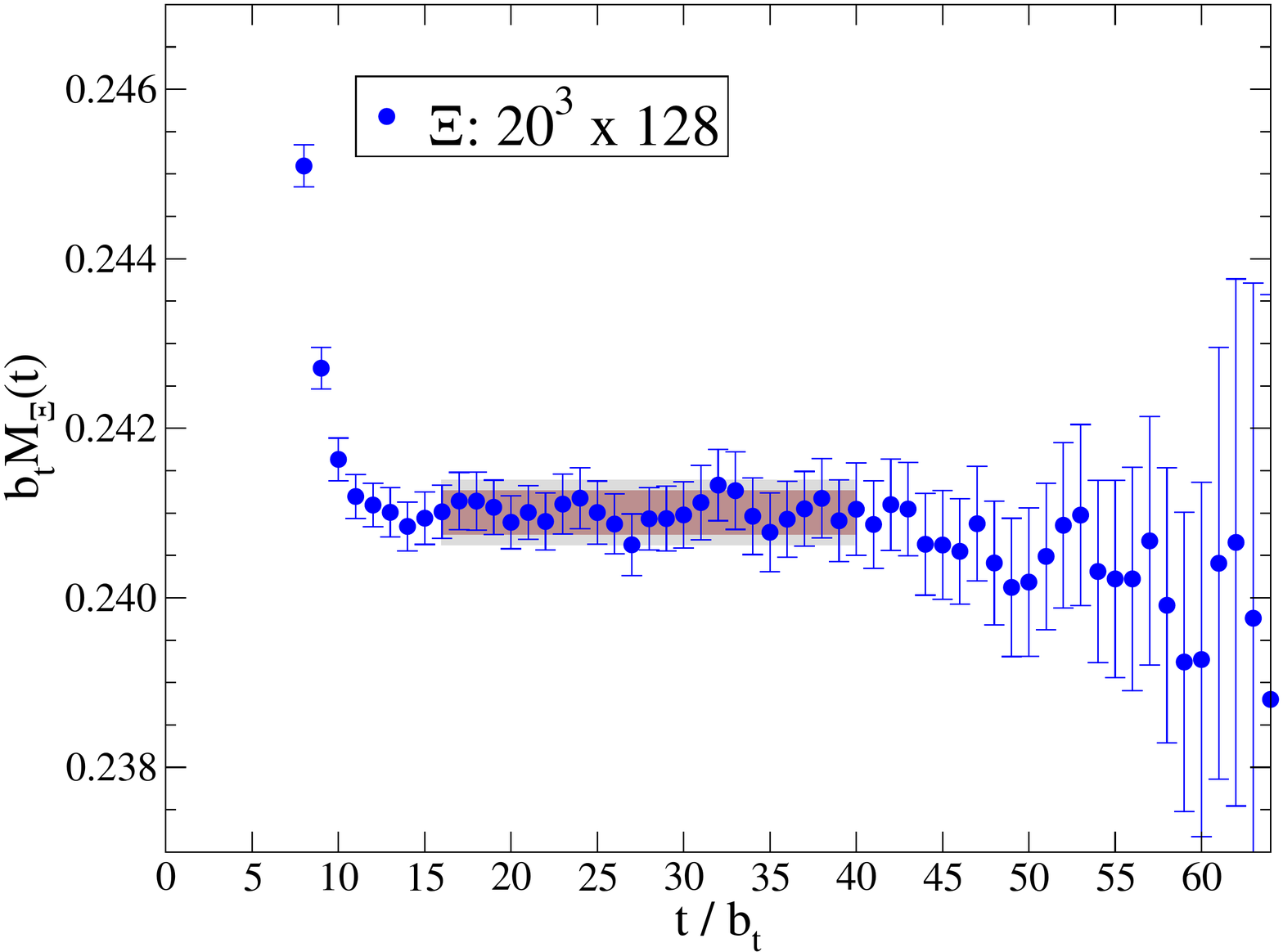} \\
     \includegraphics[width=0.49\textwidth]{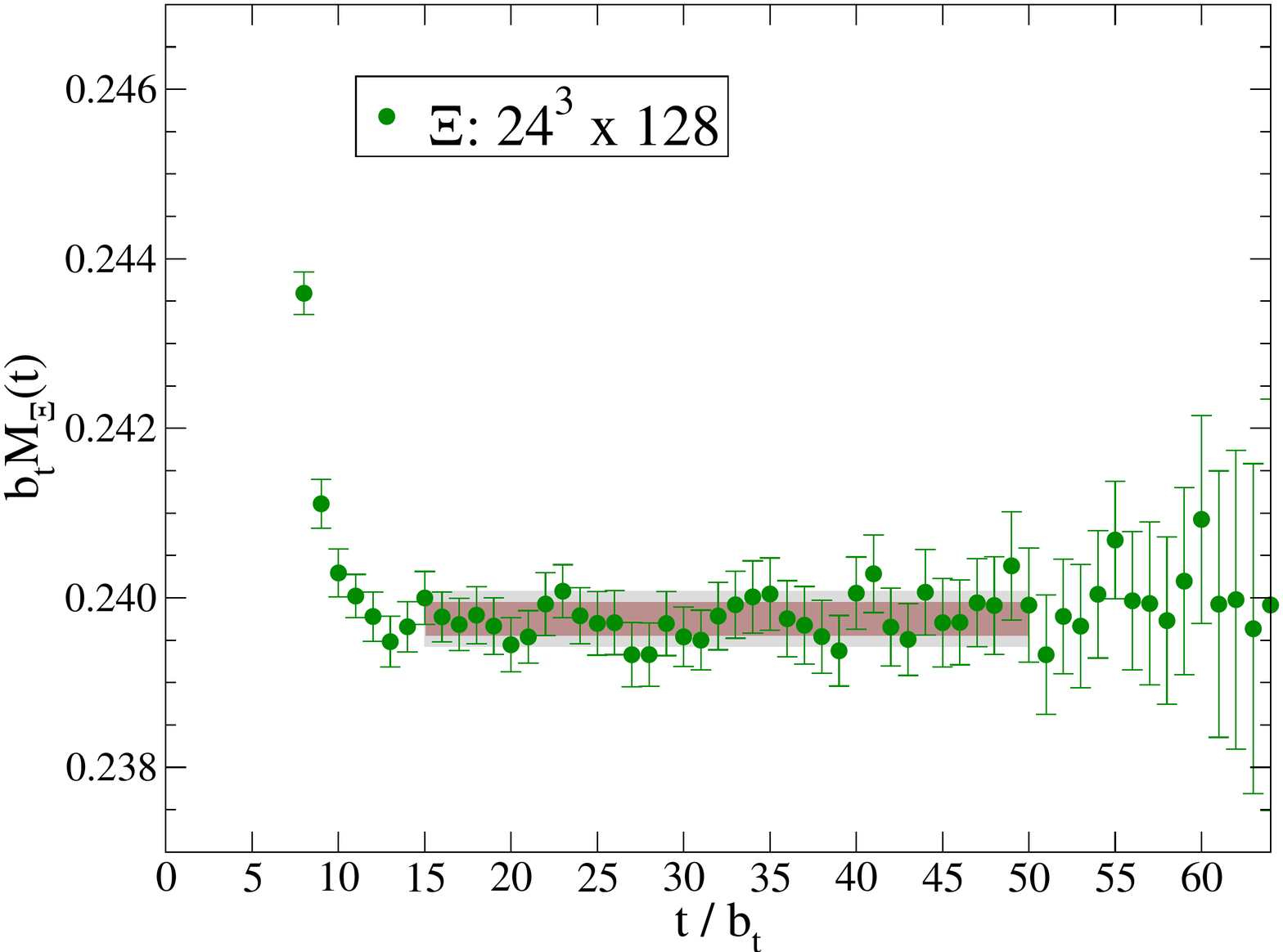} 
     \includegraphics[width=0.49\textwidth]{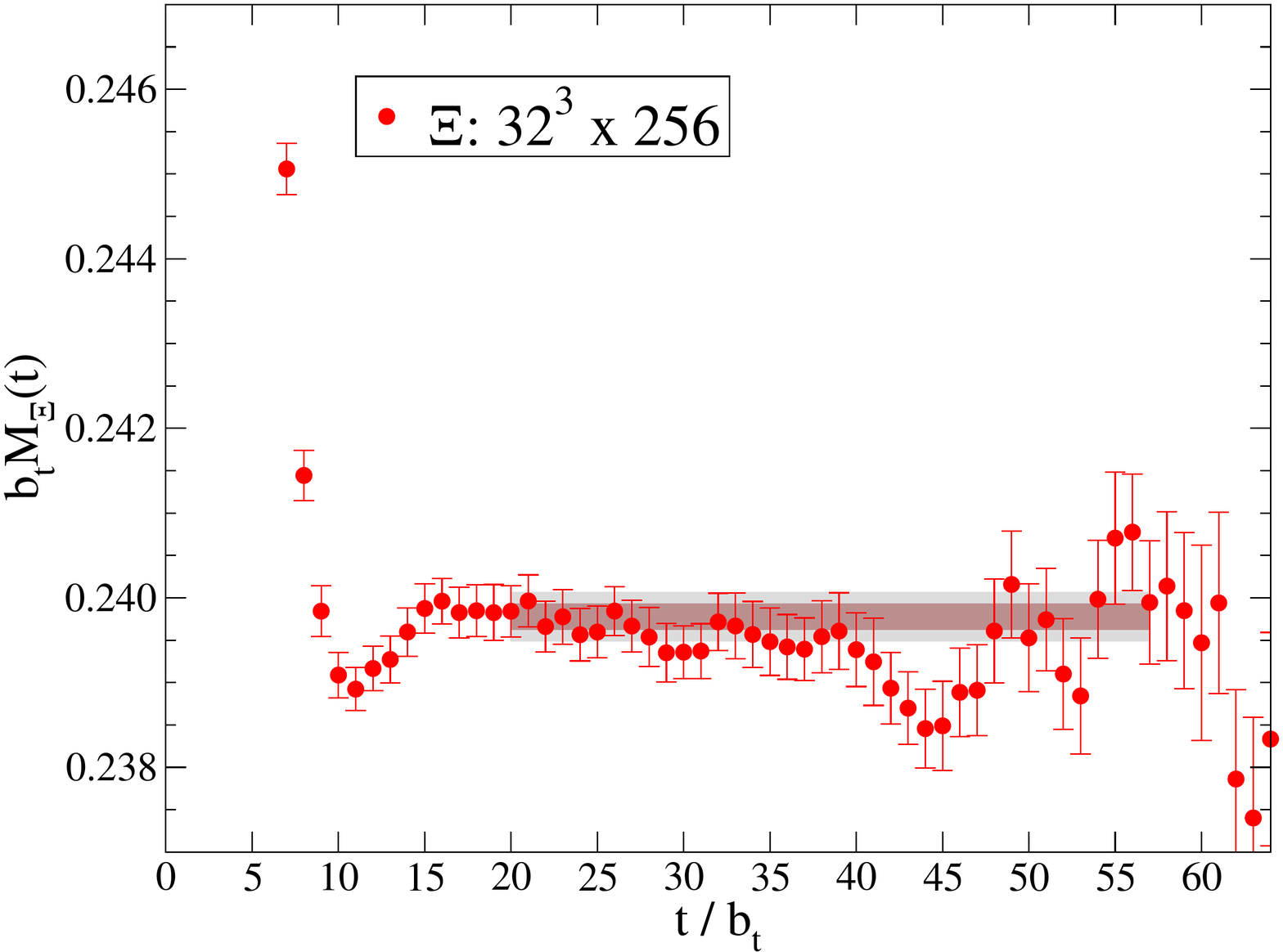}
     \caption{
The $\Xi$ EMP's determined 
on the four lattice ensembles used in this work. 
They each result from linear combinations of different correlation functions that 
optimize the plateau of the ground state.
Note that the y-axis scale is the same in all four panels.
}
  \label{fig:empXi}
\end{figure}
\begin{figure}[!ht]
  \centering
     \includegraphics[width=0.9\textwidth]{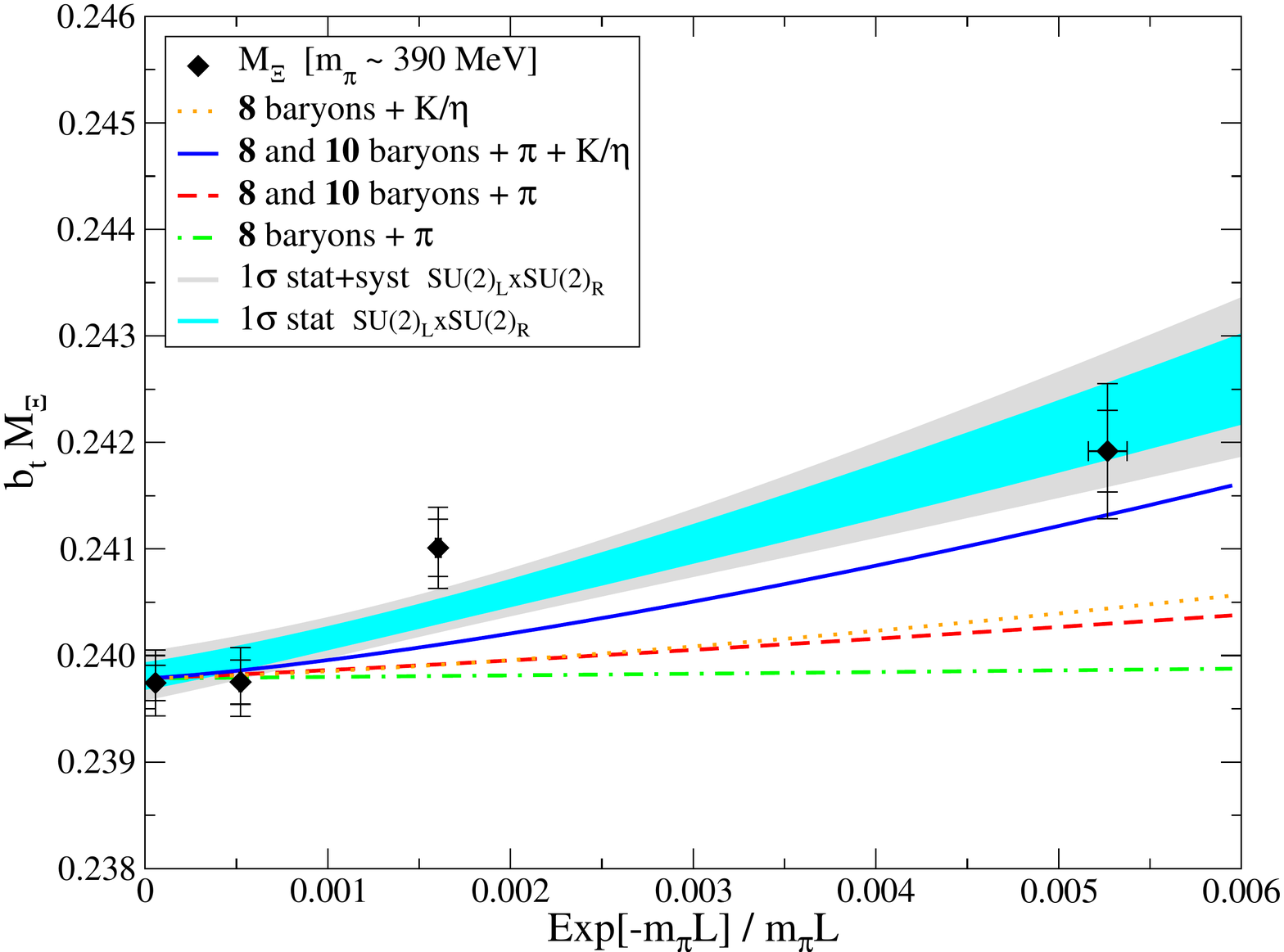}
     \caption{ The mass of the $\Xi$ as a function of $e^{-m_\pi L}/(
       m_\pi L)$.  The dark (light) grey shaded region corresponds to
       the $1\sigma$ statistical uncertainty (statistical and
       systematic uncertainties combined in quadrature) resulting from
       fitting $M_\Xi^{(\infty)}$ and $g_{\Xi^*\Xi\pi}$.  Using this
       value of $M_\Xi^{(\infty)}$, the dot-dashed curve (green)
       corresponds to the contribution from octet baryons and pions,
       the dotted curve (orange) corresponds to the contribution from
       octet baryons and kaons or an $\eta$, the dashed curve (red)
       corresponds to the contribution from octet and
       decuplet baryons and pions, and the solid curve (blue)
       corresponds to the contribution from octet and
       decuplet baryons and pions, kaons or an $\eta$.  }

  \label{fig:XiVolPlotsu3}
\end{figure}

\noindent 
The $\Xi$ EMP's and fits 
to the results of the Lattice QCD calculations
on the four lattice ensembles
are shown in fig.~\ref{fig:empXi}.
The 
fit values of the 
$\Xi$ masses in the four lattice volumes are given in  
table~\ref{tab:LQCDbaryonmasses}, and are shown as the points with
uncertainties in fig.~\ref{fig:XiVolPlotsu3}.
The shaded regions in fig.~\ref{fig:XiVolPlotsu3} show the results of 
the  ${\rm SU(2)}_L\otimes {\rm SU(2)}_R$ 
HB$\chi$PT fit to the volume dependence of the $\Xi$ mass
using eq.~(\ref{eq:FVXi}). The fit parameters are 
\begin{eqnarray}
M_\Xi^{(\infty)} & = & 0.23978(12)(18)~{\rm t.l.u}
\ \ ,\ \ 
|g_{\Xi^*\Xi\pi}|\ =\ 
2.49(23)(35)
\ \ \ .
\end{eqnarray}
If instead of fitting $g_{\Xi^*\Xi\pi}$, the estimates from other
observables are used, $g_{\Xi^*\Xi\pi} = g_{\Delta N\pi}/2 =
0.93$~\cite{Savage:1996zd}, the contribution from $\Xi$ intermediate
states and from $\Xi$ and $\Xi^*$ intermediate states are shown as the
dot-dashed (green) and dashed (red) curves in
fig.~\ref{fig:XiVolPlotsu3}, respectively.  Comparing the expectations
with the results of the Lattice QCD calculations, manifested in the
fit value of $g_{\Xi^*\Xi\pi}$ being more than twice expectations,
suggest that the pionic contributions do not dominate the
finite-volume corrections, even after considering contributions from
higher orders in  ${\rm SU(2)}_L\otimes {\rm SU(2)}_R$ 
HB$\chi$PT.  As the $\Xi$ carries two strange
quarks, one expects kaon and $\eta$ loops to make relatively larger
finite-volume contributions to the $\Xi$ mass than to the nucleon,
$\Lambda$ and $\Sigma$ masses.

The NLO  ${\rm SU(3)}_L\otimes {\rm SU(3)}_R$ 
HB$\chi$PT prediction from  eq.~(\ref{eq:FVXisu3}) gives
rise to the curves shown in fig.~\ref{fig:XiVolPlotsu3}.
The dot-dashed (green) and dashed (red) curves are the contributions
from pions, while the dotted (orange) and solid (blue) curves
correspond to the sum of contributions from pions, kaons and an $\eta$.
HB$\chi$PT predicts that it is the kaons and $\eta$ that dominate the
finite-volume contributions to the $\Xi$ mass.  The exponential
suppression of the kaon and $\eta$ contributions is not sufficient to
overcome the relatively large axial coupling constants at this pion
mass.  These results suggest that NLO  ${\rm SU(3)}_L\otimes {\rm SU(3)}_R$ 
HB$\chi$PT provides a good
description of the finite-volume modifications to the $\Xi$ mass.
Again, one expects NLO HB$\chi$PT to provide an estimate of the finite
volume effects that would be modified at the $\sim 30\%$ level by
higher orders in the expansion.

\subsubsection{Summary}
\label{subsec:Summary}

\noindent The infinite-volume values of the octet-baryon masses fit
from the Lattice QCD data using HB$\chi$PT are consistent (almost
identical) with the infinite-volume values extracted from the simple
phenomenological fits presented above (see
table~\ref{tab:baryonmassesExtrapchipt} for a summary). However, there
are several important lessons that one learns from the HB$\chi$PT
analysis of the volume dependence of the octet-baryon masses. First,
by comparing, for instance, the dashed (red) and dot-dashed (green)
curves in fig.~\ref{fig:NucVolPlotsu3},
fig.~\ref{fig:LambdaVolPlotsu3}, fig.~\ref{fig:SigmaVolPlotsu3} and
fig.~\ref{fig:XiVolPlotsu3}, one sees the relevance of the
octet-decuplet axial transitions in the description of the
finite-volume effects. We conclude that HB$\chi$PT with the decuplet
states integrated out does not provide a reliable description of the
finite-volume effects at NLO at this pion mass. Second, by comparing,
for instance, the solid (blue) curve and the dashed (red) curve in
fig.~\ref{fig:NucVolPlotsu3}, fig.~\ref{fig:LambdaVolPlotsu3},
fig.~\ref{fig:SigmaVolPlotsu3} and fig.~\ref{fig:XiVolPlotsu3}, one
sees the relative importance of fluctuations to intermediate states
involving kaons and/or $\eta$, which are not captured in the
two-flavor chiral expansion. While these effects are small in the case
of the nucleon, they are significant for the hyperons at the heavy
pion mass at which the Lattice calculations were performed. Therefore,
while ${\rm SU(2)}_L\otimes {\rm SU(2)}_R$ HB$\chi$PT is adequate for
the nucleon, ${\rm SU(3)}_L\otimes {\rm SU(3)}_R$ HB$\chi$PT is
necessary to account for the finite-volume effects experienced by the
hyperons~\footnote{ We caution that reproducing the calculated volume
  dependence using HB$\chi$PT does not imply that the quark-mass
  dependence itself is under control. Indeed it is well known that
  ${\rm SU(3)}_L\otimes {\rm SU(3)}_R$ HB$\chi$PT converges poorly for
  certain
  quantities~\cite{WalkerLoud:2008bp,Ishikawa:2009vc,Torok:2009dg}.}. And
thus, the fit hyperon axial couplings presented in
table~\ref{tab:baryonmassesExtrapchipt} should not be considered
reliable.  This is, of course, due to $m_K -m_\pi$ not being small
enough, and therefore the ${\rm SU(2)}_L\otimes {\rm SU(2)}_R$ fits
for the hyperons will improve as the physical pion mass is approached.
It is important to emphasize that we have not propagated the
uncertainties associated with the input axial couplings and baryon
mass splittings, as we have found that the resulting uncertainties in
the fit quantities are smaller than the expected size of NNLO
effects. The various curves in fig.~\ref{fig:NucVolPlotsu3},
fig.~\ref{fig:LambdaVolPlotsu3}, fig.~\ref{fig:SigmaVolPlotsu3} and
fig.~\ref{fig:XiVolPlotsu3} become bands when the input parameter
uncertainties are included.
\begin{table}[!ht]
  \caption{The results of NLO ${\rm SU(2)}_L\otimes {\rm  SU(2)}_R$  HB$\chi$PT fits 
    to the volume dependence of the baryon masses.
    $M_B^{(\infty)}$ is the infinite-volume extrapolation of the baryon mass.
    The first uncertainty is statistical, the second is the fitting
    systematic, and the third (where appropriate) is due to scale setting. The uncertainties
    in the input axial couplings and baryon mass splittings have not been propagated, as discussed in the text.}
  \label{tab:baryonmassesExtrapchipt}
  \begin{ruledtabular}
    \begin{tabular}{c||cccc}
      Hadron  &  $M_B^{(\infty)}$ (t.l.u.) & $M_B^{(\infty)}$ (MeV) &  Axial Coupling   \\
      \hline
      $M_N$  &  0.20455(19)(17) & 1151.3(1.1)(1.0)(7.5) &  $|g_{\Delta N\pi}|\ =\ 2.80(18)(21)$ \\
      $M_\Lambda$ &  0.22064(15)(19) & 1241.9(0.8)(1.1)(8.1) &  $|g_{\Sigma^*\Lambda\pi}|\ =\  2.21(16)(23)$ \\
      $M_\Sigma$ &  0.22747(17)(19) & 1280.3(1.0)(1.1)(8.3) & $|g_{\Sigma^*\Sigma\pi}|\ <\ 1.38 [1.90]$\\
      $M_\Xi$ &  0.23978(12)(18) & 1349.6(0.7)(1.0)(8.8) &  $|g_{\Xi^*\Xi\pi}|\ =\ 2.49(23)(35)$\\
      \hline
    \end{tabular}
  \end{ruledtabular}
\end{table}
%

\subsection{Combinations of Masses}
\label{subsec:Relations}

\noindent 
In addition to examining the volume dependence of the baryon masses,
it is interesting to explore the volume dependence of certain
combinations of the masses.  In order to minimize both statistical and
systematic uncertainties in determining various combinations of masses
from the Lattice QCD calculation, in each case a correlation function
is formed from products or ratios of the individual baryon correlation
functions under a jackknife or bootstrap procedure, from which the
combination of masses is extracted.

\subsubsection{The Centroid of the Octet}
\label{subsec:octetcentroid}

\begin{figure}[!ht]
  \centering
     \includegraphics[width=0.9\textwidth]{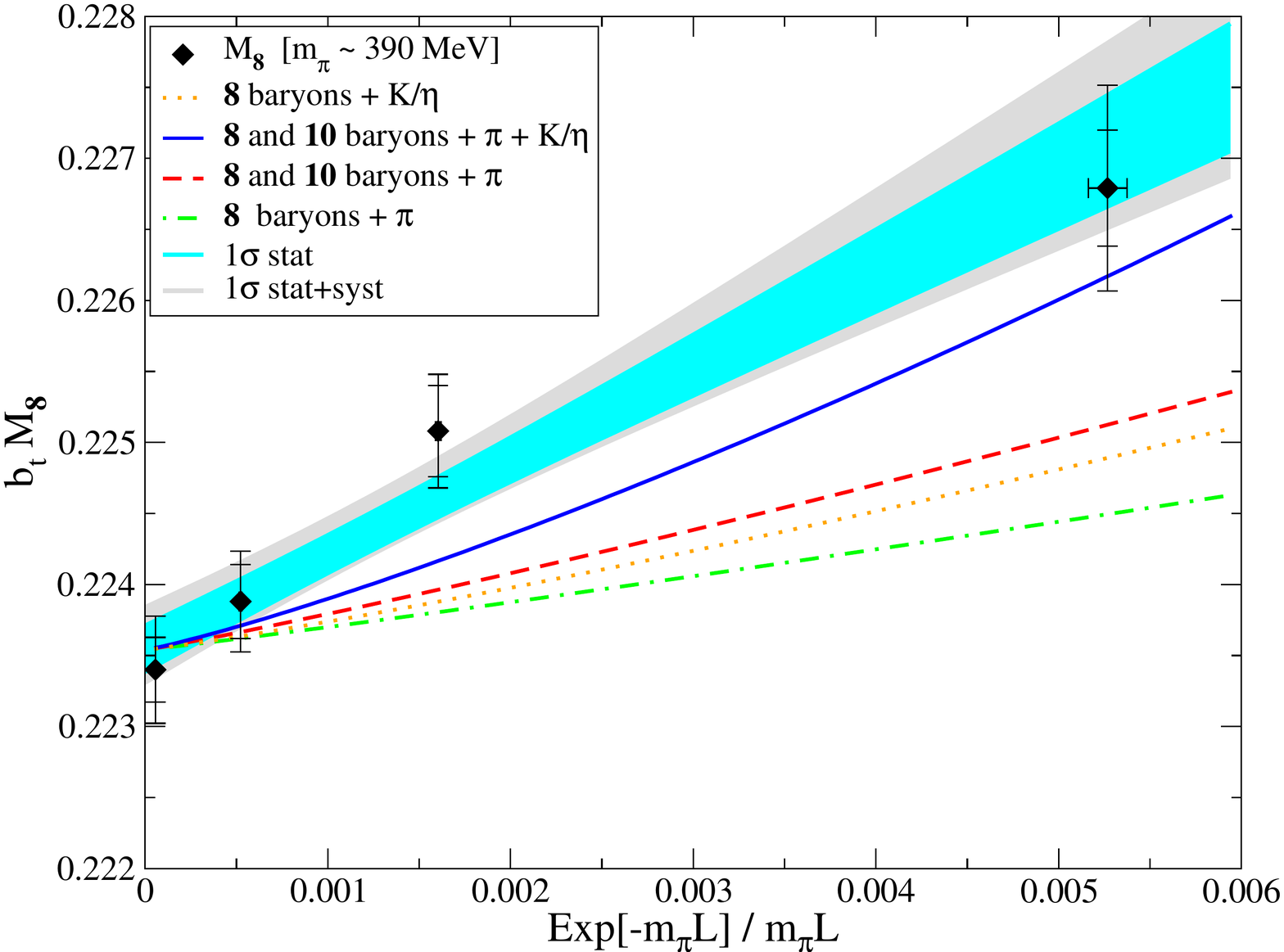}
     \caption{ The centroid of the baryon octet as a function of
       $e^{-m_\pi L}/( m_\pi L)$.  The dark (light) grey shaded region
       corresponds to the $1\sigma$ statistical uncertainty
       (statistical and systematic uncertainties combined in
       quadrature) resulting from fitting the $m_\pi L=\infty$ value
       and the coefficient of $e^{-m_\pi L}/( m_\pi L)$.  Using the
       value at $m_\pi L=\infty$, the dot-dashed curve (green)
       corresponds to the contribution from octet baryons and pions,
       the dotted curve (orange) corresponds to the contribution from
       octet baryons and kaons or an $\eta$, the dashed curve (red)
       corresponds to the contribution from octet and
       decuplet baryons and pions, and the solid curve (blue)
       corresponds to the contribution from octet and
       decuplet baryons and pions, kaons or an $\eta$.  }
  \label{fig:OCVolPlot}
\end{figure}

\noindent 
The centroid of the baryon octet is the sum of the masses weighted by the
isospin degeneracy of each state,
\begin{eqnarray}
M_{\bf 8} & = & 
{1\over 8} M_\Lambda + 
{3\over 8} M_\Sigma + 
{1\over 4} M_N + 
{1\over 4} M_\Xi 
\ \ \ .
\label{eq:octetcentroid}
\end{eqnarray}
The results of the Lattice QCD
calculations are shown in fig.~\ref{fig:OCVolPlot}, along with the results 
of a simple fit of the form described in subsection~\ref{subsec:Simple}.
The simple fit gives rise to a centroid mass of $M_8^{(\infty)} =
0.22354(17)(20)~{\rm t.l.u} = 1255.1(1.0)(1.1)(8.2)~{\rm MeV}$.  Also
shown in fig.~\ref{fig:OCVolPlot} are the predictions of both ${\rm
  SU(2)}_L\otimes {\rm SU(2)}_R$ and ${\rm SU(3)}_L\otimes {\rm
  SU(3)}_R$ NLO HB$\chi$PT resulting from the same input parameters
(not the fit parameters) as those used for the predictions of the
individual baryon masses.  In particular, the SU(3) relations between
the decuplet-octet axial coupling constants, and between the
octet-octet kaon and $\eta$ axial couplings, are employed.  Given the
overall general agreement between the leading predictions and the
individual baryon masses, it is unsurprising that the ${\rm
  SU(3)}_L\otimes {\rm SU(3)}_R$ prediction for the centroid mass of
the octet agrees reasonably well with the results of the Lattice QCD
calculation.

\subsubsection{The $\Sigma$-$\Lambda$ Mass Splitting}
\label{subsec:SLdiff}

\noindent The mass difference between the $\Sigma$ and the $\Lambda$ vanishes in the
limit of exact SU(3) symmetry.  In Nature, the splitting is found to be 
$M_\Sigma-M_\Lambda\sim 74~{\rm MeV}$, and consequently, at the pion mass of the
Lattice QCD calculations, the calculated splitting is expected to be small.
\begin{figure}[!ht]
  \centering
     \includegraphics[width=0.85\textwidth]{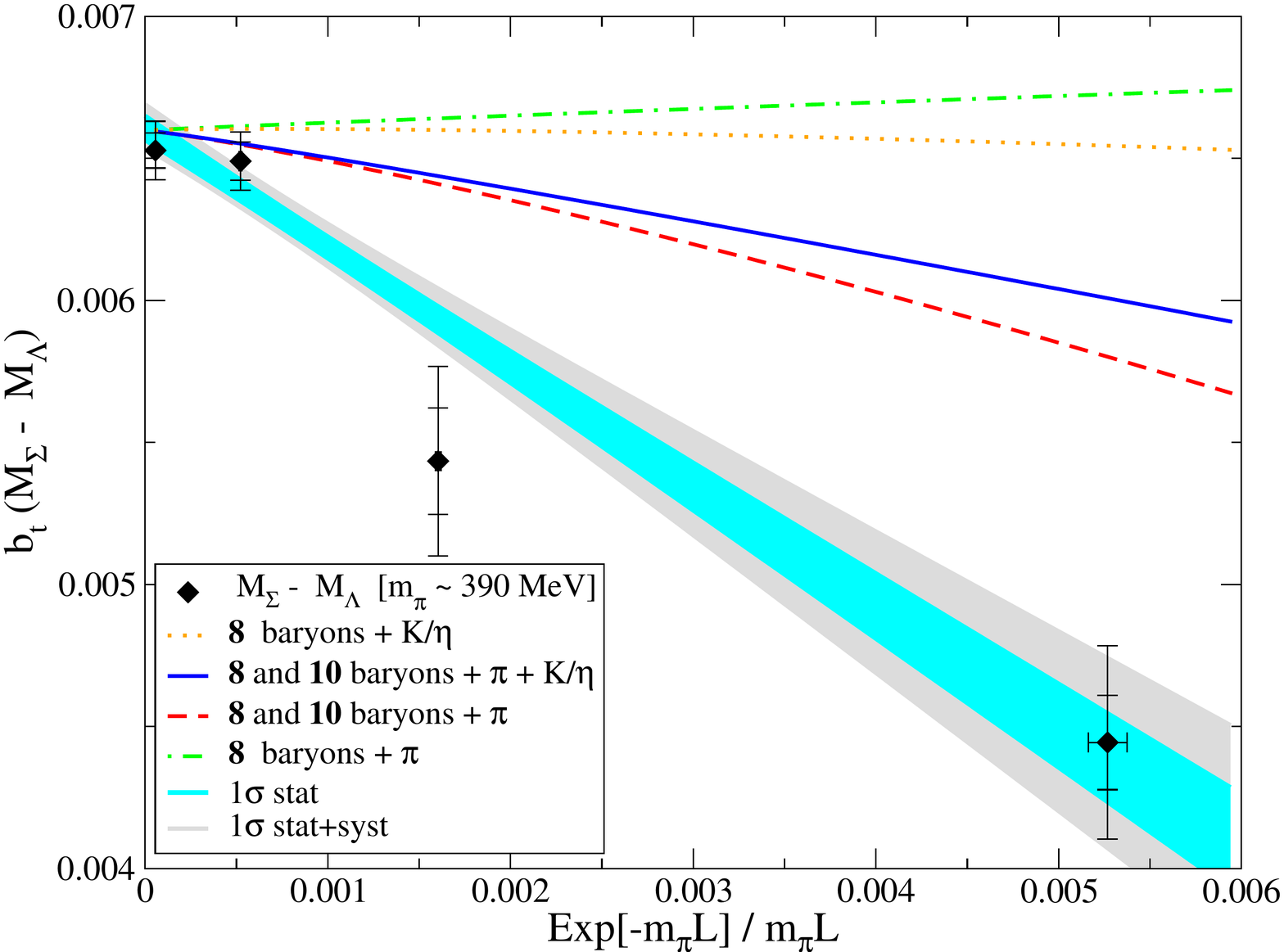}
     \caption{ The $\Sigma$-$\Lambda$ mass difference as a function of
       $e^{-m_\pi L}/( m_\pi L)$.  The dark (light) grey shaded region
       corresponds to the $1\sigma$ statistical uncertainty
       (statistical and systematic uncertainties combined in
       quadrature) resulting from fitting $(M_\Sigma -
       M_\Lambda)^{(\infty)}$ and the coefficient of $e^{-m_\pi L}/(
       m_\pi L)$.  Using this value of $(M_\Sigma -
       M_\Lambda)^{(\infty)}$, the dot-dashed curve (green)
       corresponds to the contribution from octet baryons and pions,
       the dotted curve (orange) corresponds to the contribution from
       octet baryons and kaons or an $\eta$, the dashed curve (red)
       corresponds to the contribution from octet and
       decuplet baryons and pions, and the solid curve (blue)
       corresponds to the contribution from octet and
       decuplet baryons and pions, kaons or an $\eta$.  }
  \label{fig:LambdaSigmaDIFFVolPlot}
\end{figure}
The results of the Lattice QCD calculation are shown in
fig.~\ref{fig:LambdaSigmaDIFFVolPlot}, along with the results of a
simple fit, of the form given in eq.~(\ref{eq:FVsimplefit}), shown as
the shaded regions.  The result of the simple fit gives $(M_\Sigma -
M_\Lambda)^{(\infty)} = 0.006598(48)(63)~{\rm t.l.u} =
37.05(27)(35)(24)~{\rm MeV}$, which is approximately half of its value
at the physical quark masses.  This is, in part, due to the strange
quark mass used in the calculation being somewhat lighter than its
value in Nature.  The finite-volume contributions significantly
suppress the mass splitting in smaller volumes.

The NLO expressions for the mass splitting do not describe the
observed volume dependence well.  While the full  ${\rm SU(3)}_L\otimes {\rm
  SU(3)}_R$  NLO
amplitude agrees in its sign with the volume dependence, the magnitude
is significantly smaller than the results of the Lattice QCD
calculations.  It is clear that SU(3) breaking contributions that
enter at higher orders in the chiral expansion play an important role
in the volume dependence of the $\Sigma$-$\Lambda$ mass splitting.

\subsubsection{The Gell-Mann--Okubo Mass Relation}
\label{subsec:GMO}

\begin{figure}[!ht]
  \centering
     \includegraphics[width=0.9\textwidth]{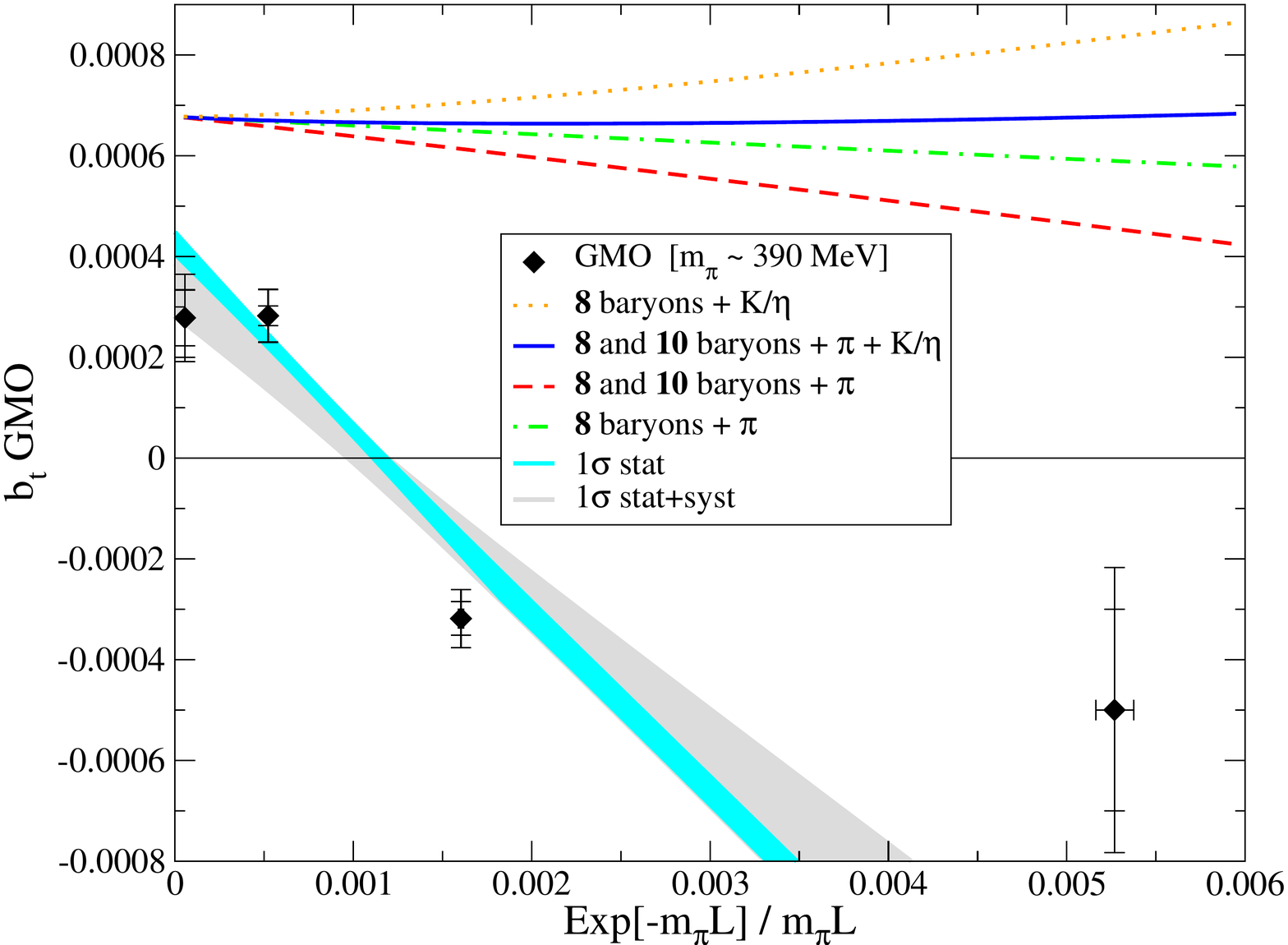}
     \caption{ The GMO relation as a function of
       $e^{-m_\pi L}/( m_\pi L)$.  The dark (light) grey shaded region
       corresponds to the $1\sigma$ statistical uncertainty
       (statistical and systematic uncertainties combined in
       quadrature) resulting from fitting the $m_\pi L=\infty$ value
       and the coefficient of $e^{-m_\pi L}/( m_\pi L)$.  The
       dot-dashed curve (green) corresponds to the contribution from
       octet baryons and pions, the dotted curve (orange) corresponds
       to the contribution from octet baryons and kaons or an $\eta$,
       the dashed curve (red) corresponds to the contribution from
       octet and decuplet baryons and pions, and the solid curve
       (blue) corresponds to the contribution from octet and
       decuplet baryons and pions, kaons or an $\eta$.  }
  \label{fig:GMOVolPlot}
\end{figure}

\noindent 
The GMO relation, 
\begin{eqnarray}
{\rm GMO} & = & 
M_\Lambda \ +\  
{1\over 3} M_\Sigma \ -\ 
{2\over 3} M_N \ -\  
{2\over 3} M_\Xi 
\ \ \ ,
\label{eq:GMOnum}
\end{eqnarray}
vanishes in the limit of exact SU(3) flavor symmetry, and also
vanishes in the limit where the SU(3) breaking transforms as an {\bf
  8} under SU(3) flavor symmetry.  Consequently, it is a valuable
probe of the structure of flavor symmetry breaking, being non-zero
only for breaking that transform in the {\bf 27} irreducible
representations of SU(3)~\footnote{Only the symmetric irreps in ${\bf
    8}\otimes {\bf 8} ={\bf 27}\oplus{\bf 10}\oplus\overline{\bf
    10}\oplus{\bf 8}\oplus{\bf 8}\oplus{\bf 1}$ are allowed, i.e. the
  ${\bf 27}$, ${\bf 8}$, and ${\bf 1}$.  By construction the ${\bf 8}$
  and ${\bf 1}$ do not contribute.  }.  The results of the Lattice QCD
calculations are shown in fig.~\ref{fig:GMOVolPlot}, along with the
results of the simple fit, of the form given in
eq.~(\ref{eq:FVsimplefit}), shown as the shaded regions.
The result of the simple fit gives ${\rm GMO}^{(\infty)} = 3.49(25)(46)\times 10^{-4}~{\rm t.l.u} =
1.96(14)(26)(01)~{\rm MeV}$.
Given the smallness of the GMO combination, it is no surprise that it
has a large volume dependence, even changing sign between the
$20^3\times 128$ and the $24^3\times 128$ lattice volumes.

As the GMO relation is sensitive only to the {\bf 27} SU(3) breaking,
both the finite-volume and infinite-volume contributions are
calculable in ${\rm SU(3)}_L\otimes {\rm SU(3)}_R$ HB$\chi$PT at one
loop (which gives a finite result).  It is straightforward to show
that the infinite-volume value of the GMO relation
is~\cite{Jenkins:1991ts}
\begin{eqnarray}
GMO^{(NLO)} & = & 
{1\over 24\pi f_\pi^2} 
\left[ 
\left( {2\over 3}D^2 - 2 F^2  \right) \left( 4 m_K^3 - 3 m_\eta^3 - m_\pi^3  \right)
- {{\cal C}^2\over 9\pi } 
\left( 4 F_K - 3 F_\eta - F_\pi \right)
\right],\hspace{.4cm}
\label{eqn:chiral}
\end{eqnarray}
where the function $F_{c}=F( m_{c},\Delta,\mu)$ is
\begin{eqnarray}
F (m,\Delta,\mu) & = & 
\left(m^2-\Delta^2\right)\left(
\sqrt{\Delta^2-m^2} \log\left({\Delta -\sqrt{\Delta^2-m^2+i\epsilon}\over
\Delta +\sqrt{\Delta^2-m^2+i\epsilon}}\right)
-\Delta \log\left({m^2\over\mu^2}\right)\ \right)
\nonumber\\
& - & {1\over 2}\Delta m^2 \log\left({m^2\over\mu^2}\right)
\ \ \ .
\label{eq:massfun}
\end{eqnarray}
We have set $f_\pi=f_K=150~{\rm MeV}$ in the GMO relation to eliminate formally
higher-order contributions that depend upon the renormalization scale,
$\mu$, from this expression.  Inserting the values of the constants
and parameters that we have used previously into
eq.~(\ref{eq:massfun}) gives a value of ${\rm GMO}^{(NLO)} \sim 6.8\times
10^{-4}~{\rm t.l.u.}$, which is a factor of two greater than the
extrapolation of the Lattice QCD results.  This is not that surprising
given the expected size of the higher-order contributions in the
HB$\chi$PT expansion, as well as the fact that this quantity is
anomalously small (suppressed beyond naive expectations based upon
SU(3) symmetry alone due to a further suppression by $1/N_c^2$ in the
large-$N_c$ limit of QCD~\footnote{If $\epsilon\sim m_s-m_{u,d}$ denotes the SU(3)
breaking parameter, then the GMO relation scales as $\sim \epsilon^{3/2}/N_c^2$
relative to the baryon masses; see, for example, eq.~(\ref{eqn:chiral}).
For a review, see Ref.~\cite{Manohar:1998xv}.}).

The NLO predictions of the volume dependence of the GMO relation in
${\rm SU(2)}_L\otimes {\rm SU(2)}_R$ and ${\rm SU(3)}_L\otimes {\rm
  SU(3)}_R$ HB$\chi$PT are shown in fig.~\ref{fig:GMOVolPlot}.  The
contribution from the kaons and the $\eta$ are of opposite sign to
that from the pion (as expected from their cancellation in the SU(3)
limit), which gives rise to a small volume dependence, even on the
scale of fig.~\ref{fig:GMOVolPlot}.  The predictions for the volume
dependence of the GMO relation are in clear contradiction with the
results of the Lattice QCD calculations.  Clearly NLO HB$\chi$PT does
not describe the higher-dimensional SU(3) breaking that provides the
finite-volume dependence of the GMO relation, and we have found that
this relation is particularly sensitive to the volume of the lattice.

The GMO relation was previously explored by some of the present
authors~\cite{Beane:2006pt,WalkerLoud:2008bp}.  At approximately the
same pion mass, and in the volume with $L\sim 2.5~{\rm fm}$, the GMO
relation was found to be positive and consistent with the loop-level
expression given in eq.~(\ref{eq:massfun}).  This is in slight
disagreement with the current determination; however, neither
calculation extracts values in the continuum limit and the difference
in this very small quantity may arise from the different
discretizations and/or the somewhat different values of the
strange-quark mass.

\subsubsection{The $R_4$ Mass Relation}
\label{subsec:R4}

\begin{figure}[!ht]
  \centering
     \includegraphics[width=0.9\textwidth]{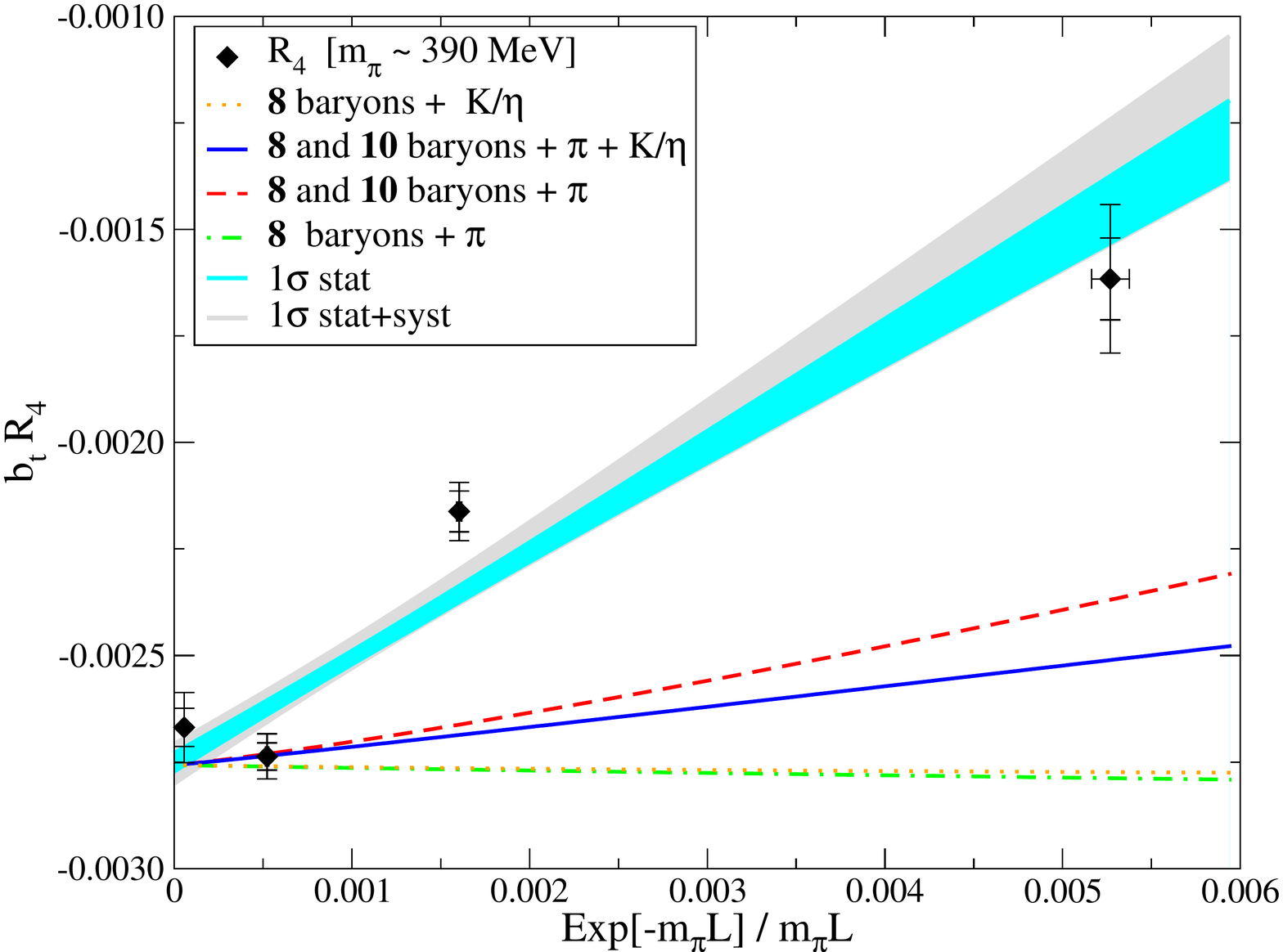}
     \caption{ The $R_4$ mass relation as a function of
       $e^{-m_\pi L}/( m_\pi L)$.  The dark (light) grey shaded region
       corresponds to the $1\sigma$ statistical uncertainty
       (statistical and systematic uncertainties combined in
       quadrature) resulting from fitting the $m_\pi L=\infty$ value
       and the coefficient of $e^{-m_\pi L}/( m_\pi L)$.  The
       dot-dashed curve (green) corresponds to the contribution from
       octet baryons and pions, the dotted curve (orange) corresponds
       to the contribution from octet baryons and kaons or an $\eta$,
       the dashed curve (red) corresponds to the contribution from
       octet and decuplet baryons and pions, and the solid curve
       (blue) corresponds to the contribution from octet and
       decuplet baryons and pions, kaons or an $\eta$.  
}
  \label{fig:R4VolPlot}
\end{figure}

\noindent 
The $R_4$ relation, defined to be 
\begin{eqnarray}
R_4 & = & {1\over 6}\ \left(\ 
M_N\ +
M_\Lambda \ +\  
M_\Xi \ -\ 
3 M_\Sigma 
\ \right)
\ \ \ ,
\label{eq:R4rel}
\end{eqnarray}
vanishes in the limit of exact SU(3) flavor symmetry, and is formally
dominated by a single insertion of the light-quark mass matrix in the
infinite volume limit.  This relation is of phenomenological interest
because not only does it vanish in the limit of exact SU(3) flavor
symmetry, but it also vanishes in the large-$N_c$ limit of
QCD~\cite{Manohar:1998xv,Jenkins:1995td,Jenkins:2009wv}, scaling as
$\sim\epsilon/N_c$ relative to the baryon masses ($\epsilon$ is
defined in the footnote in subsection~\ref{subsec:GMO}).  Recently,
this relation, along with other relations among masses were examined
with Lattice QCD calculations in work by Jenkins {\it et
  al.}~\cite{Jenkins:2009wv} using domain-wall light-quark and
strange-quark propagators generated on a number
of ensembles of improved Kogut-Susskind dynamical
quarks~\cite{Orginos:1999cr,Bernard:2001av}.  The results of the
present Lattice QCD calculations are shown in
fig.~\ref{fig:R4VolPlot}, along with the results of the simple fit, of
the form given in eq.~(\ref{eq:FVsimplefit}), shown as the shaded
regions.
The result of the simple fit gives  
$R_4^{(\infty)} = -0.002757(27)(40)~{\rm t.l.u} =
-15.48(15)(22)(09)~{\rm MeV}$, which is approximately half of its value at the
physical quark masses, $R_4^{\rm expt} = -34.5~{\rm MeV}$.  
The finite-volume contributions  significantly suppress the
mass splitting in smaller volumes.

The NLO expressions for $R_4$ do not describe the observed volume dependence 
well. 
While the full  ${\rm SU(3)}_L\otimes {\rm  SU(3)}_R$  NLO
amplitude agrees in its sign with the volume dependence, the magnitude
is significantly smaller than the results of the Lattice QCD
calculations.  SU(3) breaking contributions that
enter beyond NLO  in HB$\chi$PT play an important role
in the $R_4$ mass relation.

\section{The Volume Dependence of the Meson Masses}
\label{sec:Mvol}
\noindent
It is also interesting to explore the volume dependence of the meson
masses, specifically that of the pion and the kaon.  There has been
significantly more theoretical and numerical exploration of the meson
masses and how they depend upon the volume of the lattices used in
Lattice QCD calculations.  An overview of the theoretical status of
such finite-volume contributions can be found in
Ref.~\cite{Colangelo:2005gd}.

\begin{figure}[!ht]
  \centering
     \includegraphics[width=0.49\textwidth]{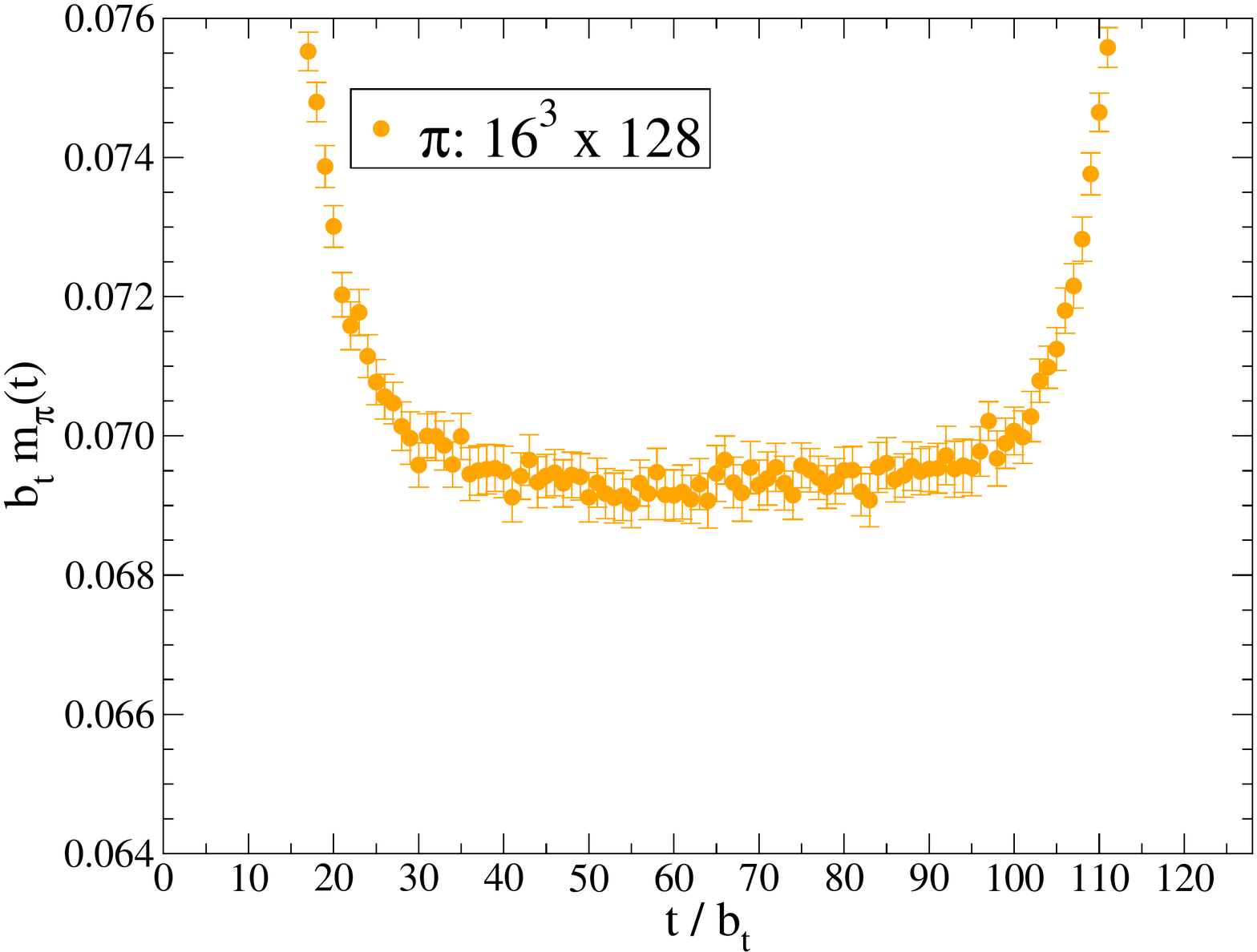} 
     \includegraphics[width=0.49\textwidth]{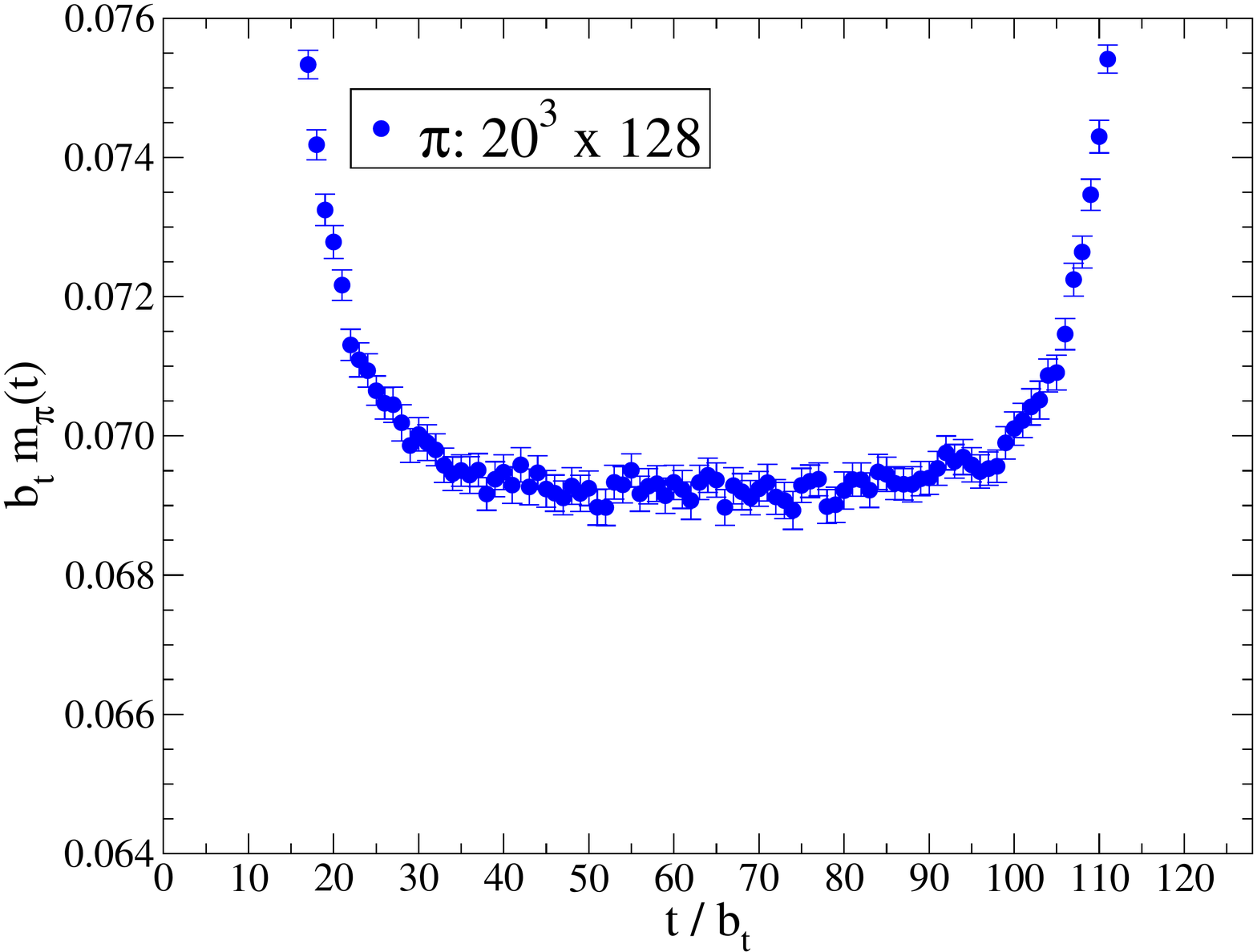} 
     \includegraphics[width=0.49\textwidth]{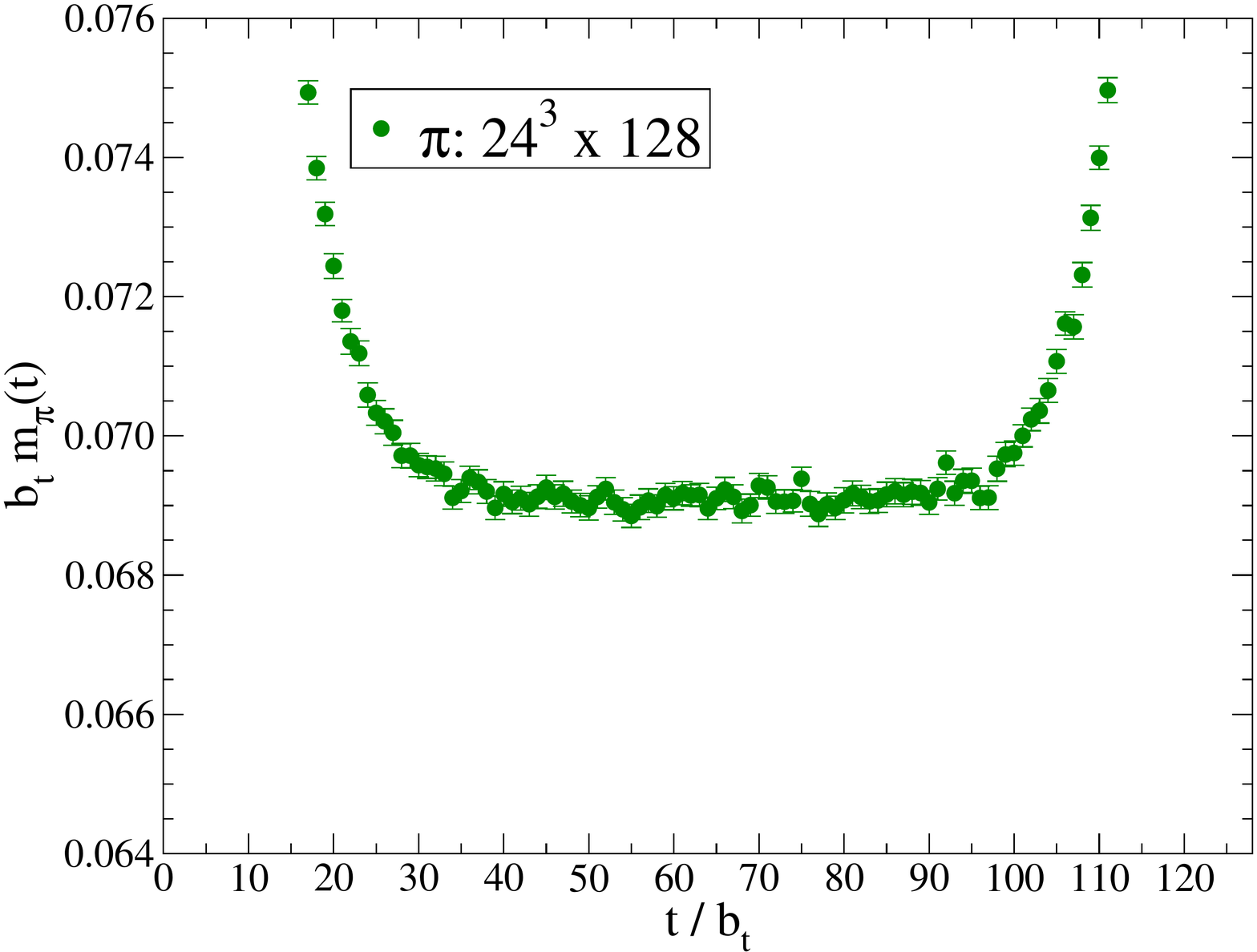} 
     \includegraphics[width=0.49\textwidth]{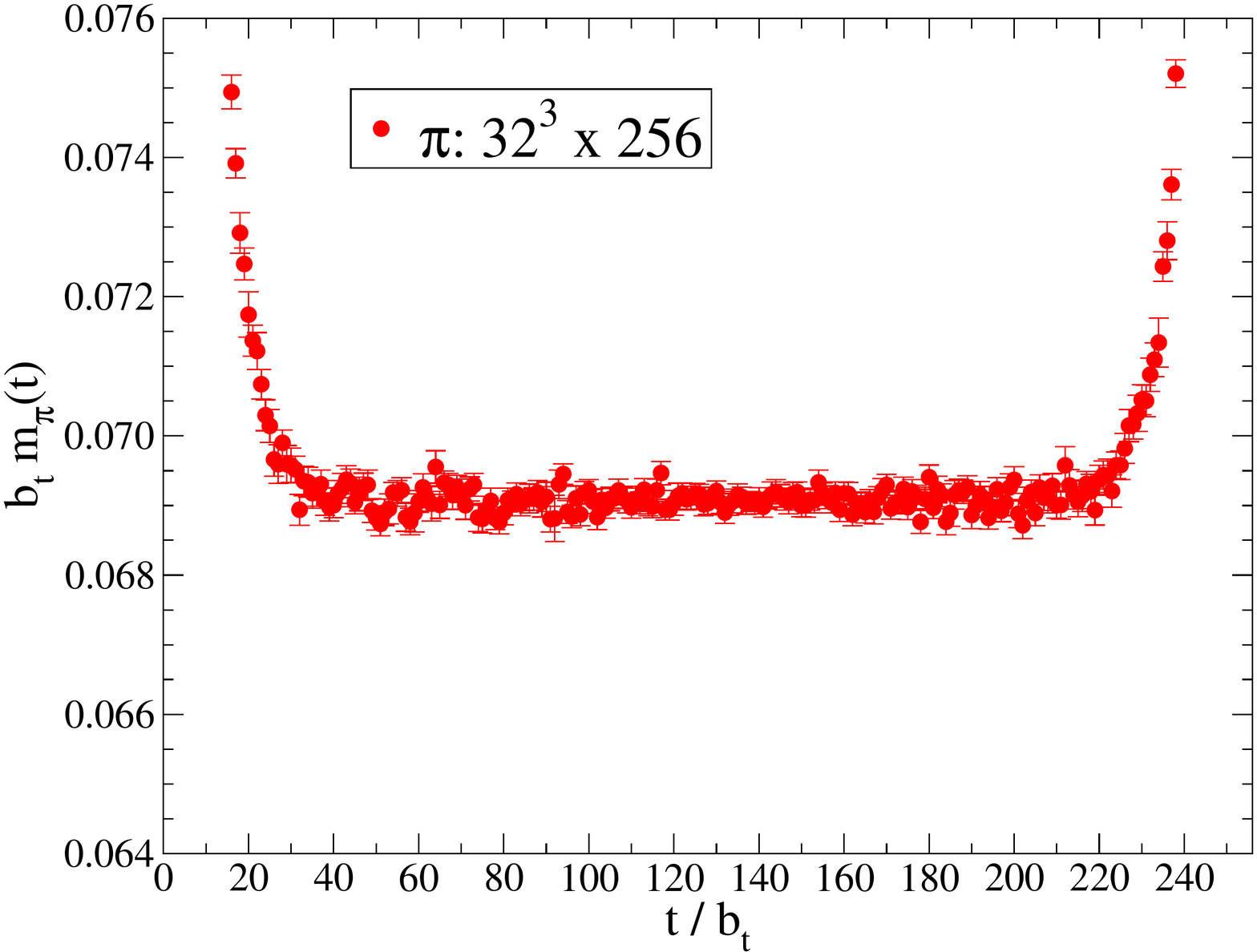} 
     \caption{
The pion EMP's determined 
on the four lattice ensembles used in this work. 
Note that the y-axis scale is the same in all four panels.
}
  \label{fig:emppions}
\end{figure}

\begin{figure}[!ht]
  \centering
     \includegraphics[width=0.49\textwidth]{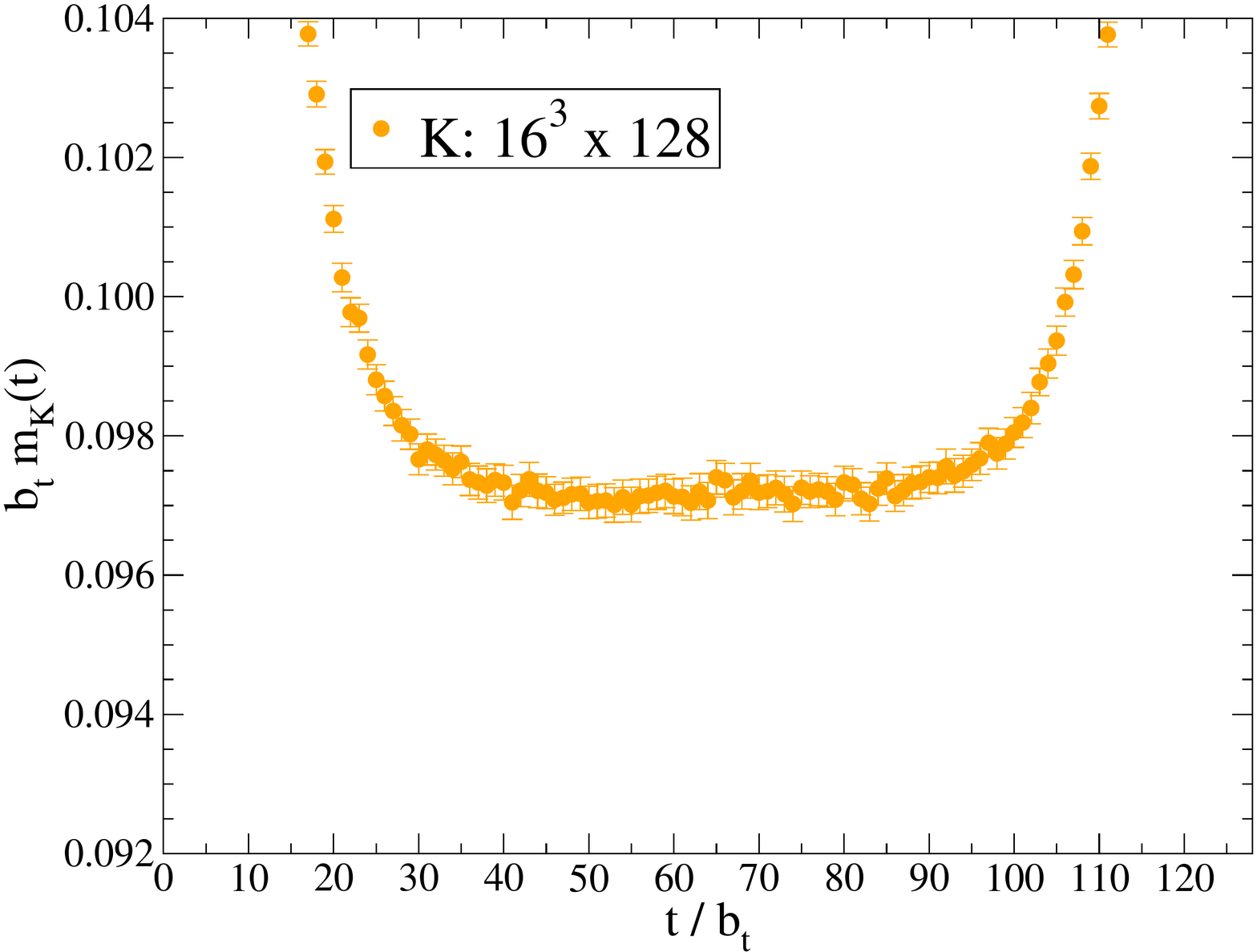} 
     \includegraphics[width=0.49\textwidth]{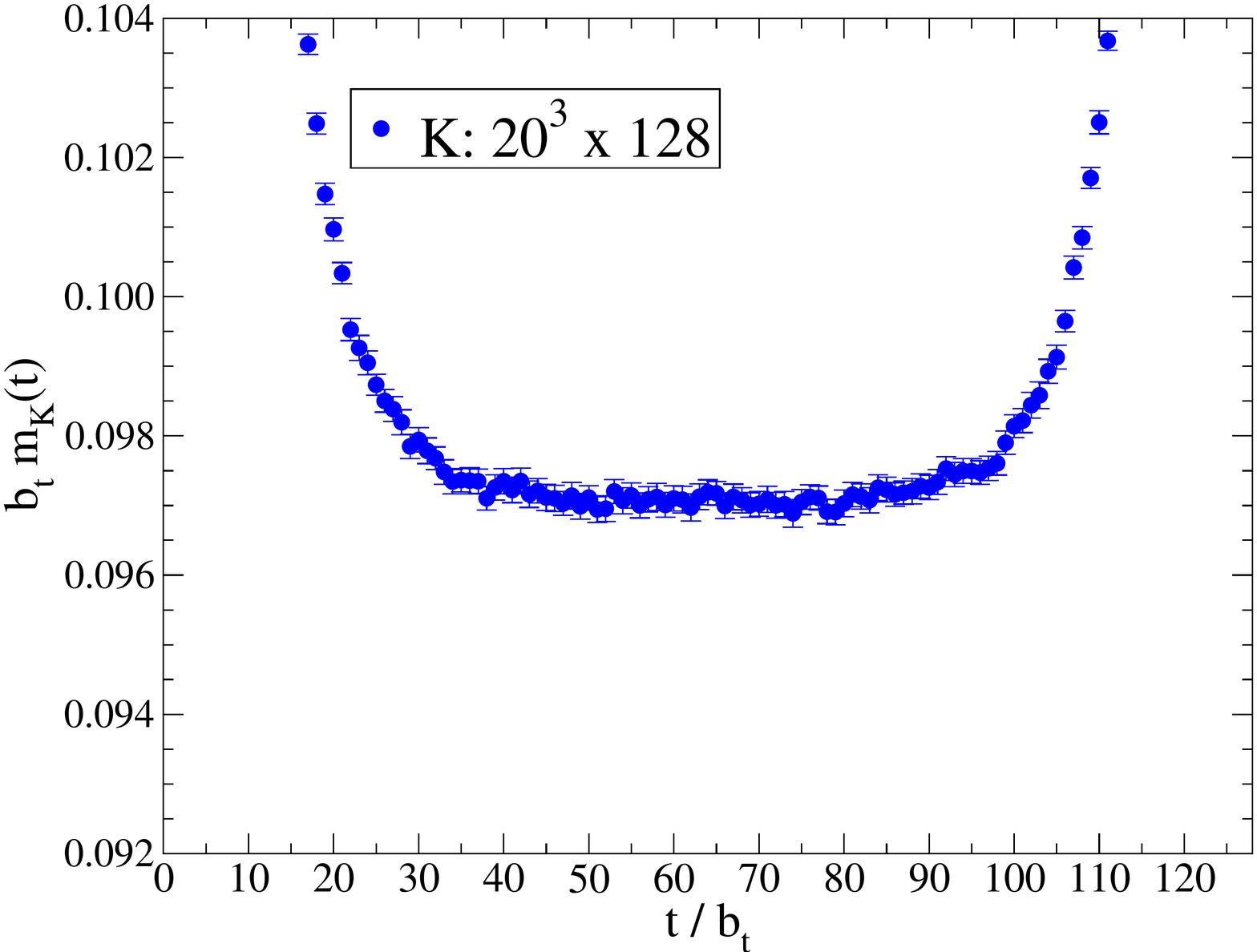} 
     \includegraphics[width=0.49\textwidth]{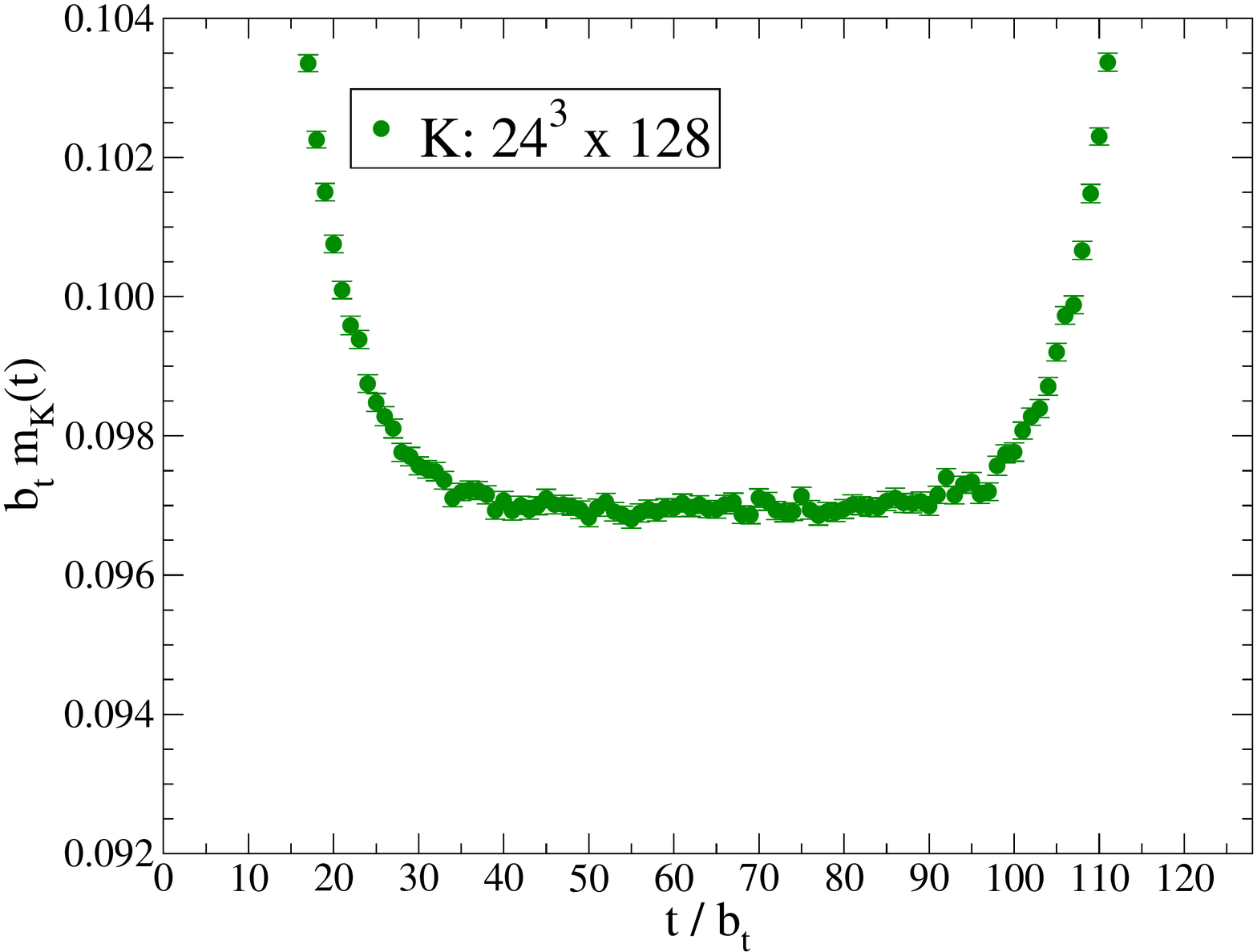} 
     \includegraphics[width=0.49\textwidth]{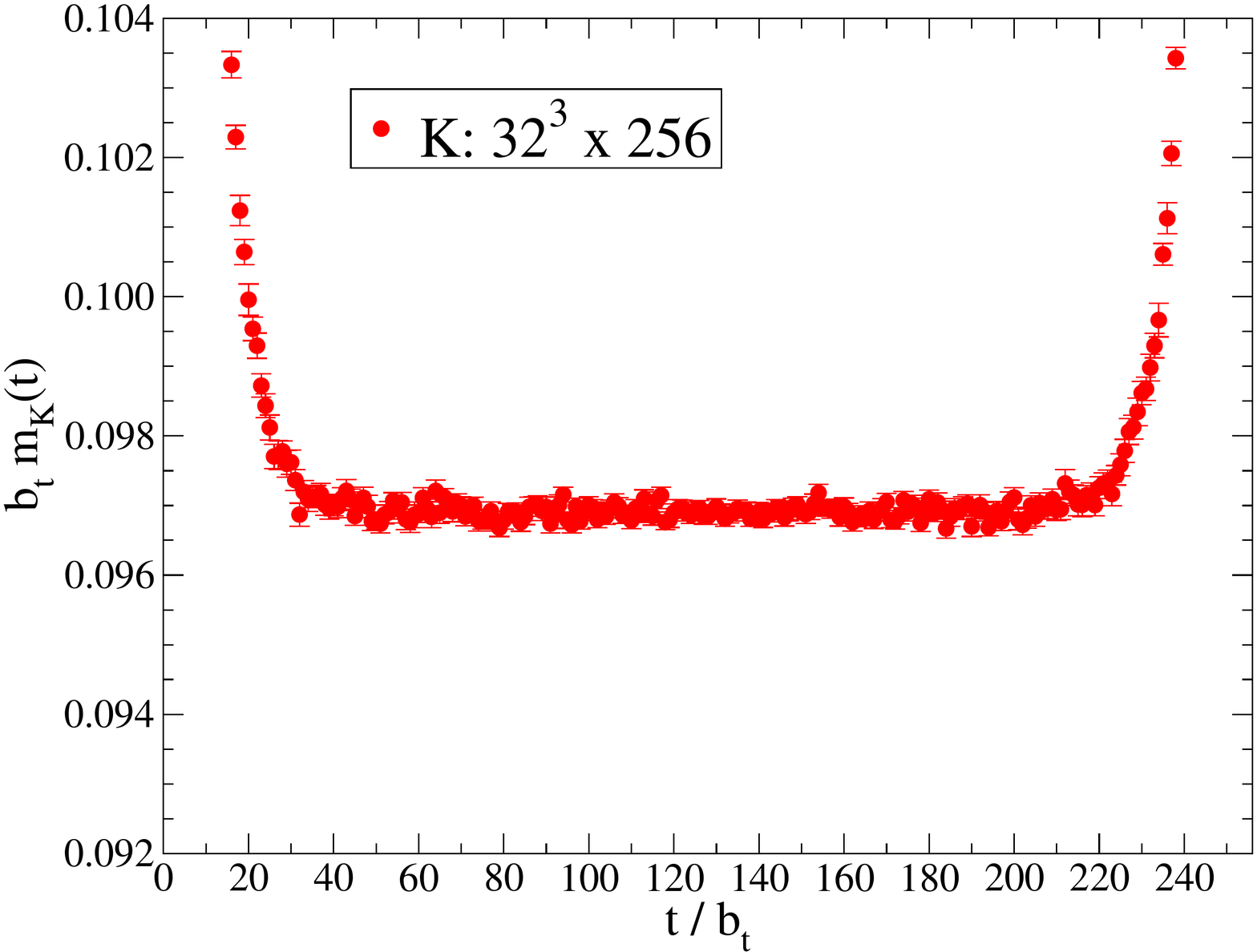} 
     \caption{
The kaon EMP's determined 
on the four lattice ensembles used in this work. 
Note that the y-axis scale is the same in all four panels.
}
  \label{fig:empkaons}
\end{figure}

The results of the present Lattice QCD calculations of the meson
masses in the four different lattice volumes are given in
table~\ref{tab:LQCDmesonmasses} and the EMP's are shown in
fig.~\ref{fig:emppions} and fig.~\ref{fig:empkaons}.
\begin{table}[!ht]
  \caption{
Meson masses from the Lattice QCD calculations in the 
four lattice volumes.
  }
  \label{tab:LQCDmesonmasses}
  \begin{ruledtabular}
    \begin{tabular}{c||cccc}
      $L^3\times T$  &  $16^3\times 128$ &  $20^3\times 128$ &  $24^3\times 128$ &
      $32^3\times 256$  \\
\hline
      $m_\pi$ (t.l.u.) &  0.06943(36)(0) & 0.06936(12)(0) & 0.06903(19)(0) & 0.069060(66)(81)\\
      $m_K$ (t.l.u.) &  0.09722(26)(0) & 0.09702(10)(03) & 0.09684(15)(01) & 0.096984(78)(60)\\
      \hline
    \end{tabular}
  \end{ruledtabular}
\end{table}

\subsection{The Pion Mass}
\label{subsec:Pion}

\begin{figure}[!ht]
  \centering
     \includegraphics[width=0.9\textwidth]{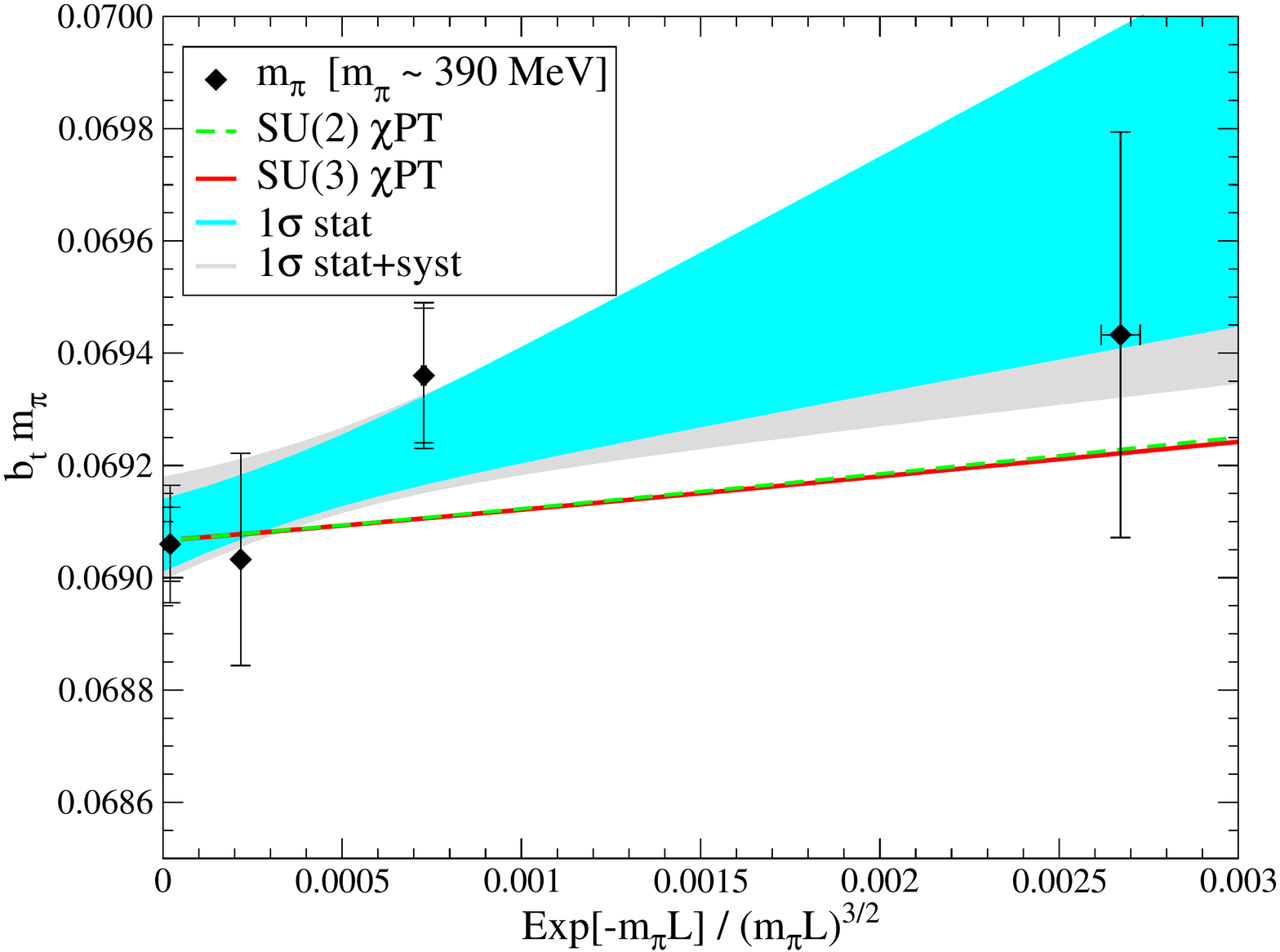}
     \caption{ The mass of the pion as a function of $e^{-m_\pi
         L}/(m_\pi L)^{3/2}$.  The points and associated uncertainties
       (blue) are the results of the Lattice QCD calculations, as
       given in table~\ref{tab:LQCDmesonmasses}.  The dark (light)
       shaded region corresponds to the $1\sigma$ statistical
       uncertainty (statistical and systematic uncertainties combined
       in quadrature) associated with a fit of the form given in
       eq.~(\protect\ref{eq:FVsimplefitmeson}).  The solid (red) curve
       corresponds to the prediction of  NLO ${\rm SU(3)}_L\otimes {\rm SU(3)}_R$ $\chi$PT, while the
       dashed (green) curve corresponds to the prediction of  NLO ${\rm SU(2)}_L\otimes {\rm SU(2)}_R$
       $\chi$PT using the value of $m_\pi^{(\infty)}$ found in the
       fit.  }
  \label{fig:Pionvol}
\end{figure}

\noindent 
The finite-volume contribution to the mass of the pion in  ${\rm SU(2)}_L\otimes {\rm SU(2)}_R$
$\chi$PT is given by~\cite{Gasser:1986vb}
\begin{eqnarray}
m_\pi(L) - m_\pi(\infty)  =  
{3 m_\pi^3\over 4\pi^2 f_\pi^2} {1\over m_\pi L}
\left[ K_1(m_\pi L) + \sqrt{2}  K_1(\sqrt{2} m_\pi L)
 + {4\over 3\sqrt{3}}  K_1(\sqrt{3} m_\pi L) + \ldots
\right]\hspace{.2cm}
\label{eq:luschermeson}
\end{eqnarray}
where $K_1(x)$ is the modified Bessel function. The meson masses have
different overall volume scaling to the baryons, due to the absence of a
three-meson vertex. 
As $K_1(z)\rightarrow e^{-z}/\sqrt{z}$, the
results of the Lattice QCD calculations are shown in
fig.~\ref{fig:Pionvol} as a function of $e^{-m_\pi L}/(m_\pi L)^{3/2}$
rather than $e^{-m_\pi L}/(m_\pi L)$ as was used for the baryons.
Consequently, the naive fit that we perform to the meson masses is of
the form
\begin{eqnarray}
m_M^{(V)} (m_\pi L) & = & m_M^{(\infty)}\ +\ 
c_M^{(V)}\ {e^{- m_\pi\  L}\over (m_\pi L)^{3/2}}
\ \ \ .
\label{eq:FVsimplefitmeson}
\end{eqnarray}
With the current precision of the Lattice QCD calculation, we cannot
distinguish between the fit forms of $e^{-m_\pi L}/(m_\pi L)$ and
$e^{-m_\pi L}/(m_\pi L)^{3/2}$ with statistical significance.  The fit
parameters are $m_\pi^{(\infty)} = 0.069073(63)(62)~{\rm
  t.l.u.}=387.8(0.4)(0.4)(2.5)~{\rm MeV}$ and
$c_\pi^{(V)}=0.23(12)(07)~{\rm t.l.u.}=(1.30(65)(39)(01))\times
10^3~{\rm MeV}$.

At NLO in  ${\rm SU(3)}_L\otimes {\rm SU(3)}_R$ 
$\chi$PT, the finite-volume corrections to the pion mass are
given by~\cite{Colangelo:2005gd}
\begin{eqnarray}
\delta m_\pi 
& = & 
{3 m_\pi^3\over 4\pi^2 f_\pi^2}\ {1\over m_\pi L}\
\left[\ K_1(m_\pi L)\ +\ \sqrt{2}  K_1(\sqrt{2} m_\pi L)
\ +\ {4\over 3\sqrt{3}}  K_1(\sqrt{3} m_\pi L)\ +\ ...
\right]
\nonumber\\
& - & 
{m_\pi m_\eta^2\over 4\pi^2 f_\eta^2}\ {1\over m_\eta L}\
\left[\ K_1(m_\eta L)\ +\ \sqrt{2}  K_1(\sqrt{2} m_\eta L)
\ +\ {4\over 3\sqrt{3}}  K_1(\sqrt{3} m_\eta L)\ +\ ...
\right]
\ .
\label{eq:FVpionsu3}
\end{eqnarray}
Using the value of $m_\pi^{(\infty)}$ found in the fit to the form in
eq.~(\ref{eq:FVsimplefitmeson}), the predicted volume dependence is
shown in fig.~\ref{fig:Pionvol} as the solid ( ${\rm SU(3)}_L\otimes {\rm SU(3)}_R$) and
dashed curves ( ${\rm SU(2)}_L\otimes {\rm SU(2)}_R$).  
The volume dependence that is found in
the Lattice QCD calculations agrees with the expectations of
NLO $\chi$PT, and is significantly smaller than that of
the baryon masses.

\subsection{The Kaon Mass}
\label{subsec:Kaon}

\begin{figure}[!ht]
  \centering
     \includegraphics[width=0.9\textwidth]{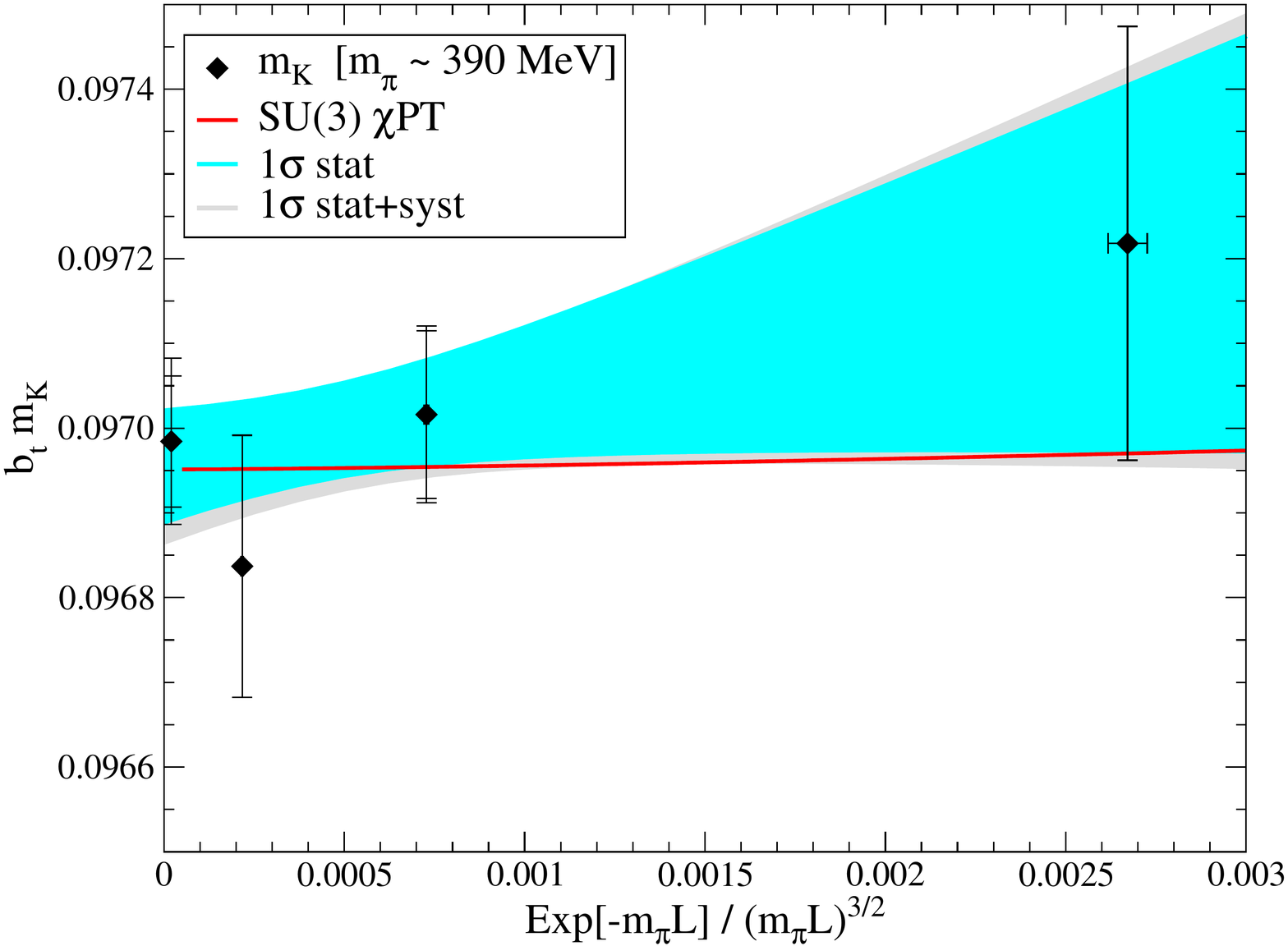}
     \caption{ The mass of the kaon as a function of $e^{-m_\pi
         L}/(m_\pi L)^{3/2}$.  The points and associated uncertainties
       (blue) are the results of the Lattice QCD calculations, as
       given in table~\ref{tab:LQCDmesonmasses}.  The dark (light)
       shaded region corresponds to the $1\sigma$ statistical
       uncertainty (statistical and systematic uncertainties combined
       in quadrature) associated with a fit of the form given in
       eq.~(\protect\ref{eq:FVsimplefitmeson}).  The solid (red) curve
       corresponds to the prediction of  NLO ${\rm SU(3)}_L\otimes {\rm SU(3)}_R$ $\chi$PT.  }
  \label{fig:Kaonvol}
\end{figure}

\noindent The formalism describing the volume dependence of the kaon
mass is analogous to that of the pion.  The NLO calculation in
 ${\rm SU(3)}_L\otimes {\rm SU(3)}_R$ $\chi$PT gives~\cite{Colangelo:2005gd}
\begin{eqnarray}
\delta m_K 
& = & 
{m_K m_\eta^2\over 2\pi^2 f_\eta^2}\ {1\over m_\eta L}\
\left[\ K_1(m_\eta L)\ +\ \sqrt{2}  K_1(\sqrt{2} m_\eta L)
\ +\ {4\over 3\sqrt{3}}  K_1(\sqrt{3} m_\eta L)\ +\ ...
\right].
\label{eq:FVkaonsu3}
\end{eqnarray}
The results of the Lattice QCD calculation, given in
table~\ref{tab:LQCDmesonmasses}, are shown in fig.~\ref{fig:Kaonvol},
along with a fit to the form given in eq.~(\ref{eq:FVsimplefitmeson}).
The resulting fit values are 
$m_K^{(\infty)} = 0.096953(68)(39)~{\rm t.l.u.}=544.4(0.4)(0.2)(3.5)~{\rm MeV}$ and 
$c_K^{(V)}=0.087(93)(44)~{\rm t.l.u.}=(4.9(5.2)(2.5)(0.0))\times 10^2~{\rm MeV}$. 
The volume dependence of the kaon is found to be very small, much
smaller than that of the baryons, and is
consistent with the absence of pion
loop contributions.

\section{Conclusions}
\label{sec:Conclusions}

\noindent 
We have performed precise Lattice QCD calculations of the
low-lying hadron masses at a pion mass of $m_\pi\sim 390~{\rm MeV}$ in
four ensembles of anisotropic clover gauge-field configurations with a
spatial lattice spacing of $b_s\sim 0.123~{\rm fm}$, an anisotropy of
$\xi=3.5$ and cubic spatial lattice volumes with extent $L\sim 2.0,
2.5, 3.0$ and $3.9~{\rm fm}$.  These calculations have allowed for a
detailed exploration of the volume dependence of the octet baryon
masses and of the pion and the kaon masses.

Our main conclusions are as follows:

\begin{itemize}
\item In order to calculate individual baryon masses with
  percent-level precision ($\pm 10~{\rm MeV}$), it is sufficient to
  work in volumes with $m_\pi L\gsim 4.3$ for $m_\pi\sim 390~{\rm
    MeV}$, and NLO HB$\chi$PT indicates that somewhat smaller values
  of $m_\pi L$ may be sufficient at lighter pion masses.

\item The expectations of NLO ${\rm SU(3)}_L\otimes {\rm SU(3)}_R$
  HB$\chi$PT (with the meson decay constants evaluated at the
  appropriate meson mass) are found to qualitatively reproduce the
  lattice results for the volume dependence of all of the baryon and
  meson masses.  The NLO ${\rm SU(3)}_L\otimes {\rm SU(3)}_R$
  predictions are sufficiently accurate to allow for meaningful
  extrapolations to lighter pion masses to be made, where the volume
  dependences are expected to be significantly smaller.  In the
  context of determinations of two-body interactions, this feature
  allows for gauge-field configurations with somewhat
  (logarithmically) smaller values of $m_\pi L$ to be used for the
  calculation of the interactions between baryons while keeping the
  exponential corrections to the L\"uscher eigenvalue relation from
  single hadron masses negligibly small. A Lattice QCD calculation at
  the physical pion mass and in a volume with $m_\pi L=5.8$ is
  predicted to generate finite-volume corrections to the nucleon mass
  of $\delta M_N\sim 100~{\rm keV}$, smaller than the typical nuclear
  excitation energies found in light nuclei.

\item The contributions from kaon and $\eta$ loops to the volume
  dependence of the nucleon mass are small in ${\rm SU(3)}_L\otimes
  {\rm SU(3)}_R$ HB$\chi$PT, and therefore a relatively stable
  determination of $|g_{\Delta N\pi}|/g_A$ has been found, which can
  be systematically improved by working at higher orders in
  HB$\chi$PT.  Given the size of the finite-volume contributions to
  the masses of the hyperons from kaon and $\eta$ loops, we have not
  made significant determinations of the hyperon axial coupling
  constants in the two-flavor chiral expansion.  However, Lattice QCD calculations at lighter pion masses
  should enable a determination of these couplings.

\item The GMO relation is found to exhibit substantial volume
  dependence, with the relation changing sign at $m_\pi L\sim 5.2$ with
  $m_\pi\sim 390~{\rm MeV}$. Perhaps this should not be surprising given the
  delicate cancellations that occur between the baryon masses to leave
  a quantity that depends only upon SU(3)-breaking in the {\bf
    27}-dimensional irreducible representation and 
vanishes in the large-$N_c$ limit as $1/N_c^2$ compared to the individual baryon masses.

\end{itemize}

\vskip0.2in

\noindent We thank G.~Colangelo for valuable conversations, K.~Roche
for computing resources at ORNL NCCS and R.~Edwards and B.~Joo for
developing QDP++, Chroma~\cite{Edwards:2004sx} and production.  We
acknowledge computational support from the USQCD SciDAC project, NERSC
(Office of Science of the DOE, Grant No.~DE-AC02-05CH11231), the UW
HYAK facility, Centro Nacional de Supercomputaci\'on (Barcelona,
Spain), LLNL, the Argonne Leadership Computing Facility at Argonne
National Laboratory (Office of Science of the DOE, under contract
No.~DE-AC02-06CH11357), and the NSF through Teragrid resources
provided by TACC and NICS under Grant No.~TG-MCA06N025.  SRB was
supported in part by the NSF CAREER Grant No.~PHY-0645570.  The Albert
Einstein Center for Fundamental Physics is supported by the
“Innovations- und Kooperationsprojekt C-13” of the “Schweizerische
Universit\"atskonferenz SUK/CRUS”.  The work of EC and AP is supported
by the contract FIS2008-01661 from MEC (Spain) and FEDER.  AP
acknowledges support from the RTN Flavianet MRTN-CT-2006-035482 (EU).
BJ was also supported by DOE Grants, No.~DE-FC02-06ER41440 and
No.~DE-FC02-06ER41449 (SciDAC USQCD).  H-WL and MJS were supported in
part by the DOE Grant No.~DE-FG03-97ER4014.  WD and KO were supported
in part by DOE Grants No.~DE-AC05-06OR23177 (JSA) and
No.~DE-FG02-04ER41302.  WD was also supported by DOE OJI Grant
No.~DE-SC0001784 and Jeffress Memorial Trust, Grant No.~J-968.  KO was
also supported in part by NSF Grant No.~CCF-0728915 and DOE OJI Grant
No.~DE-FG02-07ER41527.  AT was supported by NSF Grant No.~PHY-0555234
and DOE Grant No.~DE-FC02-06ER41443.  The work of TL was performed
under the auspices of the U.S.~Department of Energy by LLNL under
Contract No.~DE-AC52-07NA27344 and the UNEDF SciDAC Grant
No.~DE-FC02-07ER41457.  The work of AWL was supported in part by the
Director, Office of Energy Research, Office of High Energy and Nuclear
Physics, Divisions of Nuclear Physics, of the U.S. DOE under Contract
No.~DE-AC02-05CH11231.


\end{document}